\documentclass[number,twocolumn,5p]{elsarticle}
\usepackage[binary-units=true,separate-uncertainty=true]{siunitx}
\usepackage{lineno,hyperref}
\usepackage{tikz}
\usepackage{amssymb,amsmath,mathtools}
\usepackage{caption}
\usepackage{xspace}
\usepackage{booktabs}
\usepackage{textcomp}
\usepackage[cachedir=minted-cache,frozencache=true,section]{minted}
\setminted[TOML]{frame=single, framesep=2mm, baselinestretch=1.2, bgcolor=white}

\usepackage[percent]{overpic}
\usepackage[list=true,listformat=simple,margin=10pt]{subcaption}
\usepackage{hepnames}

\hypersetup{ 
    colorlinks=true,
    urlcolor=blue,
    linkcolor=blue
}

\journal{NIM A}

\newcommand{\Ap}{\texorpdfstring{\ensuremath{\mathrm{Allpix}^2}}{Allpix\textasciicircum 2}\xspace}

\newcommand{\um}{µm} 

\newcommand{\umsq}{\si{\um \squared}} 

\begin{document}
\begin{frontmatter}
  \title{Simulating Monolithic Active Pixel Sensors:\\ A Technology-Independent Approach Using Generic Doping Profiles}


\author[desy]{Håkan Wennlöf\corref{mycorrespondingauthor}}
\cortext[mycorrespondingauthor]{Corresponding author}
\ead{h.wennlof@cern.ch}

\author[cern]{Dominik Dannheim}
\author[desy,bonn]{Manuel Del Rio Viera}
\author[cern,giessen]{Katharina Dort}
\author[desy]{Doris Eckstein}
\author[desy]{Finn Feindt}
\author[desy]{Ingrid-Maria Gregor}
\author[desy]{Lennart Huth}
\author[desy,hamburg]{Stephan Lachnit}
\author[desy,bonn]{Larissa Mendes}
\author[desy,wuppertal]{Daniil Rastorguev}
\author[desy,bonn]{Sara Ruiz Daza}
\author[desy]{Paul Schütze}
\author[desy,bonn]{Adriana Simancas}
\author[cern]{Walter Snoeys}
\author[desy]{Simon Spannagel}
\author[desy]{Marcel Stanitzki}
\author[campinasFull]{Alessandra Tomal}
\author[desy]{Anastasiia Velyka}
\author[desy,bonn]{Gianpiero Vignola}


\affiliation[desy]{
    organization={Deutsches Elektronen-Synchrotron DESY},
    addressline={Notkestr. 85},
    city={22607 Hamburg},
    country={Germany}
}
\affiliation[cern]{
    organization={CERN},
    city={Geneva},
    country={Switzerland}
}
\affiliation[campinasFull]{
    organization={University of Campinas},
    addressline={Cidade Universitaria Zeferino Vaz}, 
    postcode={13083-970},
    city={Campinas},
    country={Brazil}
}

\fntext[bonn]{Also at University of Bonn, Germany}
\fntext[giessen]{Also at University of Giessen, Germany}
\fntext[hamburg]{Also at University of Hamburg, Germany}
\fntext[wuppertal]{Also at University of Wuppertal, Germany}

\begin{abstract}

The optimisation of the sensitive region of CMOS sensors with complex non-uniform electric fields requires precise simulations, and this can be achieved by a combination of electrostatic field simulations and Monte Carlo methods. This paper presents the guiding principles of such simulations, using a CMOS pixel sensor with a small collection electrode and a high-resistivity epitaxial layer as an example.
The full simulation workflow is described, along with possible pitfalls and how to avoid them.
For commercial CMOS processes, detailed doping profiles are confidential, but the presented method provides an optimisation tool that is sufficiently accurate to investigate sensor behaviour and trade-offs of different sensor designs without knowledge of proprietary information.

The workflow starts with detailed electric field finite element method simulations in TCAD, using generic doping profiles. Examples of the effect of varying different parameters of the simulated sensor are shown, as well as the creation of weighting fields, and transient pulse simulations. The fields resulting from TCAD simulations can be imported into the \Ap{} Monte Carlo simulation framework, which enables high-statistics simulations, including modelling of stochastic fluctuations from the underlying physics processes of particle interaction. Example Monte Carlo simulation setups are presented and the different parts of a simulation chain are described.

Simulation studies from small collection electrode CMOS sensors are presented, and example results are shown for both single sensors and multiple sensors in a test beam telescope configuration. The studies shown are those typically performed on sensor prototypes in test beam campaigns, and a comparison is made to test beam data, showing a maximum deviation of 4\% and demonstrating that the approach is viable for generating realistic results.



\end{abstract}


\end{frontmatter}

\setcounter{tocdepth}{2}
\tableofcontents

\section{Introduction}
\label{sec::introduction}

Monolithic active pixel sensors (MAPS) produced using commercial CMOS imaging processes are attractive in a particle physics context, as they allow for a reduced material budget and reduction of production complexity compared to most hybrid sensors. The use of commercial processes enables relatively cheap large-scale production of sensors, but it also means that precise information of the manufacturing process may not be publicly available. Predictions of sensor behaviour are thus difficult to make, as the detailed electric field configuration in the sensitive material is highly dependent on the extent and concentration of different doping regions in the silicon.

By utilising a quadruple-well technology (providing n-wells, p-wells, and deep n-wells and p-wells)~\cite{quadWell}, MAPS can be constructed with a small collection electrode, which reduces sensor capacitance and improves signal-to-noise ratio while reducing power consumption compared to sensors with larger collection electrodes. However, designs with a small collection electrode lead to a highly non-linear electric field in the pixels, further complicating sensor behaviour prediction. As prototype sensor submissions and investigations are expensive and take a long time, simulations of sensor behaviour are becoming more and more important to gain insight and speed up the design process.

This paper aims to demonstrate that by making simple assumptions and performing simulations based on the fundamental principles of silicon detectors and using generic doping profiles, performance parameters of MAPS can be inferred and compared for different sensor geometries. This will be done in context of simulations performed for the Tangerine project~\cite{tangerine1, tangerine2, tangerine3, tangerine4} and in collaboration with the CERN~EP~R\&D programme on technologies for future experiments~\cite{CERN_EPRnD}, but the methodology described is useful for many different silicon sensor simulations. The described method thus constitutes a toolbox for performing similar simulations, useful in extracting a realistic description of sensor behaviour without knowledge of proprietary information.
The efficacy of combining detailed electric field simulations with high-statistics Monte Carlo simulations has been previously demonstrated for similar silicon sensors~\cite{tcadMCcombination, transientMCstudies}, and the process described in this paper is general and applicable in multiple different cases. For sensors with non-linear electric fields, simulations like those presented here are useful for gaining a deeper understanding of the sensor performance. It is important to note that the presented simulations by no means capture the intricacies of CMOS imaging processes, but merely describe the larger features of the sensor required to model an accurate signal response.




\subsection*{Paper outline}

The paper aims to show guiding principles for performing detailed Monte Carlo simulations of silicon sensors, using basic assumptions and estimates.
In Section~\ref{sec::layout}, a general MAPS layout is described, and assumptions of the geometry and doping types used in the simulations are discussed. Then, doping concentration and electric field finite-element simulations using technology computer-aided design (TCAD) are presented in Section~\ref{sec::tcad}, with a detailed simulation procedure using generic doping profiles and assumptions based on the physics of a semiconductor sensor. Monte Carlo simulations using the \Ap{}~\cite{apsq} framework with electric fields and doping profiles from TCAD are described in detail in Section~\ref{sec::mc}, going through the simulation setup step by step. Some example results of the high-statistics Monte Carlo simulations carried out in the Tangerine project are shown in Section~\ref{sec::studies}, including in-pixel studies, transient current pulses, and simulation results from a multi-sensor setup. Finally, example comparisons of simulation results to data are shown in Section~\ref{sec::dataComparisons}.











\section{General layout and assumptions}
\label{sec::layout}

The MAPS simulated in this work consist of a high-resistivity p-doped epitaxial layer grown on an electronics-grade p-doped silicon substrate, with implanted doping wells in the epitaxial layer. The doping wells function as collection electrodes and/or shielding for the in-pixel electronics. As the epitaxial layer in the sensors is relatively thin (of the order of 10~\um{}), the sensor thickness is dominated by the substrate. Most of the visible signal is generated in the epitaxial layer, and the substrate is thus often thinned after sensor production, down to a total sensor thickness below 50~\um{}. 

The doping concentrations used in the simulations presented here are not values from a specific sensor or technology, but approximations derived from previous studies~\cite{thesis_hasenbichler, thesis-jacobus}. The substrate is assumed to have a doping concentration of $1 \cdot 10^{19}$~cm$^{-3}$, the epitaxial layer approximately $3 \cdot 10^{13}$~cm$^{-3}$, and doping wells ranging from $1 \cdot 10^{15}$~cm$^{-3}$ to $1 \cdot 10^{19}$~cm$^{-3}$, depending on their purpose.


\subsection{Doping wells}

In the centre of the pixels, an n-doped well is located. This well is simulated with a doping concentration of approximately $10^{19}$~cm$^{-3}$, is positively biased, and serves as the collection electrode. The size of the well is of the order of 1~\um{} across, and it has a square shape (when viewed from above). 

Surrounding the collection electrode is an opening without wells, and then a square deep p-well (for square pixels). This deep p-well is assumed to have a doping concentration of approximately $10^{15}$~cm$^{-3}$. While none of the doping wells in the simulations contain any internal structure or electronics, the main purpose of the deep p-well in a physical sensor is to contain both NMOS transistors and internal n-wells that contain PMOS transistors. In this way, full CMOS front-end electronics are possible in the pixels. The deep p-well shields the electronics from the sensitive region, which ensures that the n-well collection electrode is the only node electrons drift to. This also allows for a higher bias voltage to be applied to the sensor bulk without damaging the electronics. 

The extent and shape of the wells can be used to shape the electric field, and may significantly affect the charge collection properties of a sensor. For example, the effect of changing the size of the opening between the collection electrode and the p-well is explored in the work presented here. 



\subsection{Contacts and biasing}

Ohmic contacts are essential to provide bias voltages to and extract signals from a sensor, and they are achieved by having a highly-doped region in the silicon next to the metal contact. In the sensors presented here there are contacts to the collection electrode, the p-well, and the sensor substrate. In a physical sensor, the biasing of the collection electrode and the p-well are done via metal contacts, and the substrate is biased through surface contacts outside the pixel matrix. In the simulations however, the substrate is instead biased via a contact directly on the backside as guard rings and sensor edge structures are not included.

The collection electrode in the simulations presented here has a positive bias voltage of 1.2~V, whereas the p-well and substrate have bias voltages between 0~V and $-6$~V. The p-well and the substrate are commonly biased to the same voltage, but can also be biased separately.
The bias voltage that can be applied to the p-well in a physical sensor is limited by the behaviour of the NMOS transistors, as their characteristics will change and their function may cease at high voltages.


\subsection{Rectangular and hexagonal pixels}
\label{sec::rectAndHex}

Three main designs are simulated and tested in this work, labelled \emph{standard} layout~\cite{standardLayout}, \emph{n-blanket} layout~\cite{nblanketLayout}, and \emph{n-gap} layout~\cite{ngapLayout}. The \emph{standard} layout is similar to what is used in the ALPIDE sensor~\cite{alpide}, which is a MAPS used in the ALICE experiment since the ITS2 upgrade, developed in a 180\,nm CMOS imaging process. This layout has a small n-type collection electrode in a p-type epitaxial layer, and depletion grows in an approximately spherical shape from this pn-junction. This depleted region tends to not extend fully below the p-well.
The \emph{n-blanket} layout introduces a blanket layer of n-doped silicon in the p-type epitaxial layer, which forms a deep planar pn-junction, allowing for full depletion of a pixel. This layout leaves an electric field minimum under the p-well at pixel edges and corners, however, leading to slow charge collection and possible efficiency loss in these regions. This can be amended by introducing a vertical pn-junction near the edge~\cite{ngapLayout}. One way to achieve this is by leaving a gap in the blanket of n-doped silicon under the p-well, which is done in the \emph{n-gap} layout. As a pn-junction is thus formed near the pixel edges, a lateral electric field is formed there, pushing charges toward the pixel centre.
The \emph{n-blanket} and \emph{n-gap} layout modifications were originally developed for a 180\,nm CMOS imaging process, but similar developments have been implemented in a 65\,nm CMOS imaging process as well~\cite{SnoeysPIXEL}.

The three sensitive volume designs described above are applied in both rectangular and hexagonal pixel geometries in the work presented here. Using a hexagonal geometry decreases the amount of shared charge in pixel corners, as a pixel only shares a corner with two neighbours rather than three. It also reduces the maximum distance of the pixel boundary from the centre compared to rectangular geometries, while maintaining the same area. Hexagonal pixel shapes thus reduce regions with low electric fields at pixel edges. The maximum distance in a square grid between the pixel corner and the collection electrode is reduced by 12\% for the same pixel area on a hexagonal grid~\cite{fastpix}. As the pixel corner region and p-well edges have a larger opening angle, the electric fields there differ significantly compared to rectangular pixel geometries. 
The distance between collection electrodes is also the same for all adjacent pixels in a hexagonal configuration.








\section{Finite element method simulations}
\label{sec::tcad}

Technology Computer-Aided Design (TCAD) is a simulation tool that uses finite element methods to model semiconductor devices in 2D and 3D. In each node of a created mesh, calculations of the electrostatic potential and other properties are carried out by solving Poisson’s and carrier continuity equations.
This work implements TCAD simulations with generic doping profiles to study effects of layout design on the electric field of CMOS sensors, and the presented simulations have been performed in 3D with Sentaurus TCAD from Synopsys~\cite{sentaurusTCAD}. 

The body of the sensor is created initially from simple geometrical shapes, which are then adjusted to represent the different studied layouts. To obtain insight into the effects of the adjustments, iterations of layout modifications and simulation evaluations are performed. To refine the simulations, the following principles are taken into account, to ensure a physically realistic and operational sensor:

\begin{itemize}
    \item The doping concentrations in the interfaces between different doping structures (n- and p-wells, epitaxial layer/substrate) should be diffused to avoid unphysical effects, such as abrupt changes in doping concentration and the corresponding electric field. 
    \item The p-well must shield its content from the electric field in the active sensor area; the doping must thus be sufficient for the depleted region to not penetrate deep into the well.
    \item The charge carriers generated in the sensor volume have to reach the collection electrode.
    \item There should be no conductive channel between different biased structures, i.e. punch-through in the sensor should be avoided.
    \item The limitations on the operating voltages of the transistors in the readout electronics of a physical sensor should be respected.
\end{itemize}

It should be noted that no internal structure of the doping wells is simulated in this work, so no readout electronics are included. The basic principles needed to protect them (outlined above) are included, however, in order to have a realistic sensor description.

\subsection{Simulation workflow}
%
%
%
%
%
%
%
%
The simulation process starts with defining the sensor geometry using the \textit{Sentaurus Structure Editor} tool. The materials of structures are defined together with their shapes, and the materials used in these simulations are aluminium for the electrodes, silicon oxide for the dielectric material, and silicon for the sensor bulk. In order to apply electrical boundary conditions, it is necessary to define interface regions called \textit{contacts}, which correspond to physical contacts between the electrodes and the silicon bulk. For simplicity, only the top part of the sensor corresponding to the region taken up by the epitaxial layer and its interface to the substrate is used in the TCAD simulations presented here.

In addition to the geometrical definition of the sensor, doping profiles and meshing parameters
can be incorporated for different parts of the structure. 
This has been done for the epitaxial layer, the collection implant, and the p-well. Refinement/evaluation windows are defined to place the corresponding doping profiles.
Analytical doping profiles are used to emulate the well structures; the wells are formed with an error function distribution in depth and a Gaussian distribution laterally. This emulates a dopant diffusion region with an extent of $0.3-0.4$~\um{} in the depth direction, and an extent of $0.4-0.5$~\um{} in the lateral direction.
Flat doping concentrations are used across the full well structures, which is a simplifying assumption, but deemed sufficient for understanding the physical behaviour of the signal formation in a sensor.
In the interface regions between the silicon bulk and the electrodes, it is necessary to add a highly doped region to create an Ohmic contact.

Once doping regions and profiles have been defined, the refinement parameters for the mesh are established. This includes minimum and maximum mesh sizes, and the refinement function. A fine mesh provides more accurate results. However, since the number of calculation nodes can be substantial, the simulations take longer and can make subsequent calculations intractable with a finer mesh. This can be addressed by using an unstructured mesh that is refined only in certain regions. In this work, the mesh refinement is a function of the doping gradient, meaning that the mesh will be finer in places where there are significant changes in the doping concentration, e.g. at the edges of the well structures.

When the geometry has been built and the mesh is defined, the device simulations can be performed with the \textit{Sentaurus Device} tool. Simulations have been performed in both quasistationary mode, to obtain electric fields, and in transient mode to obtain the signal response to a charged particle traversing the sensor. The grid file created using the Structure Editor is imported into Sentaurus Device and the contacts are identified. Physical properties and solver properties are defined, as well as the boundary conditions of the simulation. The results of the quasistationary simulations are voltage-dependent curves and several electrical properties within the 3D volume.
The TCAD simulations presented here were performed with a collection electrode bias voltage of 1.2~V and a p-well and substrate bias voltage of $-5$~V, unless otherwise stated.
Properties studied in this work are the electric field magnitude, lateral electric field, charge current density, and depleted volume, shown in Sections~\ref{sec::dopConcScans}, \ref{sec::featureScans}, and \ref{sec::hex}.
The results of transient simulations are time-dependent curves and snapshots of the perturbed electrical properties within the 3D volume, shown in Section~\ref{sec::tcadTransient} and compared to results of Monte Carlo simulations in Section~\ref{sec::apTransient}.

\subsection{Substrate diffusion simulation}
\label{sec::tcadDiffusion}

\begin{figure}[h]
\centering
        \includegraphics[width=.85\linewidth]{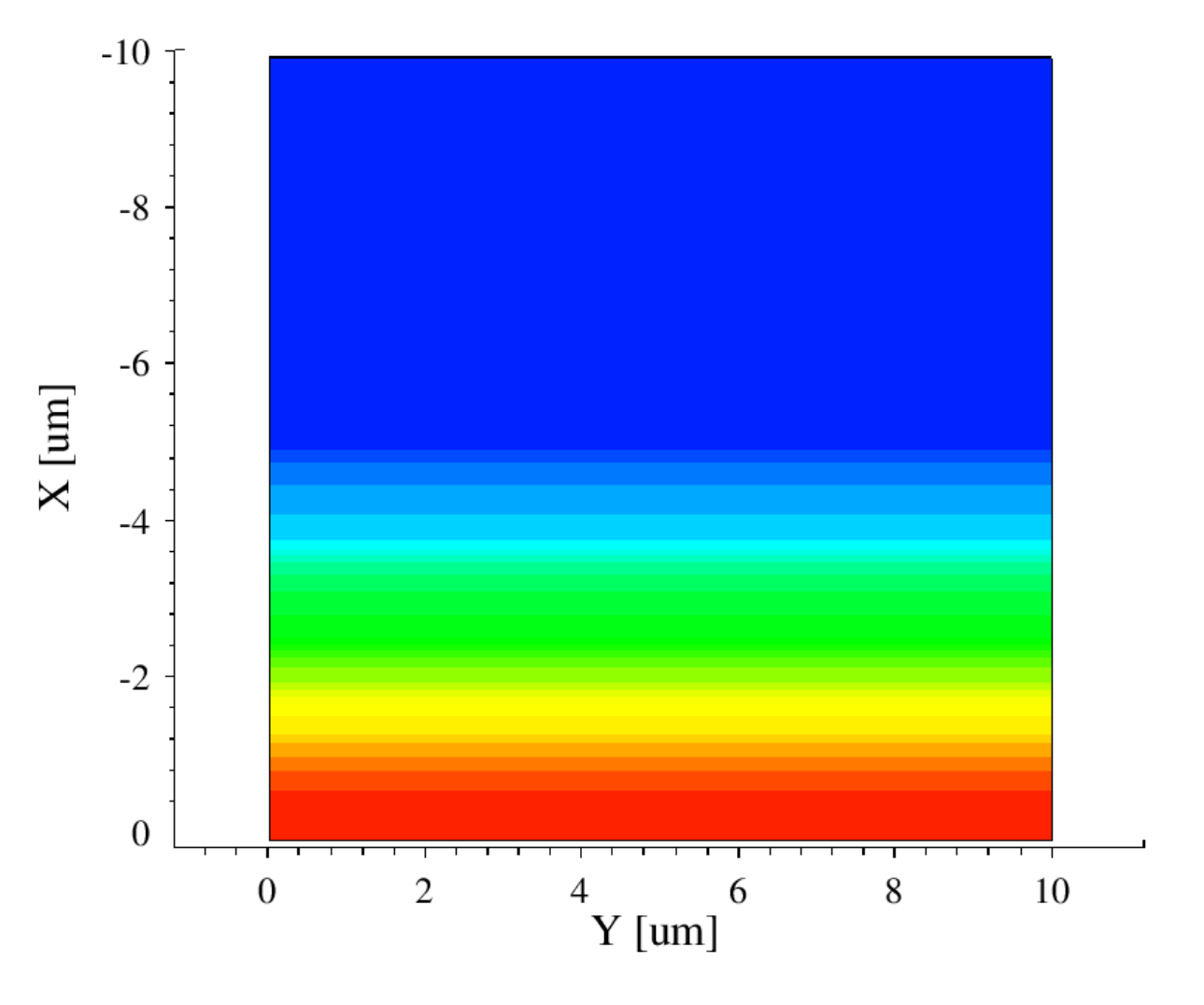}
        \caption{Simulated diffusion from the substrate to the epitaxial layer. Red indicates a higher doping concentration, and blue a lower one (by approximately six orders of magnitude).}
	\label{fig::PR}
\end{figure}

As the substrate is not expected to directly contribute to the electric field in the sensor, the TCAD simulations only include the epitaxial layer and the interface region between the epitaxial layer and the substrate. The only possible influence stems from the diffusion of p-dopants from the highly-doped substrate to the lower-doped epitaxial layer. During the process of semiconductor fabrication, a high difference in doping concentration and the high temperatures the device is exposed to is expected to produce a significant diffusion region at the interface of the epitaxial layer and the substrate.
To simulate the diffusion of dopants from the substrate to the epitaxial layer, simulations of a sensor production process were performed using the \textit{Sentaurus Process} tool of the Synopsys TCAD framework.
The simulation includes 10~minutes of a chemical vapour deposition (CVD) process on the substrate with a temperature of 1050\,$^\circ$C, which results in 10~\um{} of epitaxial silicon~\cite{thesis_velyka}.
The assumed doping concentration of the substrate in this simulation is $1 \cdot 10^{19}$~cm$^{-3}$, and $3 \cdot 10^{13}$~cm$^{-3}$ for the epitaxial layer~\cite{thesis_hasenbichler}.
All the implanted structures need to go through an annealing procedure to electronically activate the implanted ions. 
The simulated activation process involves heating of the structure to a temperature of 1100\,$^\circ$C for 240~min.
The resulting structure is converted to a one-dimensional doping profile for the epitaxial layer, which can be seen in Figure~\ref{fig::PR}. This profile is then imported for the epitaxial layer in further simulations. 

\subsection{Impact of sensor doping concentration}
\label{sec::dopConcScans}

Doping concentration is an important parameter in the design of silicon sensors, especially for the structures that constitute the junctions that shape the electric field inside the sensor. Studies conducted on changing the assumed doping concentrations on the p- and n-wells described in Section~\ref{sec::layout} for the \emph{standard} layout by an order of magnitude did not significantly change the electric field forming in the sensor. Through the studies, a value of the p-well doping concentration of $5 \cdot 10^{15}$~cm$^{-3}$ was selected as a baseline assumption for studies of the \emph{n-blanket} layout, where the pn-junction is larger and the impact presumed greater.
Studies of the impact of altering the doping concentrations of both the n-blanket and the p-well were performed using this layout, while keeping the epitaxial layer doping concentration fixed at $3 \cdot 10^{13}$~cm$^{-3}$. The limit values of the study were selected as the minimum value that would produce an effect on the depleted volume and the maximum value that would start to have an adverse effect on the surrounding doping structures (e.g. depletion deep into the p-well). The effects of altering the doping concentrations are studied by observing plots of the electric field magnitude and the depleted volume.

\begin{figure}[h]
\centering
    \begin{subfigure}{.33\columnwidth}
        \includegraphics[width=1.0\linewidth]{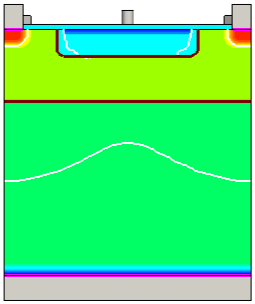}
        \caption{$1 \cdot 10^{14}$ cm$^{-3}$}
        \label{fig::blk_1e14}
    \end{subfigure}%
    \begin{subfigure}{.33\columnwidth}
        \includegraphics[width=1.0\linewidth]{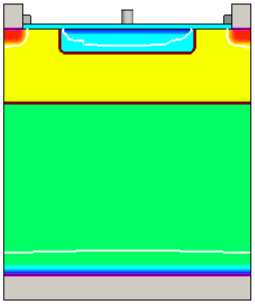}
        \caption{$9 \cdot 10^{14}$ cm$^{-3}$}
        \label{fig::blk_9e14}
    \end{subfigure}%
    \begin{subfigure}{.33\columnwidth}
        \includegraphics[width=1.0\linewidth]{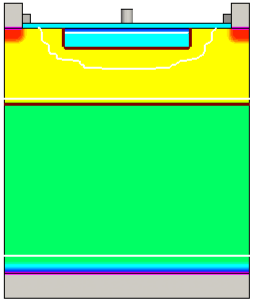}
        \caption{$4 \cdot 10^{15}$ cm$^{-3}$}
        \label{fig::blk_4e15}
    \end{subfigure}
    \caption{Different doping concentrations of the n-blanket for a 10~\um{} pitch sensor in the \emph{n-blanket} layout, represented by different colours. The colour scale corresponds to the total doping concentration (with the highest p-doped regions being blue and the highest n-doped regions red), the brown line indicates the location of a pn-junction, and the white lines delimit the depleted volume.}
	\label{fig::blk_scan}
\end{figure}

The doping concentration of the n-blanket was studied with a fixed value of the p-well concentration of $5 \cdot 10^{15}$~cm$^{-3}$, varying the n-blanket concentration between $1 \cdot 10^{14}$~cm$^{-3}$ and $4 \cdot 10^{15}$~cm$^{-3}$, shown in Figure~\ref{fig::blk_1e14} and Figure~\ref{fig::blk_4e15} respectively. The figures show a cross-section of a pixel, with half a collection electrode in the upper corners and the p-well in the centre of the image. Introduction of a low-doped n-type blanket implant creates a large pn-junction in the sensor, but a bulbous shape of the depletion region below the collection electrodes is still present. 
The highly-doped n-type blanket implant is not fully depleted, which leads to a conductive path being present between the two collection electrodes, and thus a non-functioning sensor. This can be observed by the shape of the depletion line.
The doping concentration selected for the n-blanket for use in the final simulations was $9 \cdot 10^{14}$~cm$^{-3}$, shown in Figure~\ref{fig::blk_9e14}, as it provides a good depletion at the bottom of the sensor, and the least depletion intrusion in the p-well.

\begin{figure}[h]
\centering
    \begin{subfigure}{.33\columnwidth}
        \includegraphics[width=1.0\linewidth]{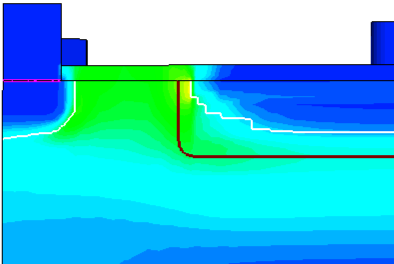}
        \caption{$5 \cdot 10^{15}$~cm$^{-3}$}
        \label{fig::pwell_5e15}
    \end{subfigure}%
    \begin{subfigure}{.33\columnwidth}
        \includegraphics[width=1.0\linewidth]{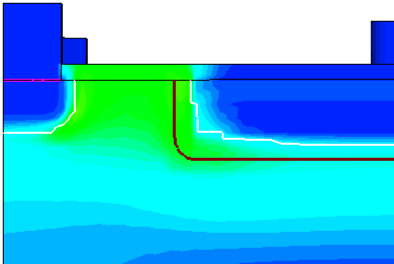}
        \caption{$1 \cdot 10^{16}$~cm$^{-3}$}
        \label{fig::pwell_1e16}
    \end{subfigure}%
    \begin{subfigure}{.33\columnwidth}
        \includegraphics[width=1.0\linewidth]{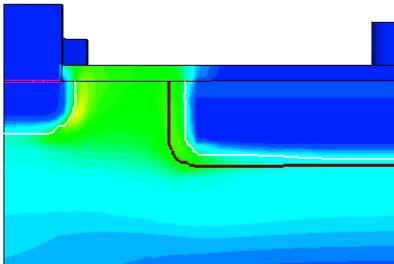}
        \caption{$5 \cdot 10^{16}$~cm$^{-3}$}
        \label{fig::pwell_5e16}
    \end{subfigure}
    \caption{Electric field magnitude for three different doping concentrations of the p-well, for a 10~\um{} pitch sensor. Close-up to the corner of the p-well. The brown line indicates the location of a pn-junction and the white line delimits the depleted volume. The electric field magnitude is given by the colour scale, where a dark blue colour indicates an electric field magnitude of zero.}
	\label{fig::pwell_scan}
\end{figure}

With the value of the n-blanket doping concentration fixed to $9 \cdot 10^{14}$~cm$^{-3}$, the p-well doping concentration was varied. The study started with the value employed in the n-blanket concentration study.
Figure~\ref{fig::pwell_scan} displays the electric field magnitude at a close-up to the edge of the p-well, with different tested doping concentrations. For a doping concentration of $5 \cdot 10^{15}$~cm$^{-3}$, there is a relatively large volume of the p-well that has a non-zero electric field, which is undesirable as that may influence the in-pixel electronics contained there in a physical sensor. A higher doping concentration than the one in the previous study should allow for a better shielding of the p-well, but a too high doping concentration produces a deeper structure than what is desired. Furthermore, Figure~\ref{fig::pwell_1e16} shows a more uniform electric field outside the p-well, when compared to the simulations using the upper and lower tested limit values shown in Figures~\ref{fig::pwell_5e15} and~\ref{fig::pwell_5e16}. The doping concentration of $1 \cdot 10^{16}$~cm$^{-3}$ was thus chosen as the value for the p-well to use in the final simulations.

\subsection{Impact of sensor geometry}
\label{sec::featureScans}

Modifications of the sensor layout can have a significant impact on the strength and extent of the electric field inside the sensor, and on the depleted volume. To investigate this impact, studies have been performed on the size of the p-well opening, which corresponds to the distance between the edge of the collection implant and the edge of the p-well, and on the gap size in the \emph{n-gap} layout. These purely geometrical features are defined by the mask design used in sensor production. The effects are studied by observing plots of the electric field magnitude, the lateral electric field strength, and the depleted volume. In the figures, the former two are represented in colour scale, while the latter one is delimited by a white line. 

\subsubsection{P-well opening}
\label{sec::pwellOpening}

The p-well opening extent was varied from 1~\um{} to 4~\um{}, at a pixel size of $20 \times 20$~\umsq{}. It was observed that increasing the p-well opening creates a stronger lateral electric field and increases the depleted volume in the \emph{standard} layout, on the order of \um{} in both width and depth. A larger depleted volume allows for more charge collection through drift, while a stronger lateral electric field provides a higher drift velocity for the free charge carriers produced in the edges of the pixels. However, increasing the p-well opening means decreasing the p-well size and hence the available space for front-end electronics.
The study was also performed using the \emph{n-blanket} layout. Here, it was observed that the increase in lateral electric field strength and depleted volume was not as significant as in the \emph{standard} layout case. A large p-well opening here leads to a larger undepleted region around the collection electrode, however, which has a negative impact on the sensor capacitance and charge collection behaviour.

\begin{figure}[h]
\centering
    \begin{subfigure}{.5\columnwidth}
        \includegraphics[width=1.0\linewidth]{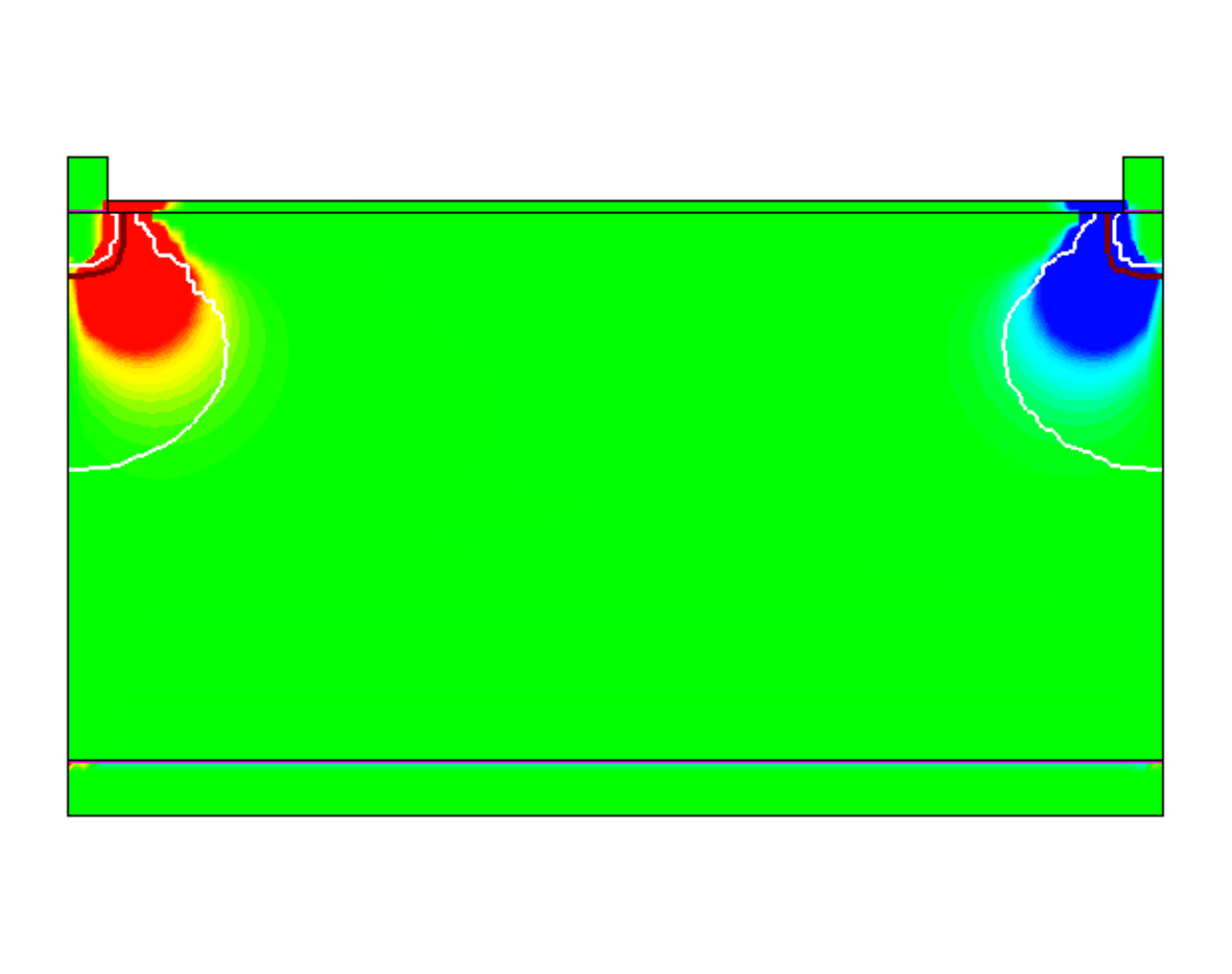}
        \caption{\emph{Standard} layout, 1~\um{} opening}
        \label{fig::std_pwell_op_1um}
    \end{subfigure}%
    \begin{subfigure}{.5\columnwidth}
        \includegraphics[width=1.0\linewidth]{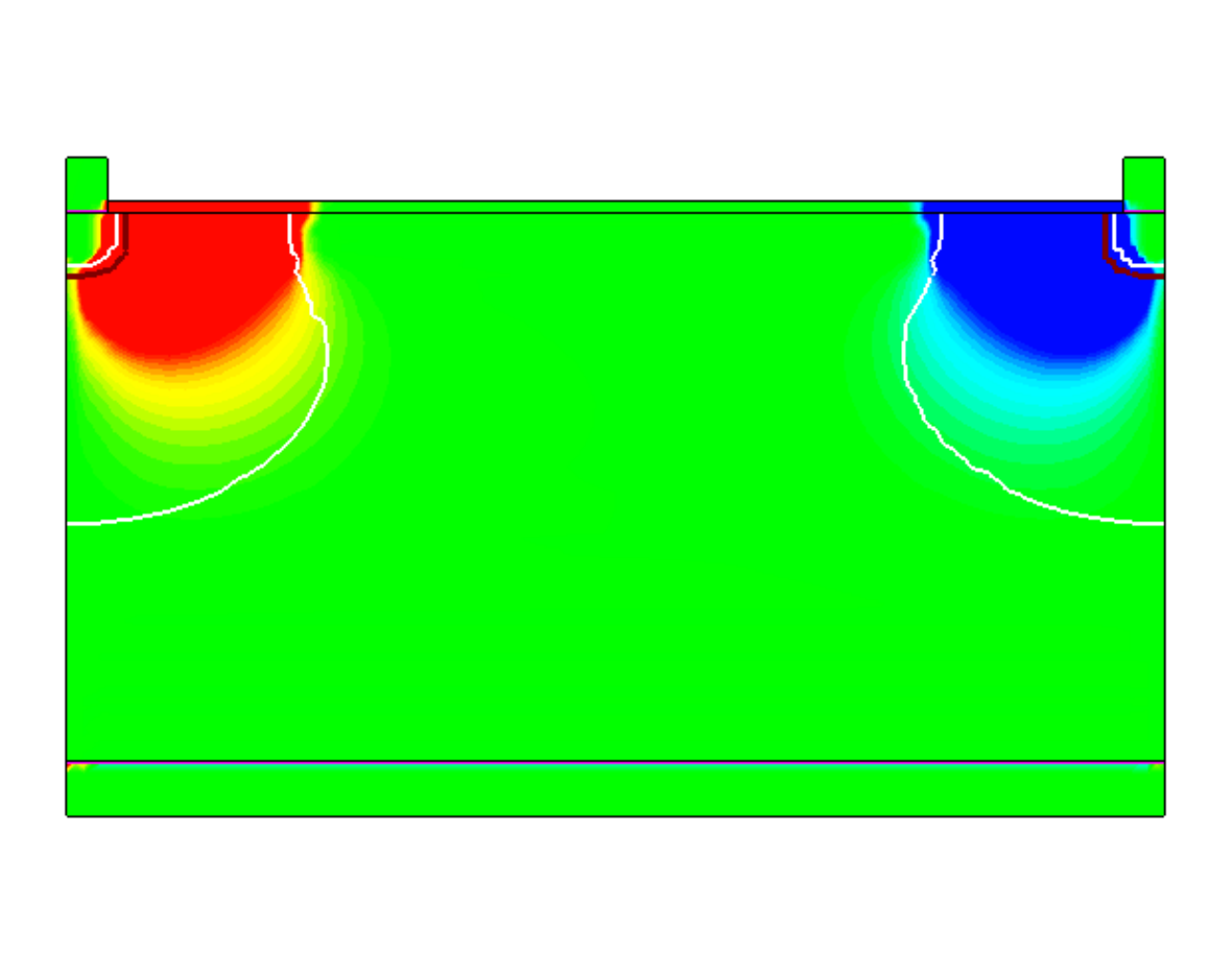}
        \caption{\emph{Standard} layout, 4~\um{} opening}
        \label{fig::std_pwell_op_4um}
    \end{subfigure}
        \begin{subfigure}{.5\columnwidth}
        \includegraphics[width=1.0\linewidth]{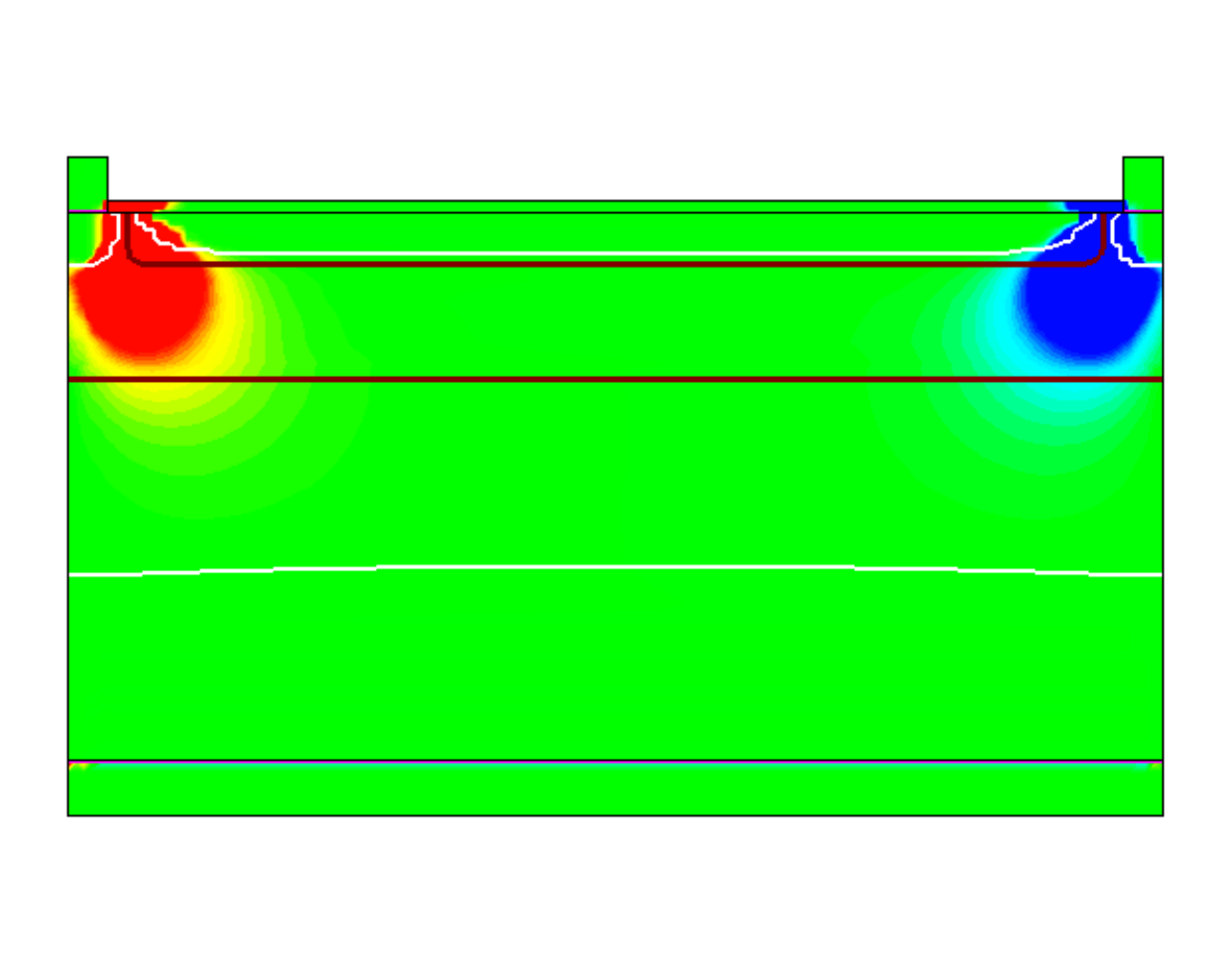}
        \caption{\emph{N-blanket} layout, 1~\um{} opening}
        \label{fig::blk_pwell_op_1um}
    \end{subfigure}%
    \begin{subfigure}{.5\columnwidth}
        \includegraphics[width=1.0\linewidth]{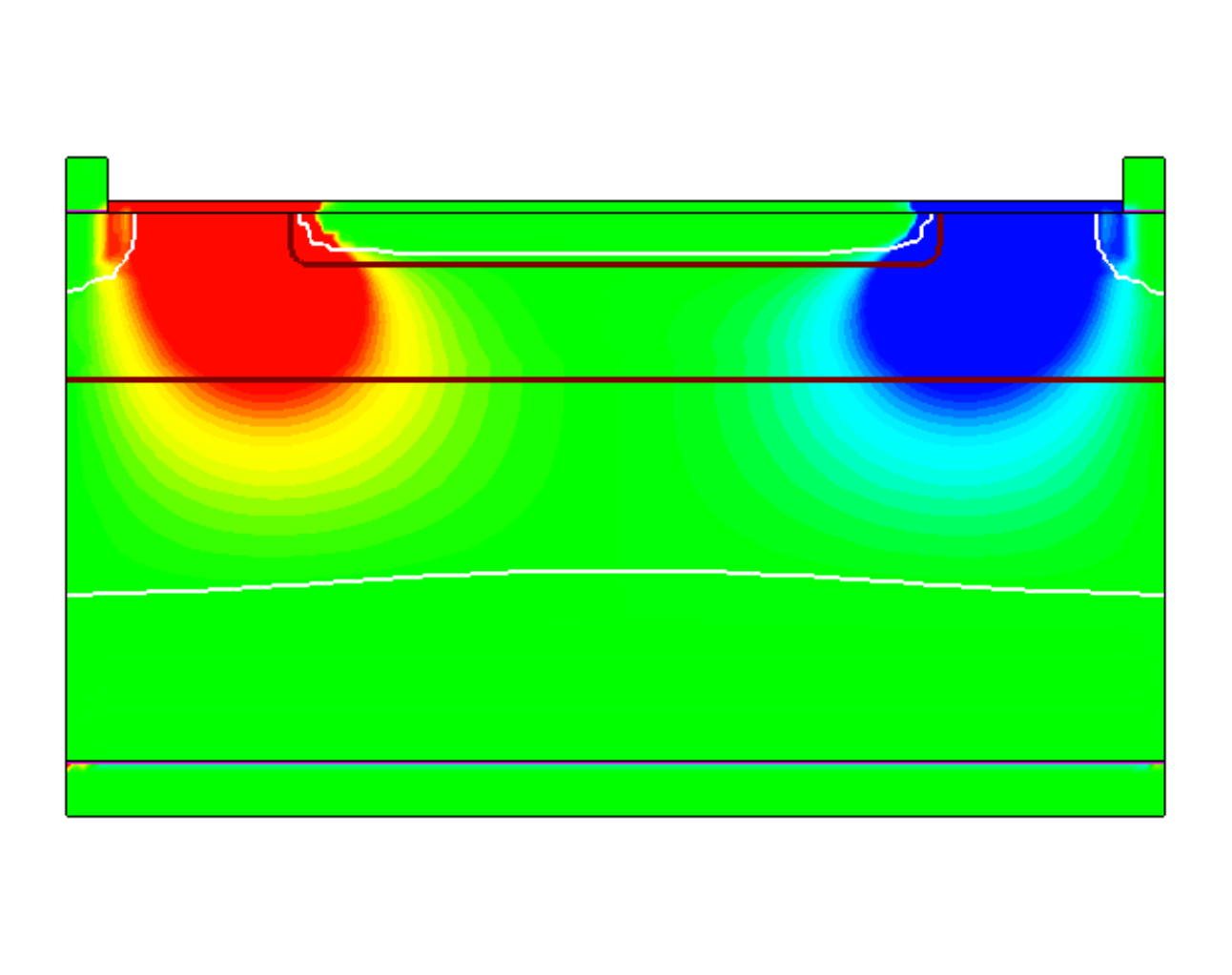}
        \caption{\emph{N-blanket} layout, 4~\um{} opening}
        \label{fig::blk_pwell_op_4um}
    \end{subfigure}
    \caption{Lateral electric field of two p-well openings for 20~\um{} pitch sensors in the \emph{standard} and \emph{n-blanket} layouts. The brown line indicates the location of a pn-junction and the white line the boundaries of the depleted volume.}
	\label{fig::pwell_op_scan}
\end{figure}

A comparison between the lateral electric field and depletion boundary for p-well openings of 1~\um{} and 4~\um{} is shown in Figure~\ref{fig::pwell_op_scan} for both the \emph{standard} and the \emph{n-blanket} layouts.
The region with a strong lateral electric field visibly increases in both layouts as the p-well opening increases.
A larger p-well opening is expected to also directly affect the sensor capacitance, but studies of this have not been carried out. An opening size of 2~\um{} is selected for use in further studies, as a balance between the increased depleted region and the total p-well size.

\subsubsection{Gap size in the n-gap layout}

\begin{figure}[htb]
\centering
    \begin{subfigure}{.5\columnwidth}
        \includegraphics[width=1.0\linewidth]{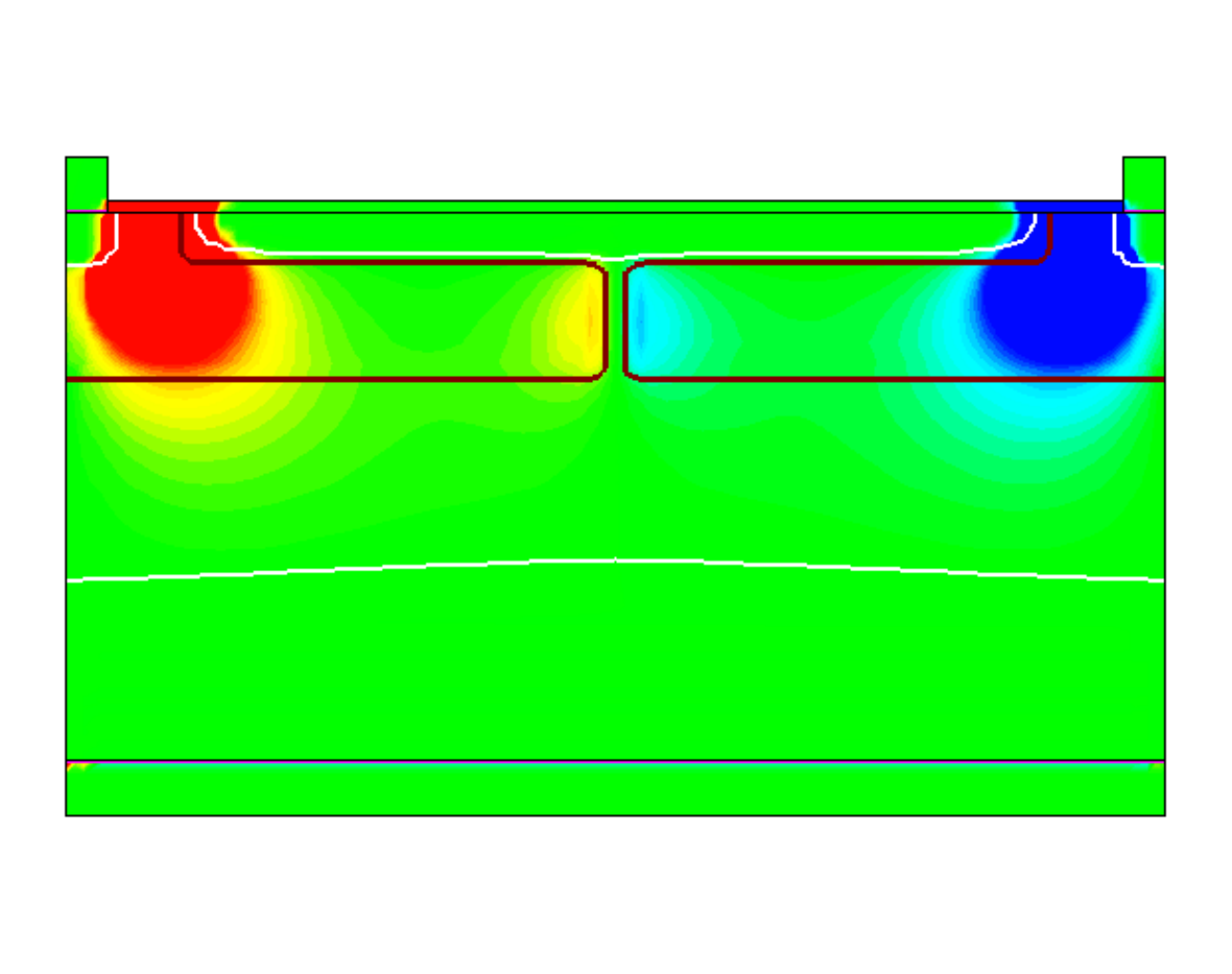}
        \caption{1~\um{} gap}
        \label{fig::gap_1um}
    \end{subfigure}%
    \begin{subfigure}{.5\columnwidth}
        \includegraphics[width=1.0\linewidth]{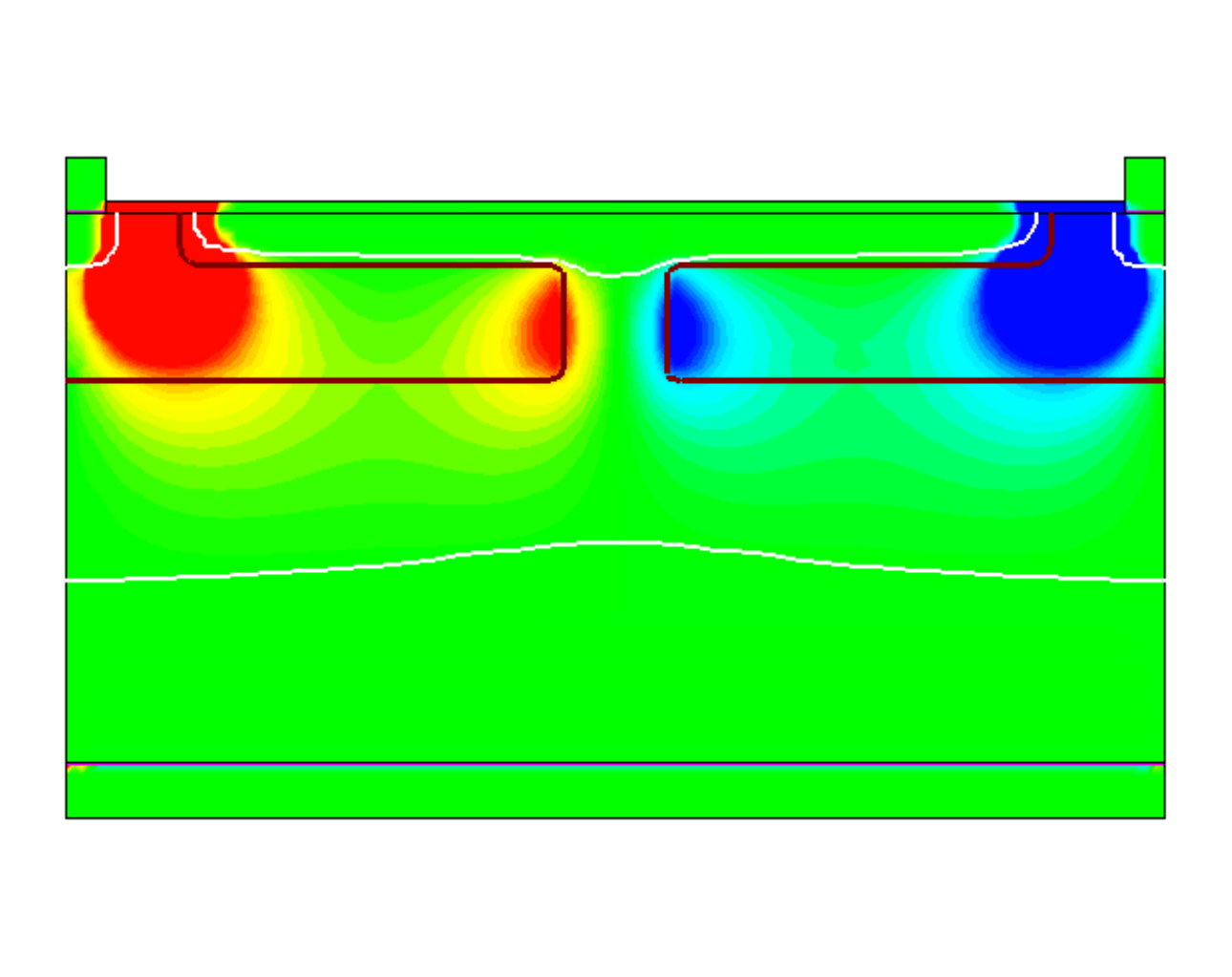}
        \caption{2.5~\um{} gap}
        \label{fig::gap_2.5um}
    \end{subfigure}
    \begin{subfigure}{.5\columnwidth}
        \includegraphics[width=1.0\linewidth]{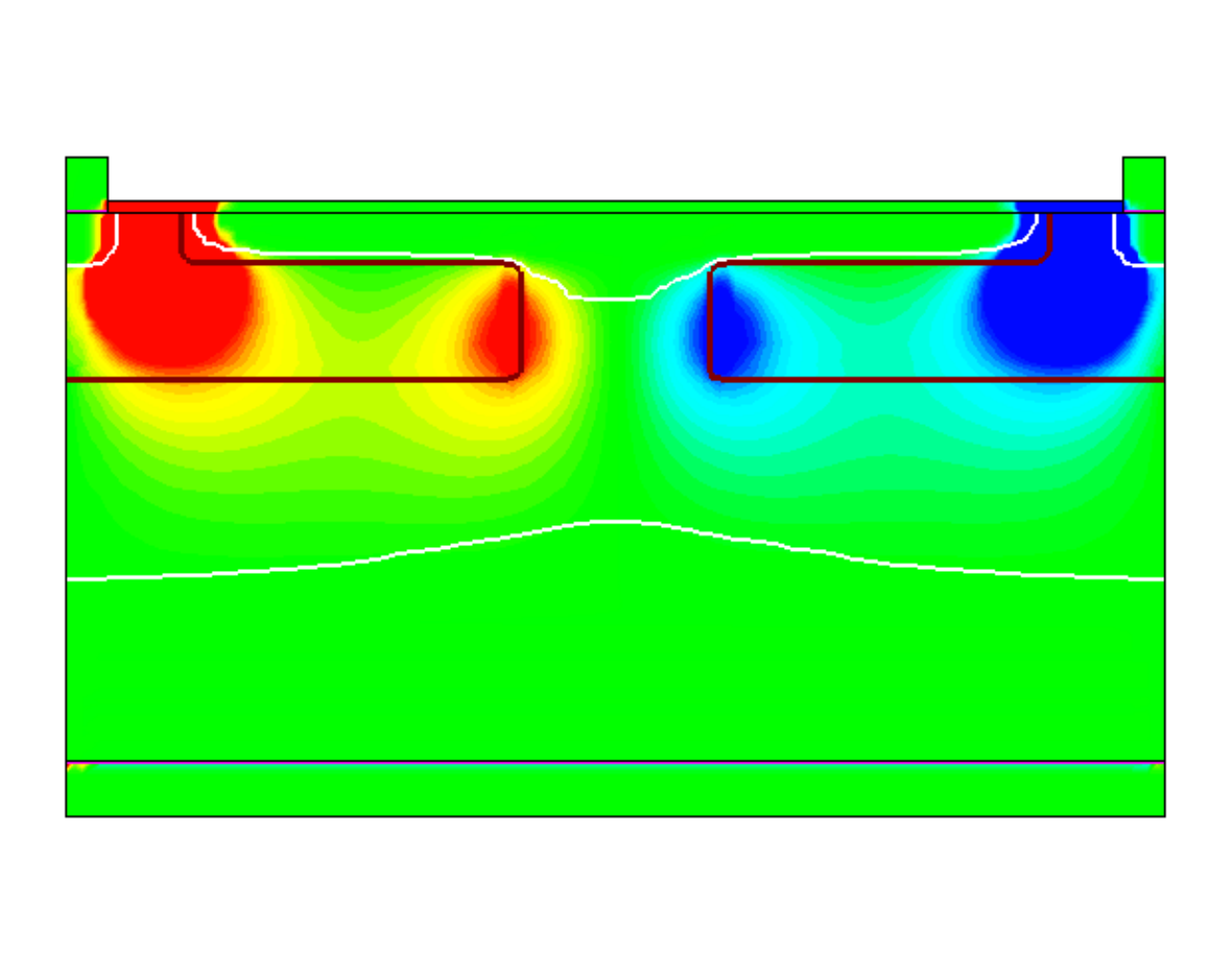}
        \caption{4~\um{} gap}
        \label{fig::gap_4um}
    \end{subfigure}%
    \caption{Lateral electric field of three n-gap sizes for a 20~\um{} pitch sensor in the \emph{n-gap} layout. The brown line indicates the location of the pn-junction and the white line delimits the depleted volume.}
	\label{fig::gap_scan}
\end{figure}

A lateral electric field is observed to appear under the p-well once the vertical junctions of the \emph{n-gap} layout are added to the sensor, as can be seen in Figure~\ref{fig::gap_scan}. The gap size was varied from 1~\um{} to 4~\um{}, and it was found that the strength of the lateral electric field is increased with increasing gap size, as the two vertical junctions move further apart. When the junctions are close, the regions of dopant diffusion will overlap, leading to a smaller lateral field. As can be seen in Figure~\ref{fig::gap_scan} the lateral field directions of the two vertical junctions are opposite, implying that they cancel out in the centre when the distance is small, so at sufficiently large gap size the lateral field strength reaches a maximum.
However, when the gap is increased, the vertical pn-junction as well as the lateral electric fields are shifted away from the pixel edges, thus reducing their usefulness in improving charge collection far from the collection electrode and leaving an electric field minimum in the gap. A gap size of 2.5~\um{} is sufficiently large to maximise the lateral field strength, while keeping the junction close to the pixel edge.

\subsection{Hexagonal pixel geometry simulation}
\label{sec::hex}

\begin{figure}[htb]
    \centering
    \includegraphics[width=\columnwidth]{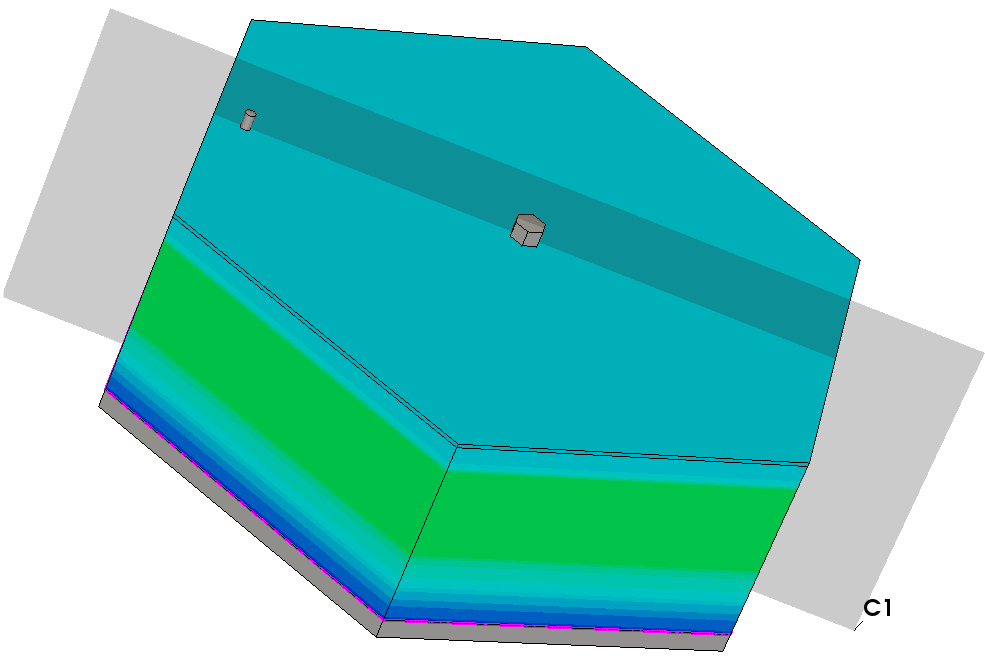}
    \caption{Simulated hexagonal pixel cell in TCAD. The colours correspond to doping level, and the plane marked C1 is a cut for display purposes. The electric field magnitude for this cut is shown in Figure~\ref{fig::efield_hexagonal_pixel}.}
    \label{fig::hex_pixel_cell_sde}
\end{figure}

\begin{figure}[htb]
    \centering
    \begin{subfigure}{0.9\columnwidth}
        \includegraphics[width=1.0\linewidth]{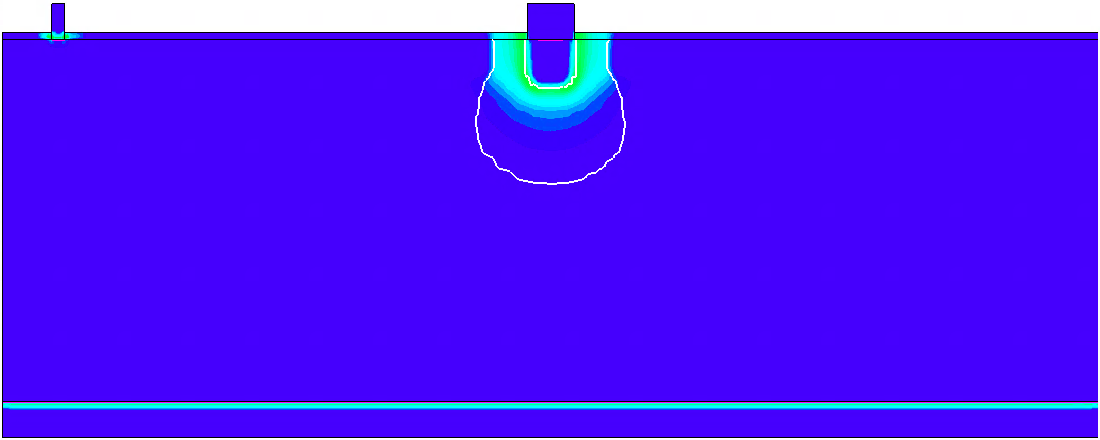}
        \caption{\emph{Standard} layout}
        \label{fig::efield_hexagonal_pixel_standard}
    \end{subfigure}
    \begin{subfigure}{0.9\columnwidth}
        \includegraphics[width=1.0\linewidth]{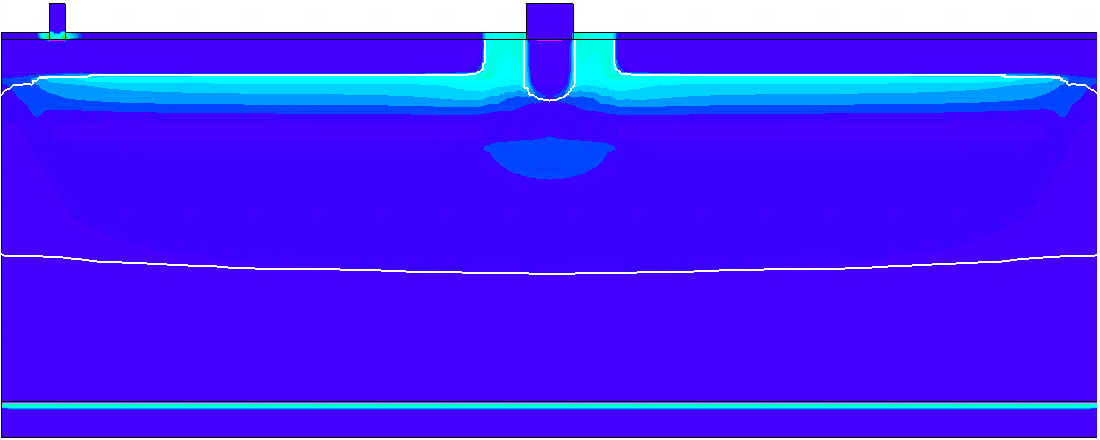}
        \caption{\emph{N-gap} layout}
        \label{fig::efield_hexagonal_pixel_ngap}
    \end{subfigure}
    \caption{Electric field magnitude as output from a TCAD simulation for a hexagonal pixel in the \emph{standard} and \emph{n-gap} layouts. The white lines denote depletion boundaries, and the colour scale denotes the magnitude of the electric field.}
    \label{fig::efield_hexagonal_pixel}
\end{figure}

Detailed investigations of hexagonal pixel designs require custom field maps from TCAD for hexagonal geometries.
One full pixel cell with the collection electrode in the centre is used in these simulations, and the p-well and substrate are biased with a voltage of $-1.2$~V. In Figure~\ref{fig::hex_pixel_cell_sde}, a pixel cell for a simulation of a sensor in the \emph{standard} layout is shown, with the colour indicating the doping level. 
The plane indicated as C1 represents a cross-section, along which the electric field magnitude is shown in Figure~\ref{fig::efield_hexagonal_pixel} for both the \emph{standard} and \emph{n-gap} layouts. The regions sticking out from the top of the sensor are the metal biasing contacts for the collection electrode and the p-well.

It can be seen that the depletion region is small for the \emph{standard} layout, only extending below the opening between the collection electrode and the p-well. For the \emph{n-gap} layout, however, it extends across the full pixel.

\subsection{Transient simulations}
\label{sec::tcadTransient}

Transient simulations were performed to estimate the shape, amplitude, and duration of a signal generated by a minimum ionising particle traversing the sensor. For these simulations, the p-well and substrate are biased with a voltage of $-1.4$~V. The particle traversing the sensor is represented by linear charge deposition along the particle track with Gaussian lateral smearing of 0.5~\um{} using the ``Heavy Ion'' charge deposition model. The mesh is adjusted for the transient simulations to have a finer cell size around the areas with a doping concentration gradient and the track of the traversing particle. For the thin sensors used in the simulations, a deposition of 63 electron-hole pairs per micrometre is assumed~\cite{SMeroli_2011}. 
A matrix of $3 \times 3$ pixels is simulated in 3D, in order to avoid edge effects.
The time step of the transient simulation is adapted to the expected shape of the signal. Two incident positions were simulated; in the centre and in the corner of a square pixel, for the three different layouts. The simulated track of the traversing particle is perpendicularly incident on the sensor for all the simulations.

\begin{figure}[hbt]
        \includegraphics[width=1.\linewidth, trim={55mm 45mm 45mm 10mm}, clip]{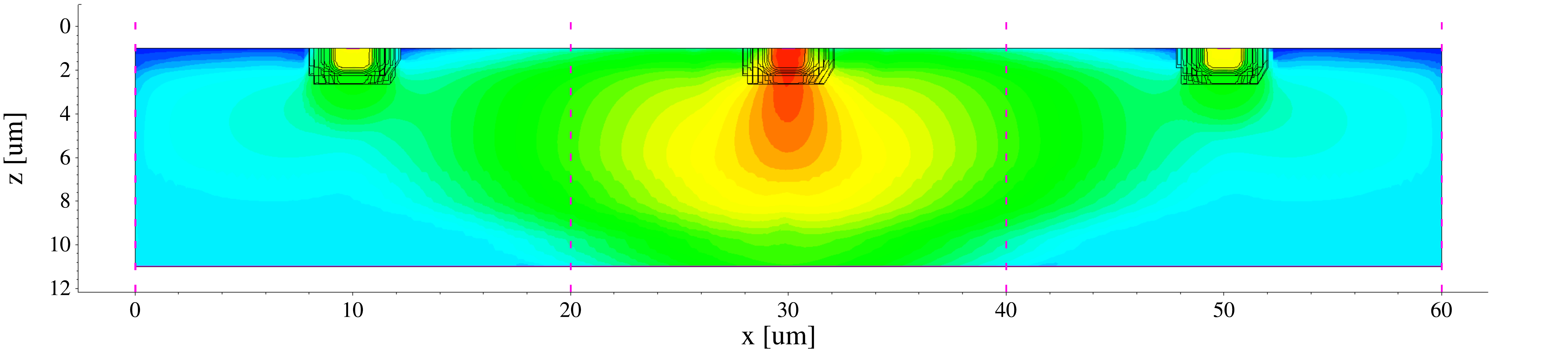}
        \caption{Absolute electron current density. Three adjacent pixels are shown, and the incident position is in the centre of the middle pixel readout implant. \emph{Standard} layout. The dashed lines indicate pixel edges.}
	\label{fig::Current}
\end{figure}

\begin{figure}[hbt]
        \includegraphics[width=1.\linewidth]{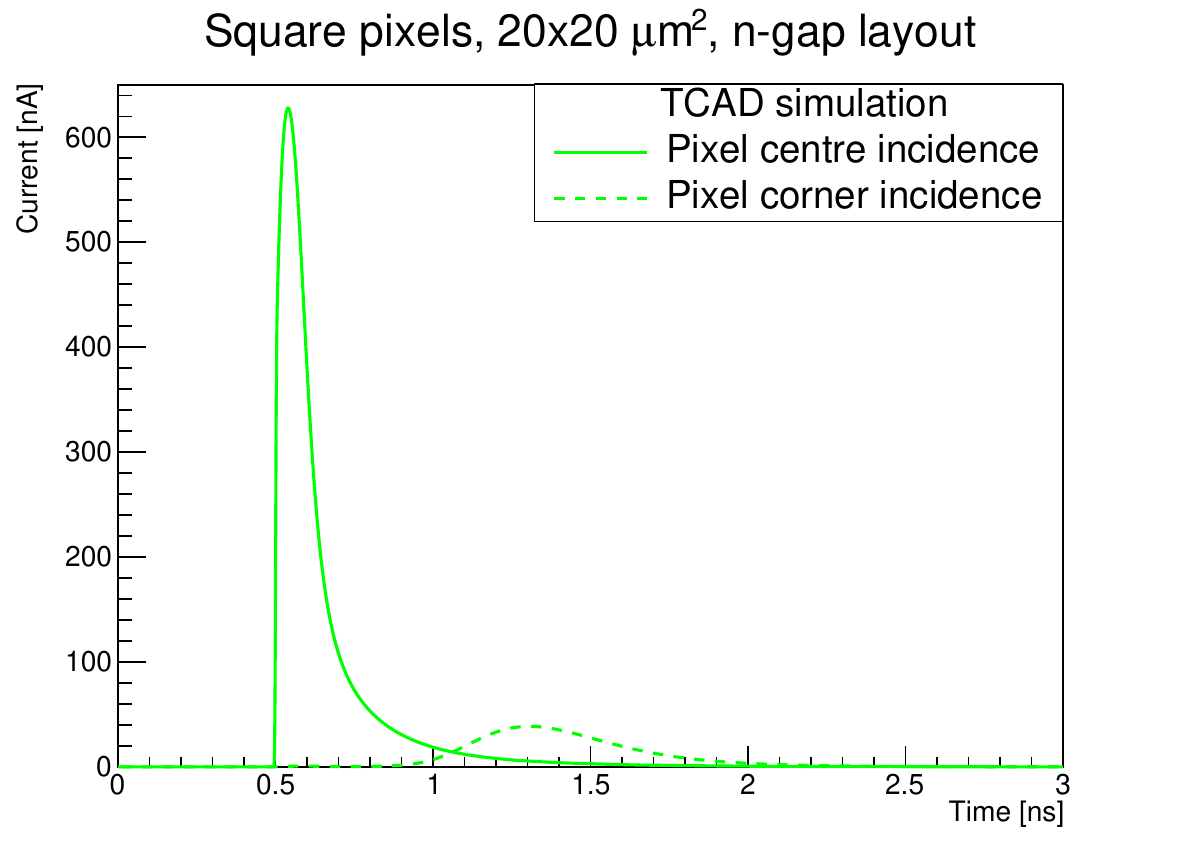}
        \caption{Signal in the \emph{n-gap} layout sensor in the centre and corner incident positions.}
	\label{fig::Gap}
\end{figure}

The absolute electron current density of the \emph{standard} layout for the incidence of a MIP in the centre of the pixel is shown in Figure~\ref{fig::Current}.
Three adjacent pixels are shown.
The depletion volume in the \emph{standard} layout is limited, and the layout allows for significant charge sharing due to diffusion.
The \emph{n-blanket} and \emph{n-gap} layouts were developed in order to improve the charge collection efficiency of the sensor in incident positions further from the readout implant~\cite{nblanketLayout, ngapLayout}. 
In Figure~\ref{fig::Gap} the signals for the centre and corner incident positions for the \emph{n-gap} layout are shown. 
The duration of the signal is dependent on the MIP incident position; the faster charge collection is observed in the centre of the pixel, due to the immediate proximity to the readout implant. The charges are deposited 0.5~ns after the start of the simulation, leading to the rising edge of the pulse.

\begin{figure}[hbt]
        \includegraphics[width=1.\linewidth]{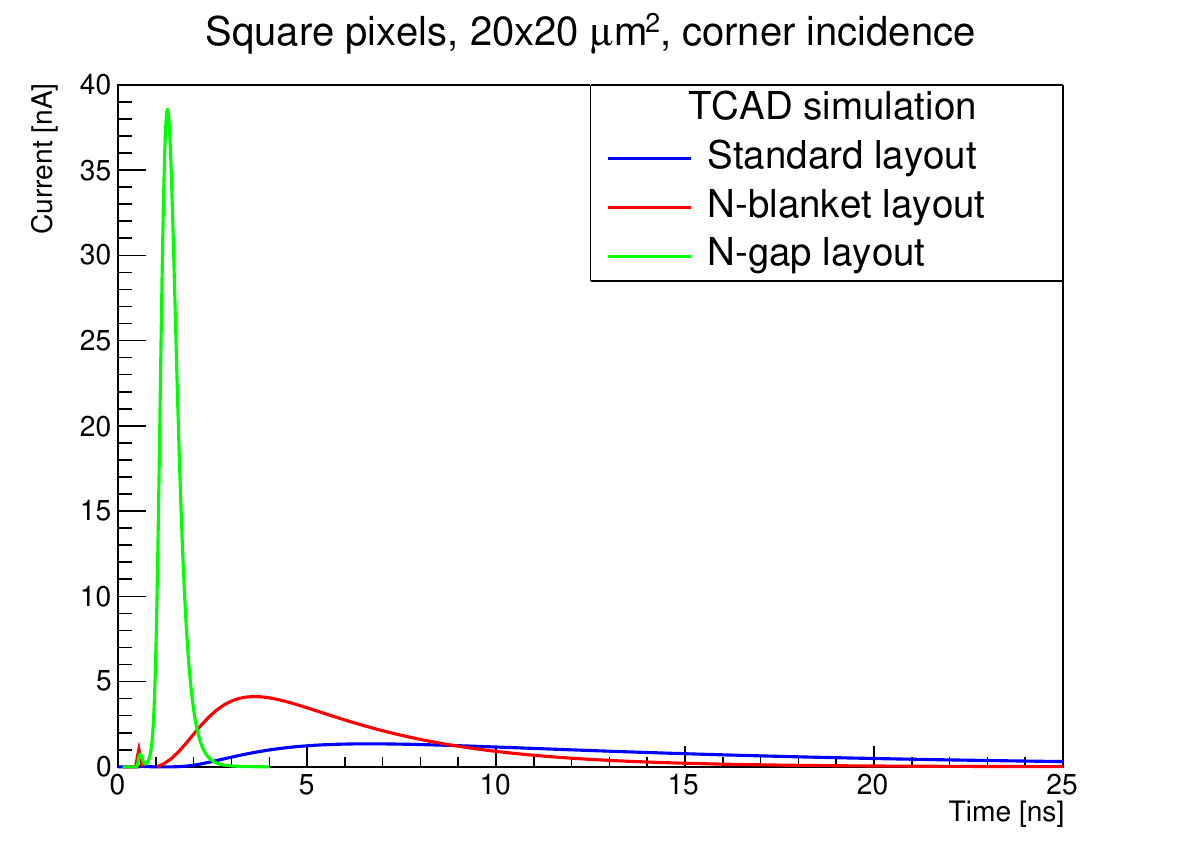}
        \caption{Signal of the sensor with the incident position in the corner of the pixel, for the \emph{standard}, \emph{n-blanket}, \emph{n-gap} layouts.}
	\label{fig::Comp}
\end{figure}

Figure~\ref{fig::Comp} shows the signals for corner incident positions, for all three layouts.
The \emph{n-blanket} has a larger depletion region compared to the \emph{standard} layout, and the \emph{n-gap} has an additional area with a stronger lateral electric field component, which improves the charge collection far from the pixel centre. As the \emph{standard} layout is undepleted in the pixel corners, charges formed there move slowly by diffusion, and the charge collection thus takes a comparatively long time.

\subsection{Generating weighting potentials}
\label{sec::weightingFieldCreation}




Transient simulations using TCAD are computationally intensive, and it can thus be beneficial to perform transient simulations using e.g. \Ap{} instead.
A weighting potential is required to be able to perform transient simulations using the Shockley-Ramo theorem~\cite{Shockley, Ramo}. The potential can be calculated by taking the difference of the electrostatic potentials arising from applying two slightly different bias voltages to one collection electrode in a sensor, keeping the other collection electrodes at a constant bias voltage.
The two required electrostatic potentials can be simulated using TCAD, using the same sensor geometry with a difference of 0.01~V in the bias voltage of a single collection electrode. By calculating the difference between the two potentials in each mesh point, and dividing the difference by the difference in collection electrode bias voltage, the weighting potential is acquired. Before utilising this weighting potential for simulations, the values should be constrained to be between 0 and 1, as this is the physical range of a weighting potential. Larger and smaller values may occur in the calculation due to numerical errors. 


\section{Monte Carlo simulations}
\label{sec::mc}


Simulation of sensor response to incident particles can be performed using TCAD, but studies with high statistical significance taking stochastic fluctuations into account are not feasible due to the long simulation time required per particle hit.
By combining the doping concentrations, electric fields, and weighting potentials generated using TCAD with the \Ap{} Monte Carlo simulation framework however, high-statistics simulations can be carried out~\cite{apsq, apsq2021}.
This section demonstrates how such simulations can be performed for the monolithic sensors described earlier, using \Ap{} version 3.0~\cite{apsq_v3p0}.

\subsection{Simulation flow}

\Ap{} is built on the concept of exchangeable modules, making it possible to flexibly change simulation aspects such as particle source and charge propagation method. The modules also constitute different steps taken in the simulation process, and parameters of each module can be controlled by configuration files with keyword-value pairs. 
When providing values to keywords that represent physical quantities in module configurations, it is important to also provide a unit in order to avoid unexpected behaviour.


\subsection{Sensor geometry and setup}
\label{sec::geometryBuilding}

A detector model in \Ap{} is defined in a configuration file, and an example can be seen in Listing~\ref{lst::detectorConfiguration}. The example shows a monolithic sensor assembly with square pixels, with a pixel size of $20 \times 20$~\umsq{} and a total sensor thickness of 50~\um{}. The sensor excess consists of sensor material without pixels, and is an important parameter to keep in mind when calculating sensor efficiency from simulation results, as particles hitting the sensor excess should not be counted as particles that should produce a signal in the sensor.

\begin{listing}[h]
\begin{minted}{TOML}
type = monolithic
geometry = pixel
sensor_material = silicon
number_of_pixels = 20 20
pixel_size = 20um 20um
sensor_thickness = 50um
sensor_excess_right = 200um

[implant]
type = frontside
shape = rectangle
size = 2.2um 2.2um 0.8um
\end{minted}
\caption{Detector model configuration example.}
\label{lst::detectorConfiguration}
\end{listing}

The \texttt{geometry} parameter can be used to select rectangular, radial strip, or hexagonal pixel geometries.
For hexagonal geometries, the pixel size is defined from corner to corner along axes $60^\circ$ apart, i.e. the maximum distance across a hexagon.

The \texttt{[implant]} section defines the x-, y-, and z-extent of the collection electrode of the sensor. In \Ap{}, this defines the volume in which charge propagation stops, and charges are counted as ``collected''. A small difference in this parameter can have a sizeable effect on the final results.


A geometry configuration file defines the full simulated geometry, and can contain several sensors and passive volumes. An example configuration for a single-sensor simulation is shown in Listing~\ref{lst::geomConfiguration}.
\begin{listing}[h]
\begin{minted}{TOML}
[dut]
type = "detectorModel"
position = 0mm 0mm 0mm
orientation = 0deg 0deg 0deg
\end{minted}
\caption{Geometry configuration example, for a single sensor without random misalignment.}
\label{lst::geomConfiguration}
\end{listing}
For each sensor or passive volume, a position and orientation has to be defined, along with an alignment precision. In the given example, the name of the detector is ``dut'', located at the centre of the global coordinate system. The alignment precision is more important to include when several sensors are involved. The detector type in the example is \texttt{detectorModel}, which is the name of a detector model configuration file such as the example shown in Listing~\ref{lst::detectorConfiguration}. 

The global coordinate system is defined by the simulated world volume, and positions of components are defined in this system. Each detector placed in the world has a local coordinate system, with an origin defined in the centre of the lower left pixel of the sensor pixel matrix.

\subsubsection{Constructing the geometry for use with Geant4}

The \texttt{[GeometryBuilderGeant4]} module constructs the geometry for use with Geant4~\cite{geant4,geant4-2,geant4-3}, which allows for detailed particle interaction simulations.
To visualise the geometry constructed by the module, the module \texttt{[VisualizationGeant4]} can be used. This opens a Geant4 graphical user interface window, with the possibility of also starting a Geant4 terminal, giving access to both a visualisation of the setup and Geant4 commands. By using the \texttt{/run/beamOn} Geant4 command, it is possible to see where particles from a defined source will hit the setup. This is useful for making sure that the source is aligned with the detectors in the desired way, but the command cannot be used to perform a proper simulation.

\subsection{Importing results from TCAD simulations}
\label{sec::importFromTCAD}

Electric fields and doping concentrations can be imported from TCAD simulations, which gives access to more detailed fields than the built-in parametric models. To be usable in \Ap{} the TCAD mesh has to be adapted into a regularly-spaced grid, and this process is performed using the \texttt{Mesh converter} tool. The tool either performs a barycentric interpolation of values for each point in the new regularly-spaced grid or uses the value of the closest TCAD mesh point without interpolation. The second case is particularly useful for conversion of large field maps.

An example of a configuration file for the mesh converter is shown in Listing~\ref{lst::meshConverterConfiguration}. This file is used for converting the electric field of the region named ``epitaxial'' of a $20 \times 20$~\umsq{} pixel. The \texttt{model} keyword defines the output format, and the units of the converted observable have to be provided; for electric fields, it is typically V/cm, and for doping concentrations cm$^{-3}$ (written \texttt{/cm/cm/cm} in the \Ap{} configuration files).

\begin{listing}[h]
\begin{minted}{TOML}
model = "APF"
region = "epitaxial"
observable = ElectricField
observable_units = "V/cm"
divisions = 300 300 100
xyz = x y -z
\end{minted}
\caption{Mesh converter configuration example, for the electric field in the ``epitaxial'' region of a sensor.}
\label{lst::meshConverterConfiguration}
\end{listing}

The \texttt{divisions} parameter defines the number of points used in the regularly-spaced grid, and thus the granularity of the field map when imported into the framework. The number of grid points used can have a significant impact on the final observables, so the values should be chosen with care so as not to have too coarse a grid. How fine it needs to be depends on the sensor and the desired studies, but having a fine grid has no negative impact on the simulation speed, as it is only loaded into memory once and then used for looking up values. The conclusion of studies varying the grid granularity for the sensors described in Section~\ref{sec::layout} is that the grid needs to be finer than the expected geometrical features of the sensor volume, in order to not skew the results.

The keyword \texttt{xyz} is used to define the orientation of the conversion, and in the example, the z-axis is inverted when importing the field. In \Ap{} simulations, the collection electrode should be located at the top of the sensor.

To convert fields with non-rectangular shapes (e.g. hexagons), not all points in a rectangular bounding box should be filled with values. To allow for this, the keyword \texttt{allow\_failure} can be used. This sets a grid point value to zero in case no close neighbouring points in the TCAD field are found, which will happen to points outside the TCAD mesh.

The modules \texttt{[ElectricFieldReader]} and \texttt{[DopingProfileReader]} are used to include the converted electric field and doping concentration in the \Ap{} simulation. An example configuration for the two modules can be seen in Listing~\ref{lst::fieldImportConfiguration}. TCAD commonly simulates a single pixel, part of a pixel, or a small number of pixels, and the \Ap{} modules make sure that the imported fields are correctly mapped for each pixel across a full sensor using the \texttt{field\_mapping} keyword.

\begin{listing}[h]
\begin{minted}{TOML}
[ElectricFieldReader]
model = "mesh"
file_name = "ElectricField.apf"
field_mapping = PIXEL_FULL_INVERSE
field_depth = 10um

[DopingProfileReader]
model = "mesh"
file_name = "DopingConcentration.apf"
field_mapping = PIXEL_FULL_INVERSE
doping_depth = 10um
\end{minted}
\caption{Example configuration for reading electric field and doping concentration into \Ap{} from external files converted from TCAD using the \texttt{Mesh converter} tool.}
\label{lst::fieldImportConfiguration}
\end{listing}

The \texttt{mesh} model is used for importing fields from external sources, and the file names in this case are binary APF-format output fields from the \texttt{Mesh converter} tool.
When importing fields and profiles, the diagnostic plots from the modules (activated by using the \texttt{output\_plots} keyword) are useful for determining that everything is correctly imported and mapped to the sensor.

The depth used when importing fields is defined by the \texttt{field\_depth} and \texttt{doping\_depth} keywords. These values should match the depth of the actual imported fields and profiles, as it will otherwise stretch or compress them to fit the given parameter value. At depths larger than the given \texttt{field\_depth}, the electric field is set to zero. For the doping concentration, the concentration deeper than the given \texttt{doping\_depth} is set to the same value as the last value within the depth. In the given example, the imported electric field and doping profile are 10~\um{} thick, but the full sensor is 50~\um{} thick. The doping concentration in the TCAD simulation has been made with the substrate concentration at the edge, that then gets extrapolated into the remaining 40~\um{} of the sensor after importing. This creates the effect of an epitaxial layer grown on a higher-doped silicon substrate.


\subsection{Charge carrier generation}

Charge carrier creation in \Ap{} can be performed either by direct charge injection in given points, or by using the deposited energy from a physics process. 
%
%
%
The \texttt{[DepositionGeant4]} module provides a direct interface to Geant4 from within \Ap{}, and makes it possible to generate Geant4 particles that traverse the simulated setup and deposit energy and scatter. This enables energy deposition and charge carrier creation that takes stochastic effects into account, e.g. Landau fluctuations, particle decays, and secondary particles. The module comes with a number of pre-defined sources and source shapes, but also allows the use of Geant4 macros for the General Particle Source~\cite{geant4-gps}. An example configuration can be seen in Listing~\ref{lst::depositionGeant4Configuration}. Here, the physics list \texttt{FTFP\_BERT\_EMZ} is used, along with the photo-absorption ionisation model (PAI)~\cite{PAI}.

\begin{listing}[h]
\begin{minted}{TOML}
[DepositionGeant4]
physics_list = FTFP_BERT_EMZ
enable_pai = true
particle_type = "e-"
source_type = "beam"
source_energy = 5GeV
beam_size = 0.3mm
beam_direction = 0 0 1
source_position = 0um 0um -50mm
number_of_particles = 1
max_step_length = 1.0um
\end{minted}
\caption{Geant4-based charge deposition configuration example.}
\label{lst::depositionGeant4Configuration}
\end{listing}

The \texttt{beam} model emulates a beam of particles. In the example, a beam of electrons is used, with a Gaussian profile with a width of 0.3~mm and an energy of 5~GeV. A single electron is fired per event.
By changing the \texttt{max\_step\_length} parameter, the maximum length of a simulation step of a particle is altered. 
The Geant4 processes are invoked over each step, and the difference in the state of the simulated particle between the pre- and post-step points is the result of the physics process calculations.
A finer step size thus gives a finer granularity in the energy deposition calculations.
While only the maximum value can be set, and the checks can thus happen at any length below this value as well (determined by the active physics processes), the parameter may strongly impact the distribution of deposited charges in a sensor. Studies performed using relatively thin sensors (with a total thickness of 50~\um{}) have shown that using a step size larger than 5~\um{} has a significant impact on final observables such as cluster size. Using a smaller value has no noticeable impact on the simulation time for these sensors, however, so a smaller value is chosen as it provides a higher degree of realism.

The \texttt{[DepositionPointCharge]} module directly deposits electron-hole pairs, rather than produce them from energy deposited. This is useful in detailed sensor behaviour investigations, as it removes statistical fluctuations in the deposition. It can thus be used as a tool for determining how different parts of the sensor react to different inputs, and the module is useful for comparison with transient TCAD simulations where charge is deposited similarly. 
%

\subsection{Charge carrier propagation}

The \Ap{} framework has different ways of propagating the created charge carriers in a sensor, and the modules used in the work presented here are \texttt{[GenericPropagation]} and \texttt{[TransientPropagation]}.
Transient simulations are discussed in more detail in Section~\ref{sec::transientAP2}.
The \texttt{[GenericPropagation]} module is the preferred choice for simulations without transient information, and an example configuration is shown in Listing~\ref{lst::genericPropagationConfiguration}.

\begin{listing}[h]
\begin{minted}{TOML}
[GenericPropagation]
mobility_model = "masetti_canali"
recombination_model = "srh_auger"
charge_per_step = 5
timestep_min = 0.5ps
timestep_max = 0.05ns
integration_time = 25ns
propagate_electrons = true
\end{minted}
\caption{Propagation module configuration example.}
\label{lst::genericPropagationConfiguration}
\end{listing}

A recombination model is used to simulate the finite lifetime of charge carriers. By combining the Shockley-Read-Hall~\cite{SRHrecomb1, SRHrecomb2} and Auger~\cite{augerRecomb, FossumLeeRecomb} recombination models, a finite lifetime estimation can be achieved over a large range of doping concentrations. As this is the case for sensors simulated in the given example, the combined model is activated by the \texttt{srh\_auger} value.

The \texttt{charge\_per\_step} parameter determines the number of charges from a single energy deposit point that are propagated together as a group, and increasing this value can reduce simulation time. 
However, if the number is set too high, there will be a significant impact on the final simulation observables due to the loss of accuracy. A balance thus has to be reached, but the best value depends heavily on the sensor geometry and the energy deposition. A value of 1 is the most accurate and most computationally intense. 

The discussed propagation modules calculate the motion of charge carriers in electric and magnetic fields using a fourth-order Runge-Kutta-Fehlberg method~\cite{fehlberg}. The motion is calculated over a timestep, which is calculated internally and constrained by the parameters \texttt{timestep\_min} and \texttt{timestep\_max} in the \texttt{[GenericPropagation]} module. 
A smaller timestep means higher precision, but a longer simulation time required to propagate the charge. The \texttt{integration\_time} parameter sets the simulated duration of charge carrier propagation from the start of the event. This can be used to stop the simulation when sensors should no longer collect more charge, decided by e.g. readout speed requirements.


Linegraphs show the full paths travelled by the charge carriers in an event, and can be activated for all propagation modules. These graphs are a useful tool for checking the origin of particular sensor behaviours, but their creation takes a long time, so they should only be used to check a few key events.
To capture the charge carrier movement in sufficient detail, it is useful to reduce the propagation timestep. For thin silicon MAPS, a timestep size between 0.05~ps and 0.5~ps has been found to give good results that show the detail of the diffusion and drift paths.
By creating linegraphs of the same event with different values for the \texttt{integration\_time} parameter, the time evolution of collected charge can be visualised~\cite{tcadMCcombination}. This can be useful in determining charge collection time from different parts of a pixel.

\subsubsection*{Mobility models}
\label{sec::mobilityModels}





Mobility models can be selected in the propagation modules, and the default is the Jacoboni-Canali model~\cite{jacoboniModel}. The extended Canali model (\texttt{masetti\_canali}) is used in the example given in Listing~\ref{lst::genericPropagationConfiguration}. This model combines the Jacoboni-Canali model with the Masetti model~\cite{masettiModel}, to give a doping-dependent mobility parametrisation valid for both low and high electric fields.
From the available mobility models, two groups can be distinguished: doping-dependent and doping-independent models.
It is advisable to utilise doping-dependent models when a significant doping concentration is present in the simulated sensor. If no doping information is available, or the doping concentration is low, doping-independent models (e.g. the Jacoboni-Canali model) suffice.

\begin{figure}[h]
\centering
    \begin{subfigure}{.5\columnwidth}
        \includegraphics[width=1.0\columnwidth, trim={20mm 19mm 55mm 18mm}, clip]{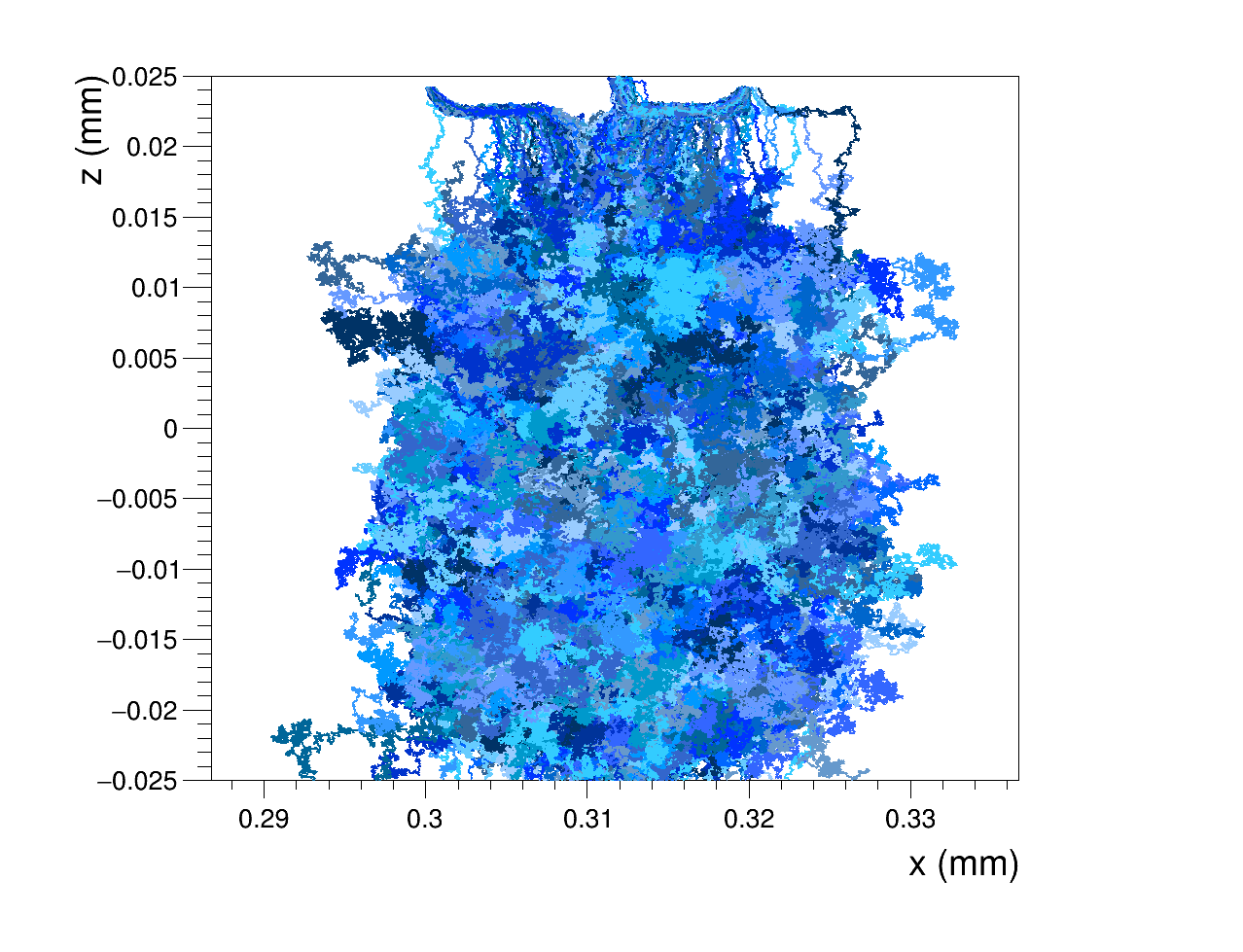}
        \caption{Doping-independent mobility model}
        \label{fig::mobility_jacobonicanali}
    \end{subfigure}%
    \begin{subfigure}{.5\columnwidth}
        \includegraphics[width=1.0\columnwidth, trim={20mm 19mm 55mm 18mm}, clip]{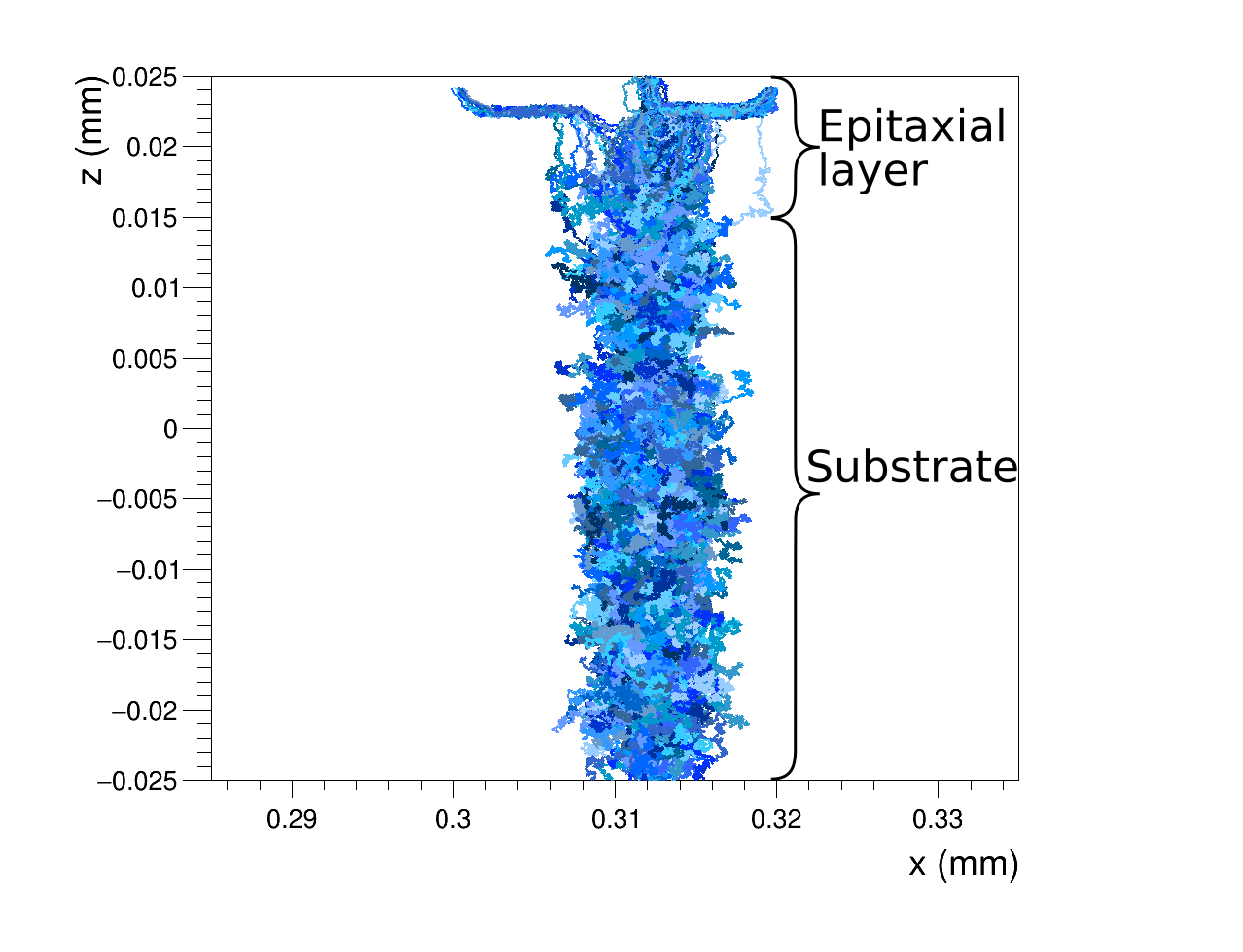}
        \caption{Doping-dependent mobility model}
        \label{fig::mobility_masetticanali}
    \end{subfigure}
    \caption{Linegraphs of a single particle event, showing a significant difference between mobility models in the substrate region.}
	\label{fig::mobility_comparison_linegraphs}
\end{figure}

Figure~\ref{fig::mobility_comparison_linegraphs} shows linegraphs of electron propagation, using a doping-independent and a doping-dependent mobility model. A single MIP traversing the sensor is simulated, and the integration time is set to 5~ns to more clearly show the differences in diffusion. Each line shows the path of a single electron, in a sensor of the type described in Section~\ref{sec::layout} with a highly-doped substrate located at the bottom 40~\um{} of the sensor.

There is a significant difference in charge propagation behaviour in the substrate region for the two mobility models. The charge cloud is wider for the doping-independent model, due to the larger diffusion in the highly-doped substrate. This leads to unphysical effects, as more charge from the substrate region reaches the epitaxial layer than in a real sensor. As the mobility is higher in the substrate for the doping-independent mobility model, the probability of electrons being recombined within the substrate region is also decreased.

\subsection{Charge transfer}
\label{sec::chargeTransfer}

After propagation, charge is transferred to the readout electronics simulation stage. The \texttt{[SimpleTransfer]} module makes a direct mapping of the final positions of the propagated charges to the nearest pixel, and an example configuration for this module is presented in Listing~\ref{lst::simpleTransferConfiguration}. Using this module, the notion of ``collected charge'' is used. This is based on that the total charge that reaches the collection electrode is equivalent to the total induced charge on the electrode from charge movement, in accordance with the Shockley-Ramo theorem~\cite{Shockley, Ramo}.

\begin{listing}[h]
\begin{minted}{TOML}
[SimpleTransfer]
collect_from_implant = true
\end{minted}
\caption{Charge transfer module configuration example.}
\label{lst::simpleTransferConfiguration}
\end{listing}

The keyword \texttt{collect\_from\_implant} makes the module only transfer charges with a final propagated position within the implant volume, which is defined in the detector configuration file (see Listing~\ref{lst::detectorConfiguration}). 
When performing non-transient simulations, the implant size should be set to slightly larger than the actual collection electrode size in a corresponding physical sensor or imported TCAD field, to make sure that all relevant charges are collected. If the size is set too small, charges that should have been collected may escape the defined volume in a final diffusion step before collection occurs. When performing transient simulations, however, the size should be set as close as possible to the size of the undepleted part of the electrode in the used weighting potential, to make sure that the induced charge moves the correct distance within the potential.
In the presented work, using the method of transferring charge that is close to the collection electrode is a good approximation for the total charge collected in a real situation, as the weighting field is strongest near the collection electrode. That implies that the bulk of charge induction will happen by movement of charge carriers close to the collection electrode.

\subsection{Signal digitisation}

The \texttt{[DefaultDigitizer]} module can be used to translate the collected charges into a signal by simulating a basic sensor front-end, including simulation of noise contributions from readout electronics by adding a random Gaussian offset, adding a gain with an arbitrary function to the signal, and setting a threshold value.
An example digitiser configuration is shown in Listing~\ref{lst::simpleDefaultDigitizerConfiguration}.

\begin{listing}[h]
\begin{minted}{TOML}
[DefaultDigitizer]
threshold = 200e
threshold_smearing = 5e
electronics_noise = 10e
\end{minted}
\caption{Digitiser configuration example.}
\label{lst::simpleDefaultDigitizerConfiguration}
\end{listing}

The threshold is set to 200 electrons here, with a dispersion of 5 electrons.
By running the full simulation except the digitisation stage and saving the results using the \texttt{[ROOTObjectWriter]} module, it is possible to investigate the effect of threshold variations without re-running the full simulation chain~\cite{apsq, tcadMCcombination}.

\subsection{Transient simulations}
\label{sec::transientAP2}






Transient simulations give access to the current pulses induced by the movement of charge carriers, and require the use of the \texttt{[TransientPropagation]}, \texttt{[WeightingPotentialReader]}, and \texttt{[PulseTransfer]} modules. The \texttt{[WeightingPotentialReader]} module can read in weighting potentials created using TCAD information, as described in Section~\ref{sec::weightingFieldCreation}.

For each simulation step, the induced charge on the pixel collection electrodes within a given distance from the moving charge carrier is calculated via the Shockley-Ramo theorem~\cite{Shockley, Ramo} by taking the difference in weighting potential between the current position and the previous position of the charge carrier.
The resulting pulses are stored for every set of charge carriers individually, and after propagation has finished the pulses are combined for each individual pixel using the \texttt{[PulseTransfer]} module.


The \texttt{[PulseTransfer]} module can also be used to produce plots containing the induced current and accumulated charge by pixel or matrix. This greatly increases memory consumption and simulation time, however, so to produce plots of induced current it is recommended to save the simulation result using the \texttt{[ROOTObjectWriter]} module and use an external analysis script instead.

Listing~\ref{lst::transientPropagationConfiguration} shows an example configuration for the \texttt{[TransientPropagation]} module. An important parameter to take into account while using this module is \texttt{timestep}; a larger timestep value can reduce the simulation time, but information about the pulse may get lost.

\begin{listing}[h]
\begin{minted}{TOML}
[TransientPropagation]
temperature = 293K
charge_per_step = 1
distance = 1
timestep = 7ps
mobility_model = "masetti_canali"
recombination_model = "srh_auger"
integration_time = 40ns
\end{minted}
\caption{Transient propagation configuration example.}
\label{lst::transientPropagationConfiguration}
\end{listing}

The \texttt{distance} keyword defines on how many pixels the induction should be calculated on. If it is set to 1, the pixel the charge carrier is located in and all its nearest neighbours are included. For a rectangular geometry, this means that the induced current is calculated on 9 pixels in total.



\subsection{Effect of dopant diffusion in electric fields}
\label{sec::eFieldComparison_MC}

\begin{figure}[h]
\centering
    \begin{subfigure}{.5\columnwidth}
        \includegraphics[width=1.0\columnwidth, trim={26mm 26mm 60mm 20mm}, clip]{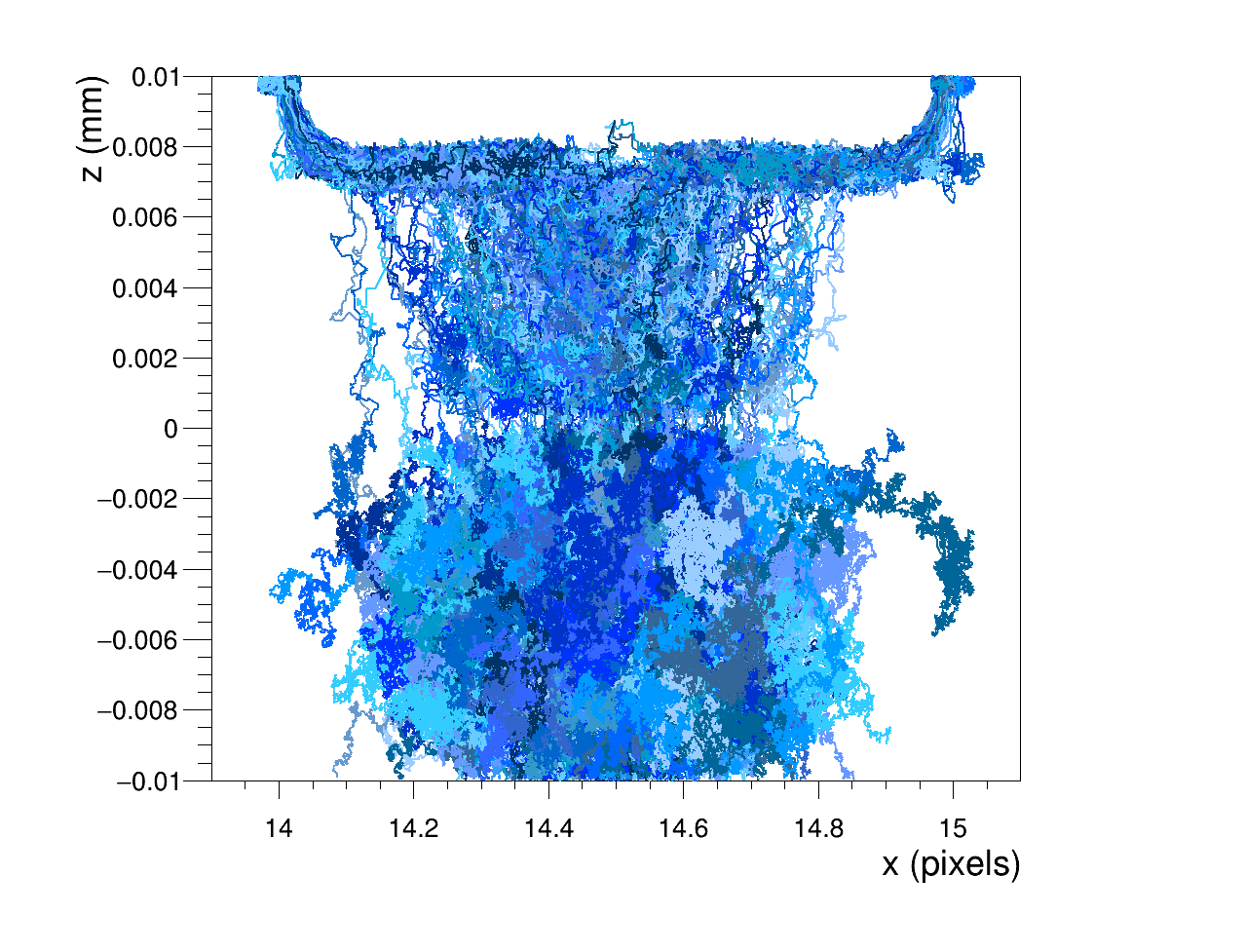}
        \caption{Without dopant diffusion}
        \label{fig::simpleFieldLinegraph}
    \end{subfigure}%
    \begin{subfigure}{.5\columnwidth}
        \includegraphics[width=1.0\columnwidth, trim={26mm 26mm 60mm 20mm}, clip]{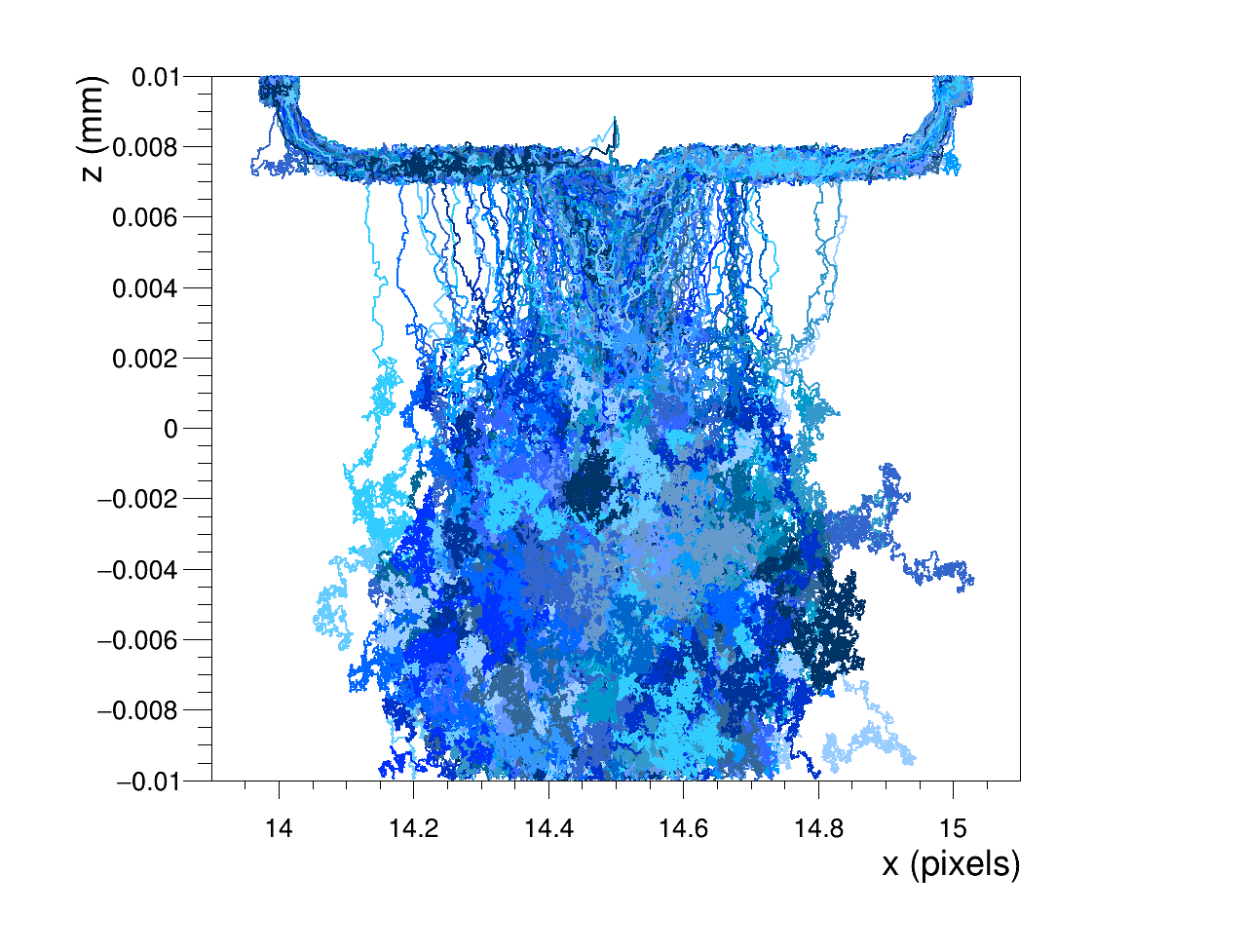}
        \caption{With dopant diffusion}
        \label{fig::complexFieldLinegraph}
    \end{subfigure}
    \caption{Linegraphs showing the paths of electrons, for two different electric field models in a 20~\um{} thick sensor.}
	\label{fig::simpleVsComplexLinegraphs}
\end{figure}

Studies were carried out both with and without dopant diffusion between the substrate and the epitaxial layer. Without dopant diffusion, the cluster size extracted from the simulations was lower than expected. A linegraph of a simulation using such an electric field for the \emph{n-gap} layout is shown in Figure~\ref{fig::simpleFieldLinegraph}, with only electrons being propagated. The sensor has a thickness of 20~\um{} in this case, and the simulated epitaxial layer thickness is 10~\um{}. At the border between the epitaxial layer and the substrate (at $z = 0$~mm), there is a ``gap'' in the lines. In this apparent ``gap'', the lines are straight, implying charge motion primarily by drift.
Without simulation of dopant diffusion, there is a discrete step in doping concentration between the highly-doped substrate and the much lower-doped epitaxial layer, which creates a depleted region with an electric field after diffusion of free charge carriers. This is unphysical, and a smooth transition region is expected, as described in Section~\ref{sec::tcadDiffusion}.

Figure~\ref{fig::complexFieldLinegraph} shows a linegraph using an electric field with dopant diffusion added between the substrate and the epitaxial layer.
There is no longer a strong electric field at $z = 0$~mm, and charge movement by drift instead starts dominating at around $z = 3$~\um{}. This situation results in a larger cluster size, as the region where charge moves primarily by diffusion is extended.




\subsection{Simulation parameter optimisation}

The sensitivity of the simulation to different parameters has been investigated, and the results can be used to optimise the accuracy and performance of the Monte Carlo simulations performed with \Ap{}.
The conclusions of such studies depend on the simulated geometry and the desired accuracy of the final observables, and results do thus not necessarily translate between different simulated setups, but it is beneficial to carry out the parameter optimisation before performing a large simulation campaign. The details of which parameters to alter, and their impact on simulation accuracy, have to be determined for each individual simulation case. Studies of the impact of parameter variation can also be used to determine systematic uncertainties in the simulations.
A few selected results of simulation parameter optimisation for the sensors studied in this project are shown below.

\subsubsection{Collection electrode size}
\label{sec::collElSizeStudy}

\begin{figure}[tb]
    \centering
        \includegraphics[width=1.0\linewidth]{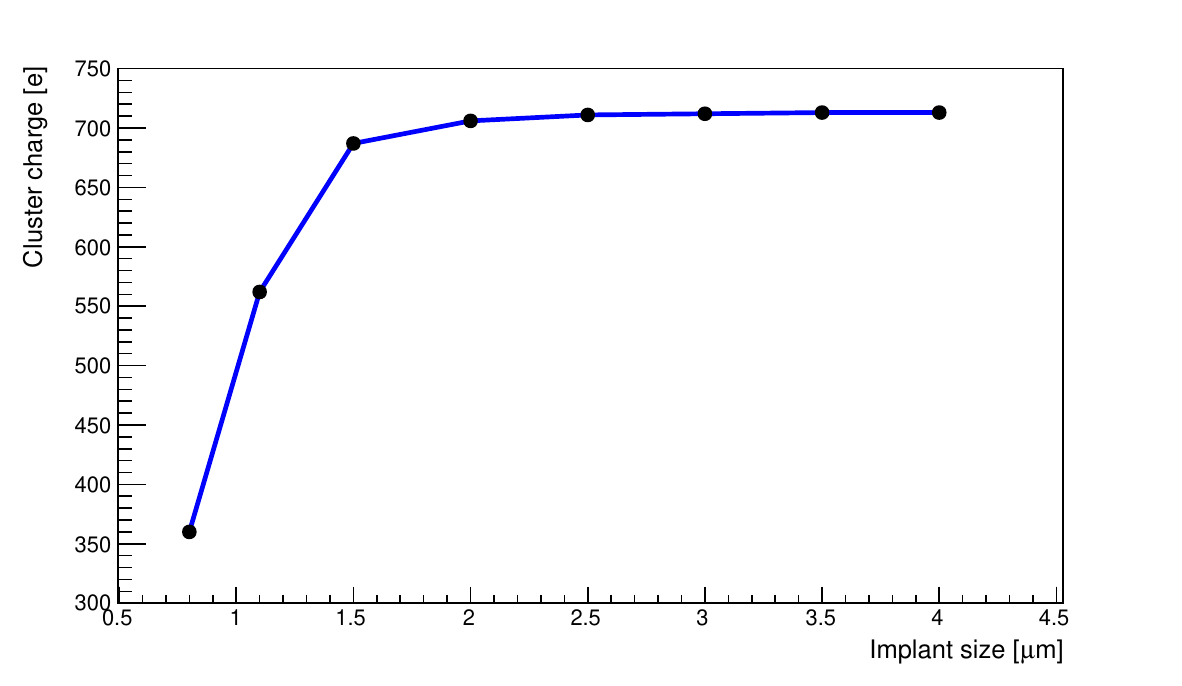}
    \caption{Mean cluster charge collected per event as a function of the implant size in x and y in \Ap{}.}
	\label{fig::collElSize}
\end{figure}

To accurately collect charge, the collection electrode size in \Ap{} has to be set correctly, as discussed in Section~\ref{sec::chargeTransfer}. A study of the impact on the total collected charge of setting it to different sizes is shown in Figure~\ref{fig::collElSize}, for a sensor with a nominal collection electrode size of $1 \times 1$~\umsq{} in the x- and y-directions before diffusion is applied in the TCAD simulations. The \texttt{[GenericPropagation]} module is used in this study, and the mean value of the cluster charge for all events is shown.

It can be seen that for implant sizes in \Ap{} set to values smaller than 1.5~\um{}, not all charge with a final position near the collection electrode is counted as collected. After dopant diffusion is applied, the effective collection electrode size in TCAD is approximately $2 \times 2$~\umsq{} for this sensor, and the value used in \Ap{} should be slightly larger to prevent charge carriers being excluded from collection after their final diffusion step. In the simulations presented in Section~\ref{sec::studies}, the size is thus set to $2.2 \times 2.2$~\umsq{}.

\subsubsection{Photo-absorption ionisation model}
\label{sec::paiModelStudy}

\begin{figure}[h]
\centering
    \begin{subfigure}{.5\columnwidth}
        \centering\text{10~\um{} thick}
        \includegraphics[width=1.\linewidth]{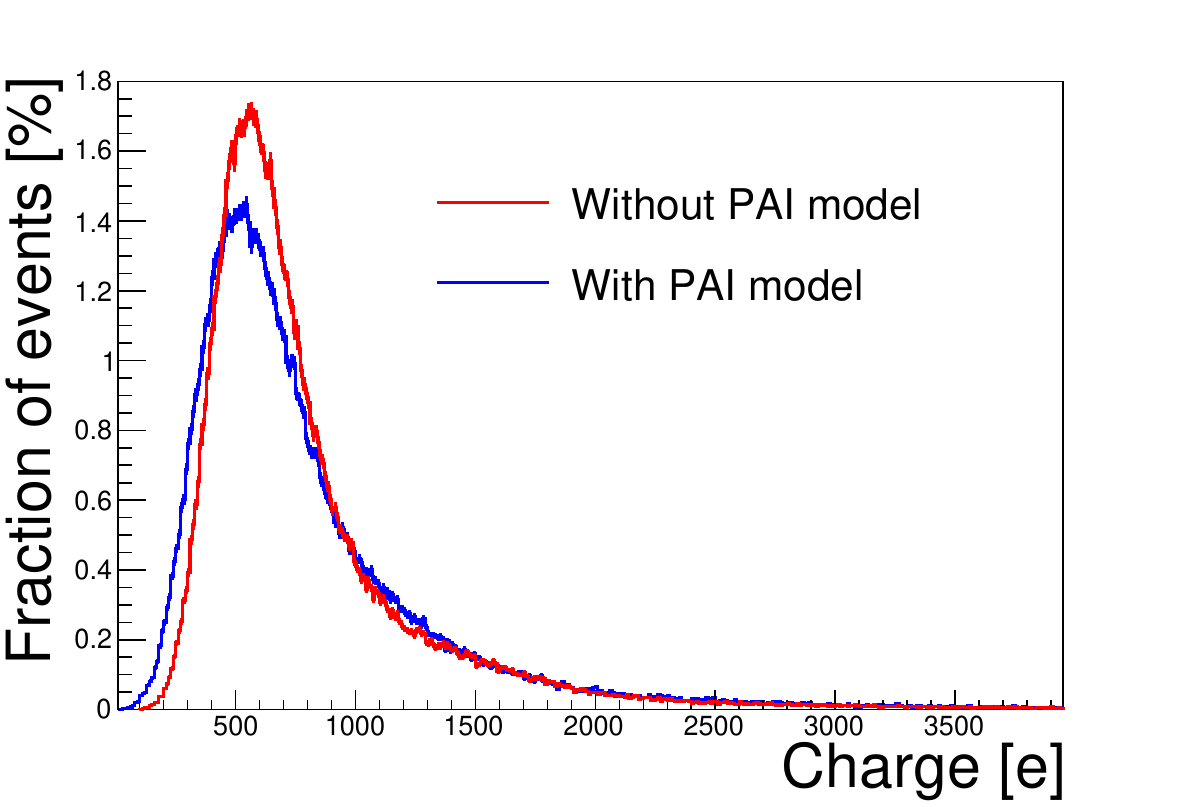}
    \end{subfigure}%
    \begin{subfigure}{.5\columnwidth}
        \centering\text{50~\um{} thick}
        \includegraphics[width=1.\linewidth]{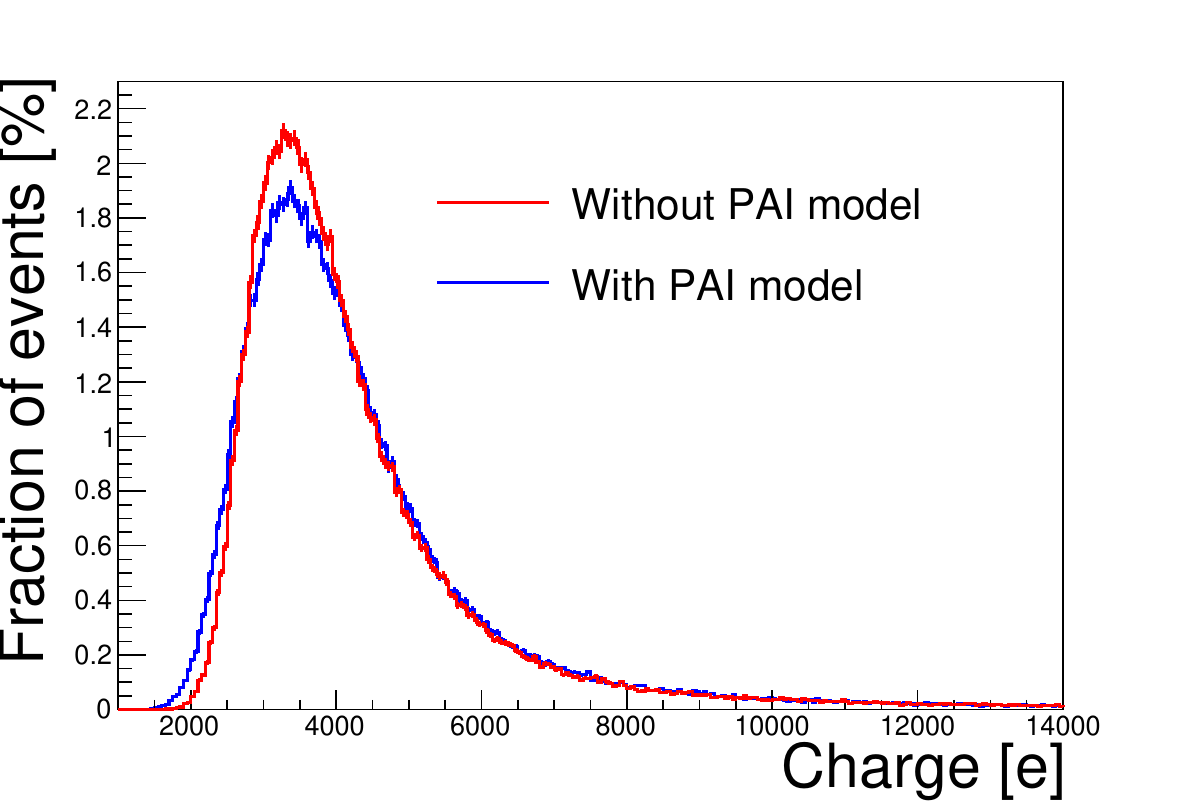}
    \end{subfigure}
    \begin{subfigure}{.5\columnwidth}
        \centering\small\text{500~\um{} thick}
        \includegraphics[width=1.\linewidth]{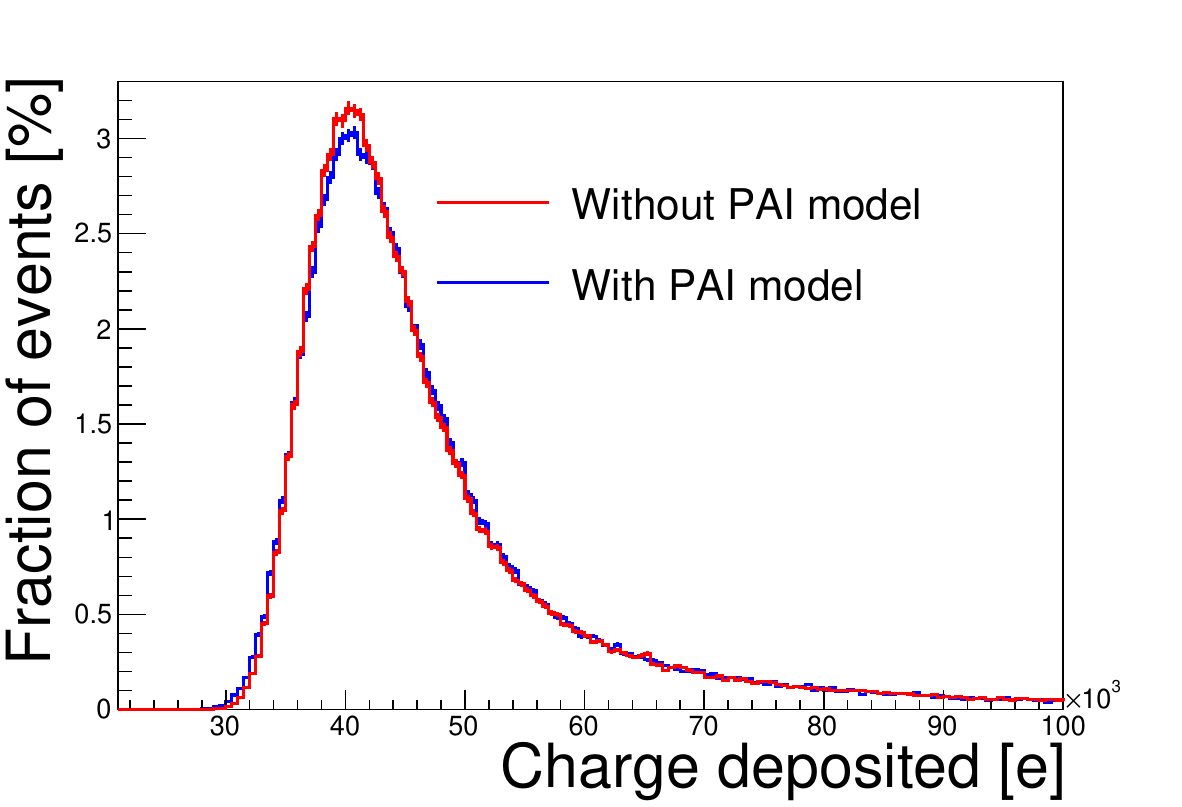}
    \end{subfigure}%
    \caption{\small Charge generated by deposited energy (in electrons) for a 5~GeV single-electron beam in 10~\um{}, 50~\um{}, and 500~\um{} of silicon. The simulations are shown with the Geant4 PAI model enabled and disabled.}
	\label{fig::paiModelStudy}
\end{figure}

For thin sensors, the PAI model of Geant4~\cite{PAI} can be used to improve the accuracy of energy deposition. This increases simulation time, so it should be determined whether it is necessary for a setup before simulating a large number of events. A study of the number of deposited charges was carried out with three different sensor thicknesses, using energy deposited by a 5~GeV electron beam with the PAI model activated or deactivated. The results of the study are shown in Figure~\ref{fig::paiModelStudy}, with 250~000 events per curve.

It can be seen that for a 500~\um{} thick sensor, the PAI model has little impact on the deposited charge. At thicknesses below 50~\um{}, however, the impact becomes increasingly significant. The sensors studied in this project are most commonly thinned, and simulated with a total thickness of 50~\um{}. The PAI model is therefore activated in all simulations presented in Section~\ref{sec::studies}.





\section{Sensor performance studies}
\label{sec::studies}

Detailed studies of sensor behaviour can be performed by using the simulation procedure outlined above. In this section, some example results of simulation studies are shown. The simulated sensors are of all three layouts described in Section~\ref{sec::rectAndHex} (the \emph{standard}, \emph{n-blanket}, and \emph{n-gap} layouts), and performance comparisons are made between them. The gap size in the n-layer for the presented \emph{n-gap} layout studies is always 2.5~\um{}.




\subsection{Cluster size and total charge}
\label{sec::clusterChargeAndTotalCharge}

\begin{figure}[h]
\centering
\begin{subfigure}{.5\columnwidth}
    \includegraphics[width=\linewidth]{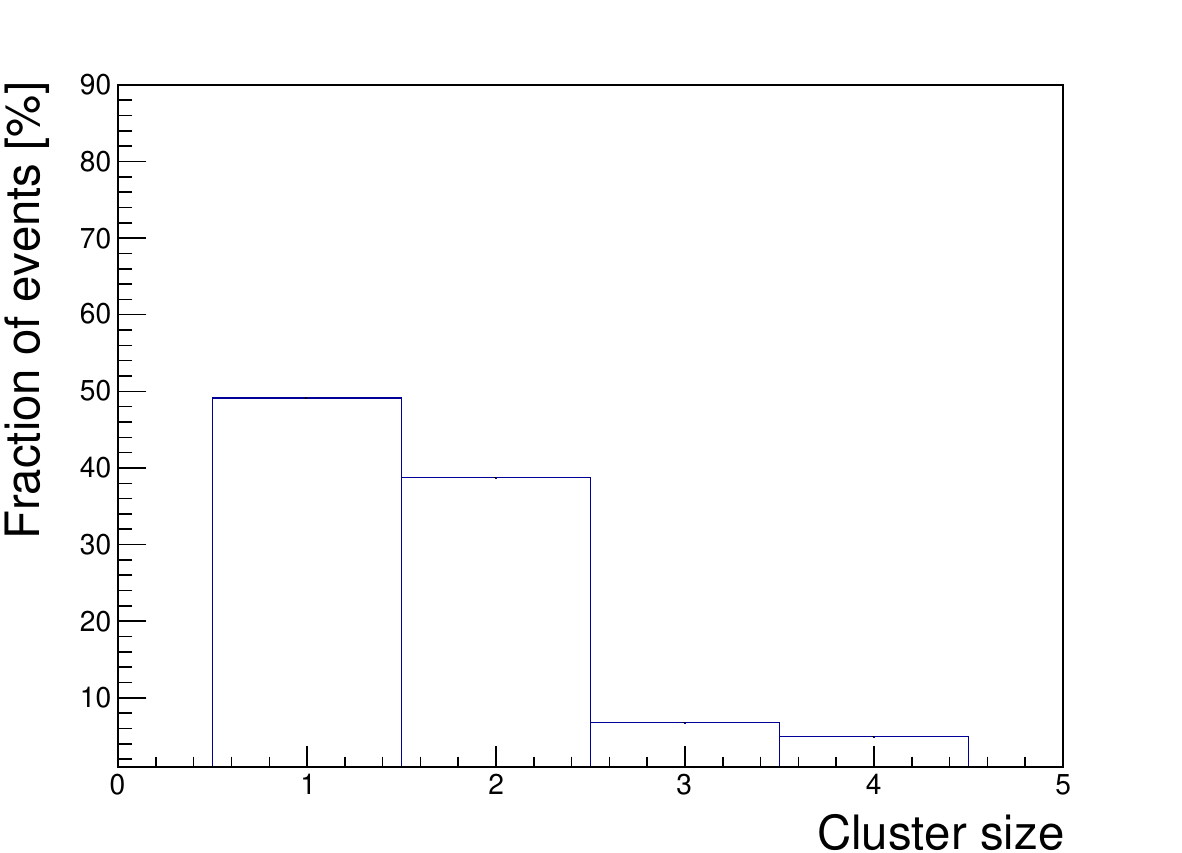}
  \caption{\emph{Standard} layout}
  \label{fig::clusterSizeStd}
\end{subfigure}%
\begin{subfigure}{.5\columnwidth}
  \centering
  \includegraphics[width=\linewidth]{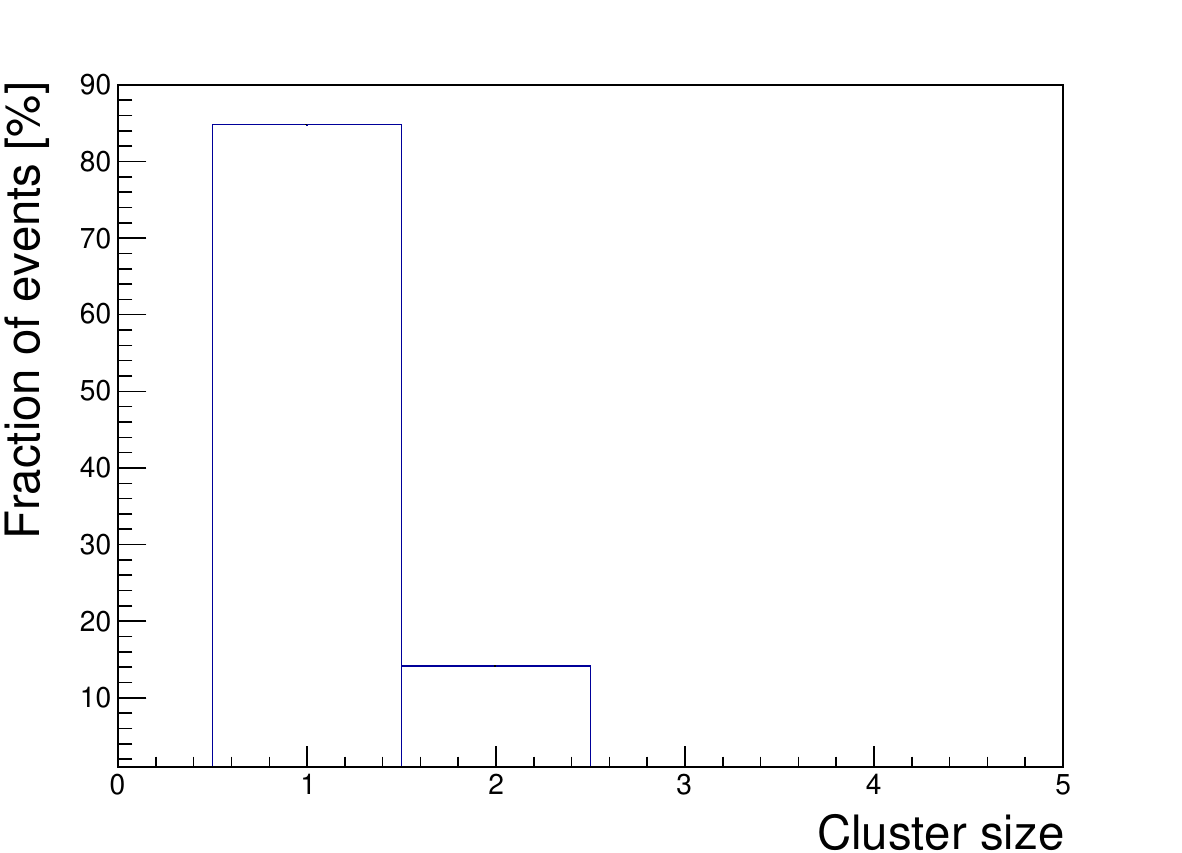}
  \caption{\emph{N-gap} layout}
  \label{fig::clusterSizeNgap}
\end{subfigure}
\caption{Cluster size distributions for the \emph{standard} and \emph{n-gap} layouts, with a $25 \times 25$~\umsq{} pixel size, at a threshold of 100~electrons.}
\label{fig::clusterSizeComp_4p8V}
\end{figure}

The epitaxial layer in the investigated sensors is thin (approximately 10~\um{}), so the collected charge from an event where a minimum ionising particle traverses the sensor is expected to be relatively small. In the presented simulations, a beam of electrons with an energy of 4~GeV is used as a particle source. Pixels with a collected charge exceeding the configured threshold register a hit, and adjacent pixels that register hits in an event are grouped together in a cluster. Figure~\ref{fig::clusterSizeComp_4p8V} shows the cluster size distributions for the \emph{standard} and \emph{n-gap} layouts, for a $25 \times 25$~\umsq{} pixel size at a threshold of 100~electrons. The bias voltage used in this study is $-4.8$~V, and the histograms are the result of 500~000 single-particle events.

For the \emph{n-gap} layout, the majority of events have only one pixel per cluster, while the fraction of larger clusters is larger for the \emph{standard} layout. This is expected, as the \emph{standard} layout has an undepleted region under the p-well, which leads to more charge movement by diffusion and thus more charge sharing between pixels. The gap in the \emph{n-gap} layout generates a lateral electric field that pushes charges towards the pixel centre, reducing the charge sharing and thus also the cluster size.

\begin{figure}[h]
\centering
\begin{subfigure}{.5\columnwidth}
    \includegraphics[width=\linewidth]{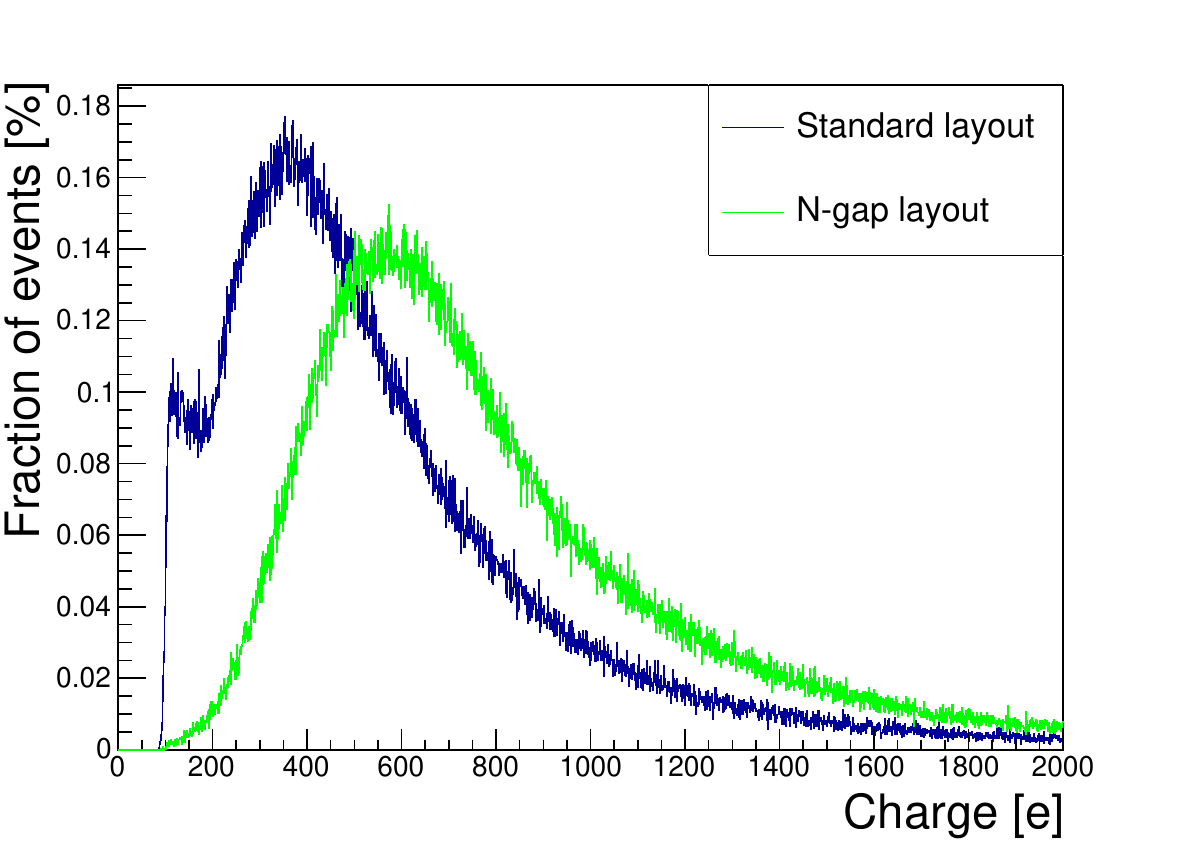}
  \caption{Cluster charge}
  \label{fig::clusterCharge}
\end{subfigure}%
\begin{subfigure}{.5\columnwidth}
  \centering
  \includegraphics[width=\linewidth]{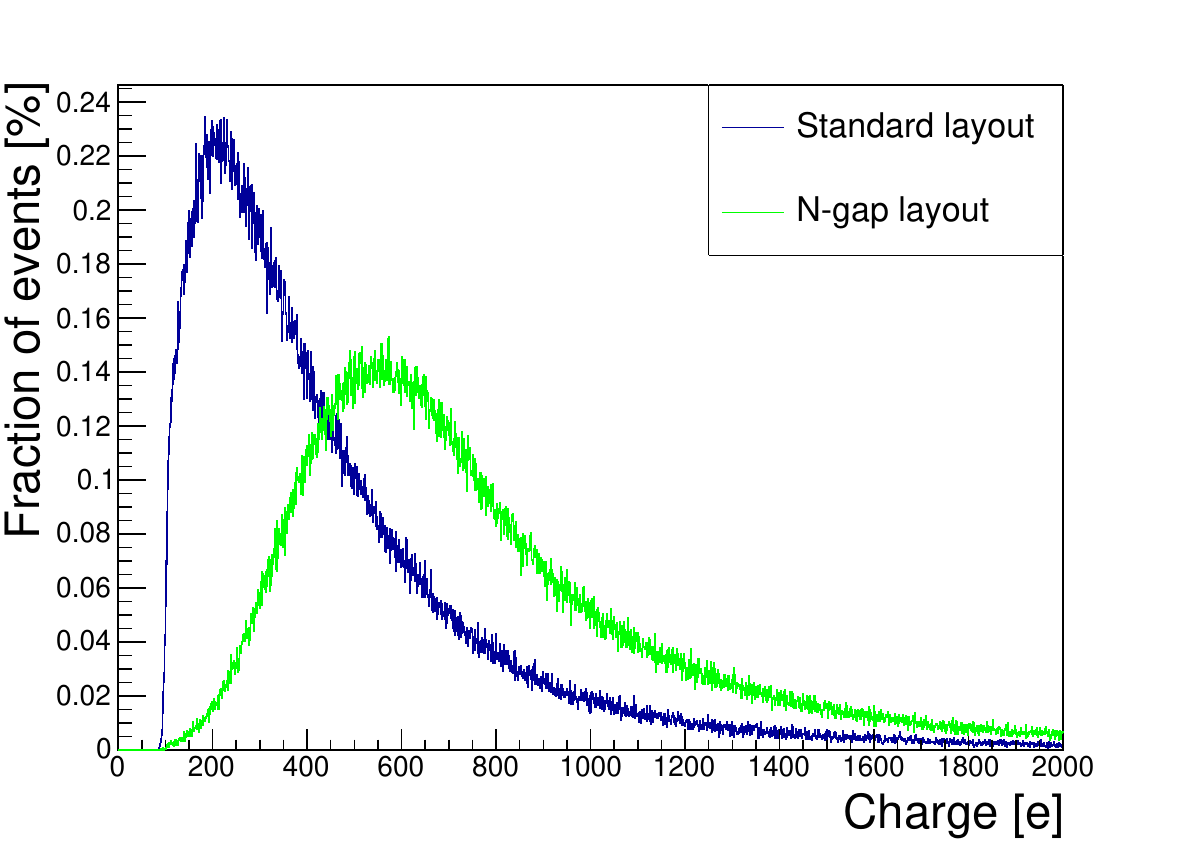}
  \caption{Cluster seed charge}
  \label{fig::clusterSeedCharge}
\end{subfigure}
\caption{Cluster charge distributions for the \emph{standard} and \emph{n-gap} layouts, with a $25 \times 25$~\umsq{} pixel size, at a threshold of 100~electrons.}
\label{fig::clusterChargeComp_4p8V}
\end{figure}

Figure~\ref{fig::clusterChargeComp_4p8V} shows the cluster charge and cluster seed charge distributions for sensors in the \emph{standard} and \emph{n-gap} layouts in the same configuration. The cluster seed charge is the highest charge of a single pixel in a cluster, while the cluster charge is the sum of charges above threshold of all pixels in the cluster.

The distributions all have a clear Landau-like shape. The ``double-peak'' structure in the cluster charge of the \emph{standard} layout is an effect of the threshold and cluster size; as the threshold is 100~electrons, cluster charge values below $2 \times 100$~electrons can only come from events where only the signal of a single pixel exceeds the threshold, and single-pixel clusters with such a low charge are rare.
The cluster charge for the \emph{n-gap} layout is generally higher than the one for the \emph{standard} layout, indicating a more complete charge collection. The cluster seed charge is also higher, which is a compound effect of more complete charge collection and reduced charge sharing between pixels. The difference between the layouts is expected, and corresponds to results for similar sensor layout modifications in other CMOS imaging technologies~\cite{ngapLayout}.

\subsection{In-pixel studies}

\begin{figure}[h]
\centering
\begin{subfigure}{0.5\columnwidth}
    \centering
    \includegraphics[width=\columnwidth]{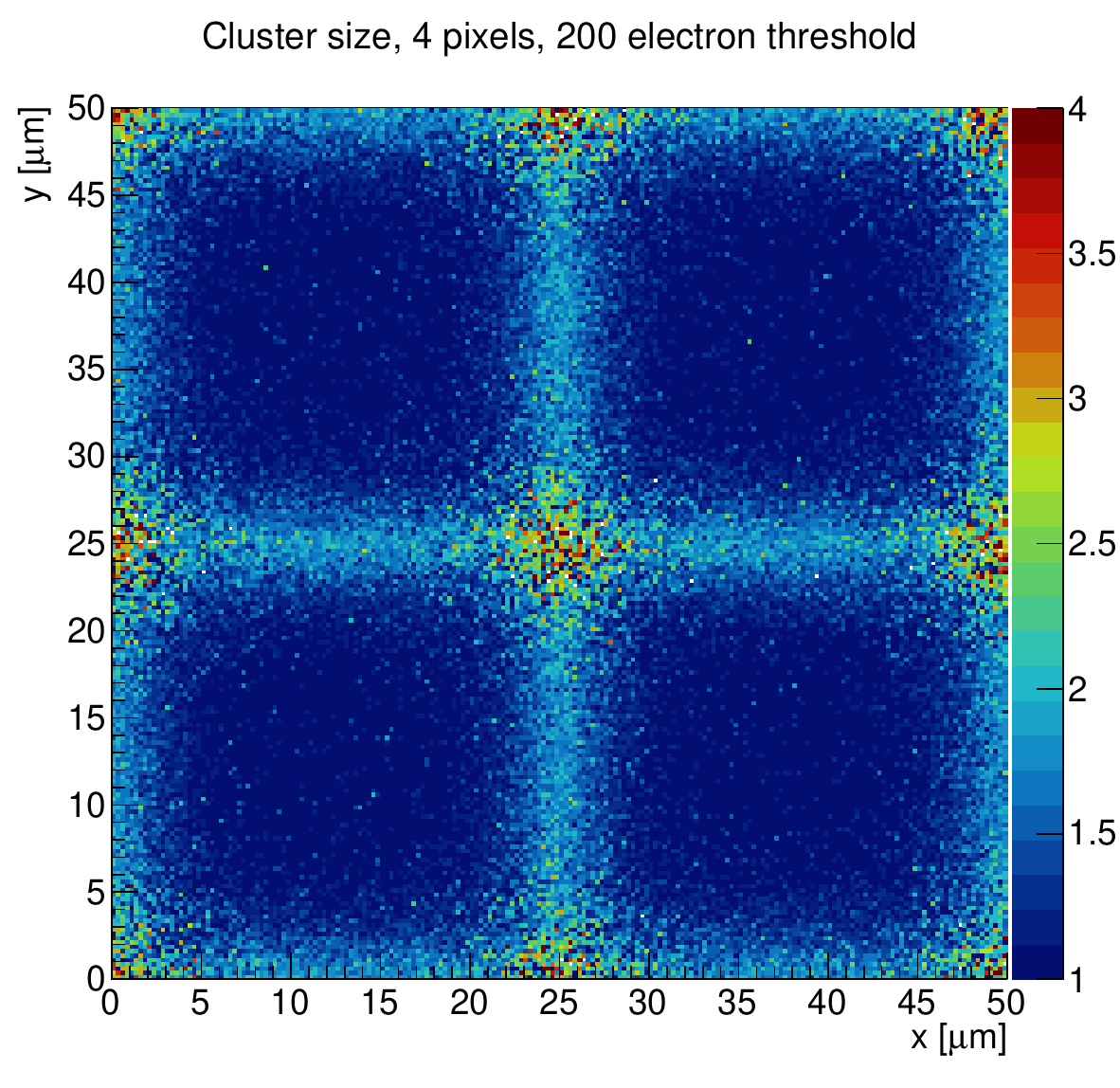}
    \caption{\emph{Standard} layout}
    \label{fig::inPixelClusterSize_std_4p8V}
\end{subfigure}%
\begin{subfigure}{0.5\columnwidth}
    \centering
    \includegraphics[width=\columnwidth]{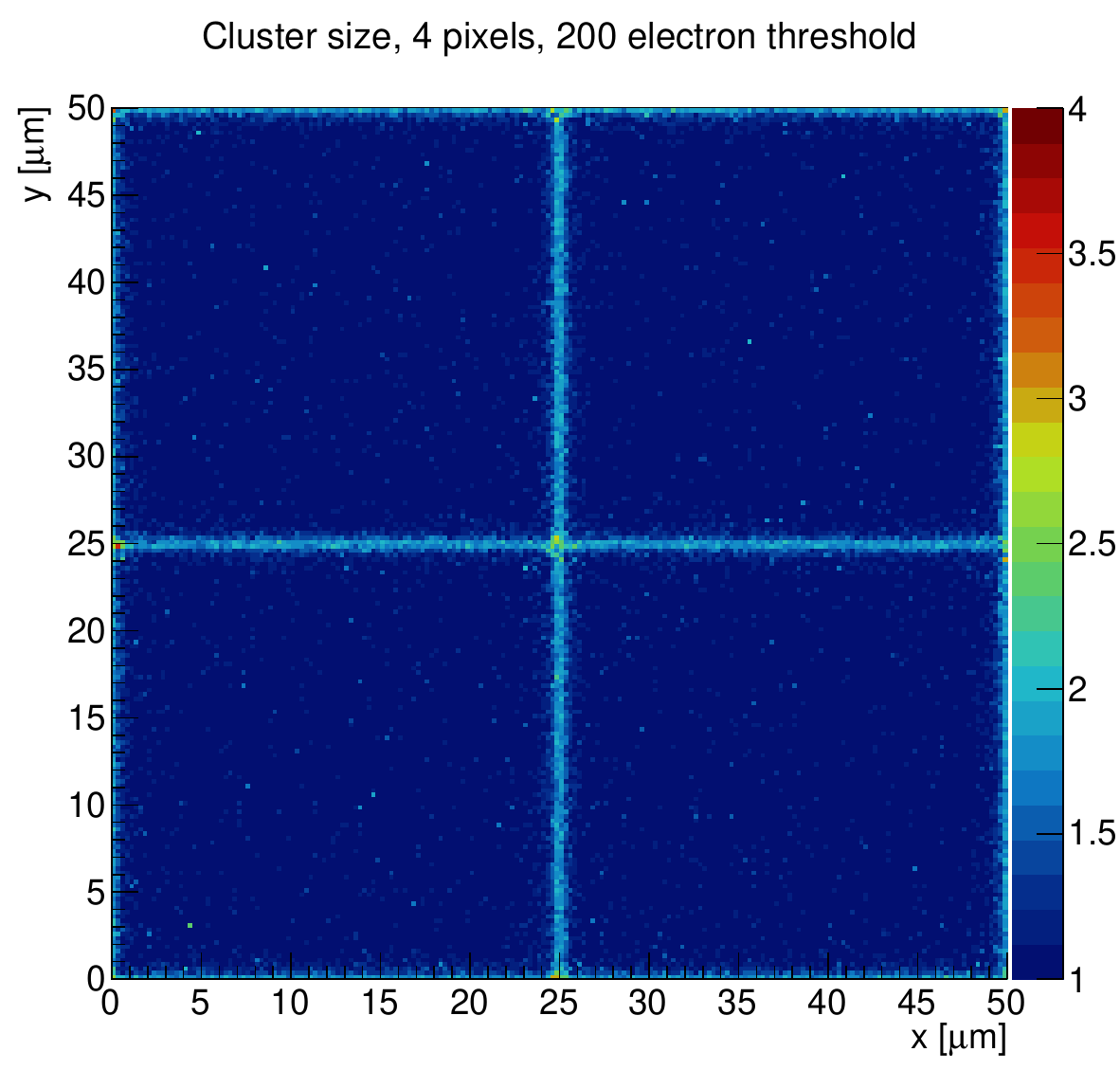}
    \caption{\emph{N-gap} layout}
    \label{fig::inPixelClusterSize_ngap_4p8V}
\end{subfigure}
\caption{In-pixel cluster size for four adjacent pixels with a pixel size of $25 \times 25$~\umsq{}, at a bias voltage of $-4.8$~V and a threshold of 200~electrons.}
\label{fig::inPixelClusterSize_4p8V}
\end{figure}

The combination of TCAD and Monte Carlo simulations allows for a large number of events to be simulated, enabling high-statistics studies of observables for different in-pixel particle hit positions. By using the Monte Carlo truth position of an impinging particle, detailed in-pixel response maps can be produced. 
Figure~\ref{fig::inPixelClusterSize_std_4p8V} shows a map of four adjacent pixels in the \emph{standard} layout, with the mean cluster size for each hit position shown on the z-axis. The pixel size is $25 \times 25$~\umsq{}, and the displayed threshold is 200 electrons.
Figure~\ref{fig::inPixelClusterSize_ngap_4p8V} shows the corresponding result for the \emph{n-gap} layout.

The cluster size for the \emph{n-gap} layout is 1 in the majority of the displayed area. Charge sharing only occurs in the region closest to the pixel edges, due to the lateral electric field introduced by the gap in the n-layer pushing charges from the edges towards the collection electrodes. The \emph{standard} layout has a much larger region where charge sharing occurs, leading to a mean cluster size above 1. 

The larger mean cluster size is a contributing factor to a reduction of seed pixel charge (cf. Figure~\ref{fig::clusterSeedCharge}). It is also a contributing factor to a reduction of the efficiency. A higher cluster size enables an improved spatial resolution, however, due to the possibility to interpolate the reconstructed position between all pixels in the cluster.

\begin{figure}[htb]
\centering
\begin{subfigure}{0.5\columnwidth}
    \centering
    \includegraphics[width=\columnwidth]{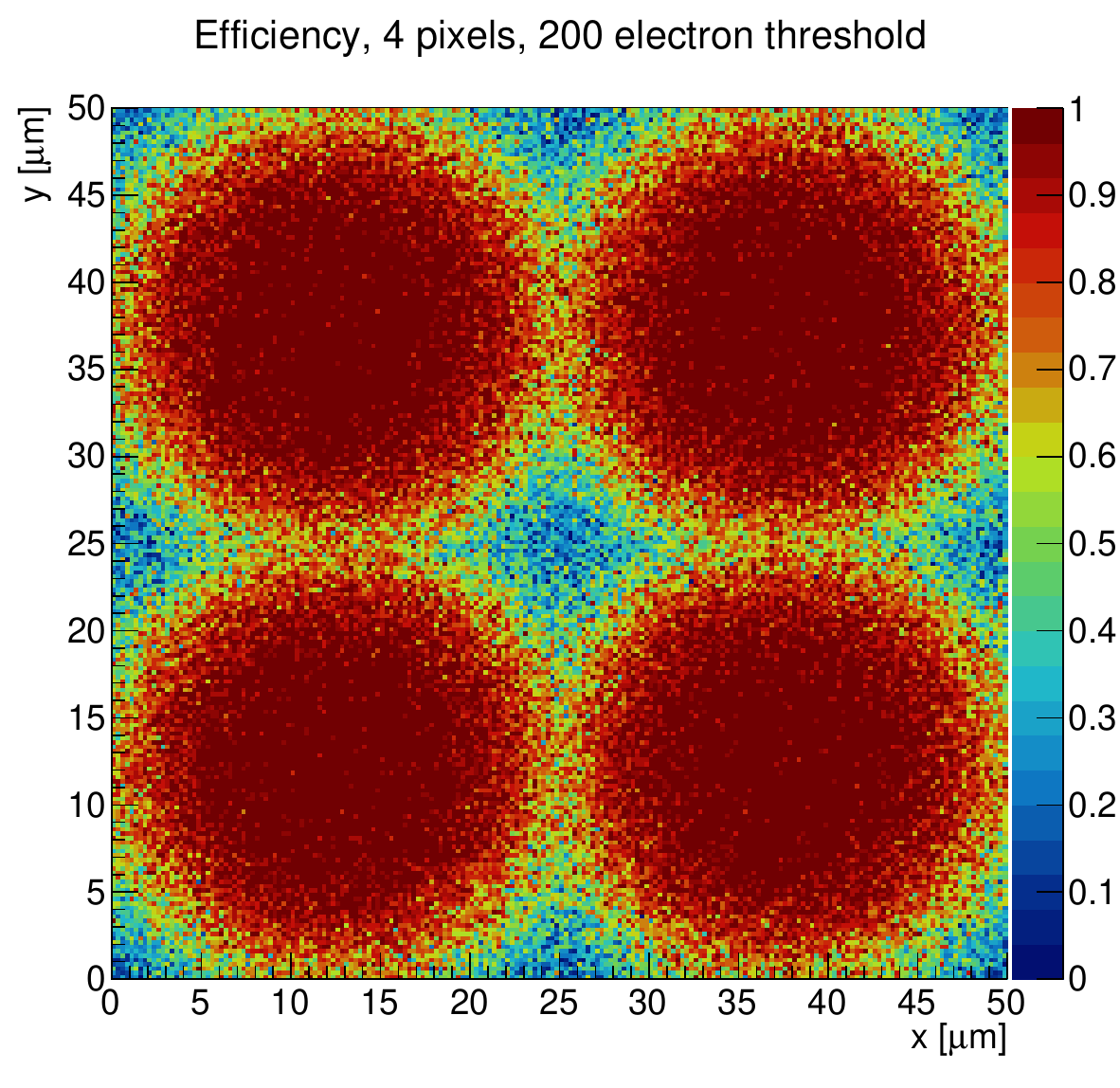}
    \caption{\emph{Standard} layout}
    \label{fig::inPixelEfficiency_std_4p8V}
\end{subfigure}%
\begin{subfigure}{0.5\columnwidth}
    \centering
    \includegraphics[width=\columnwidth]{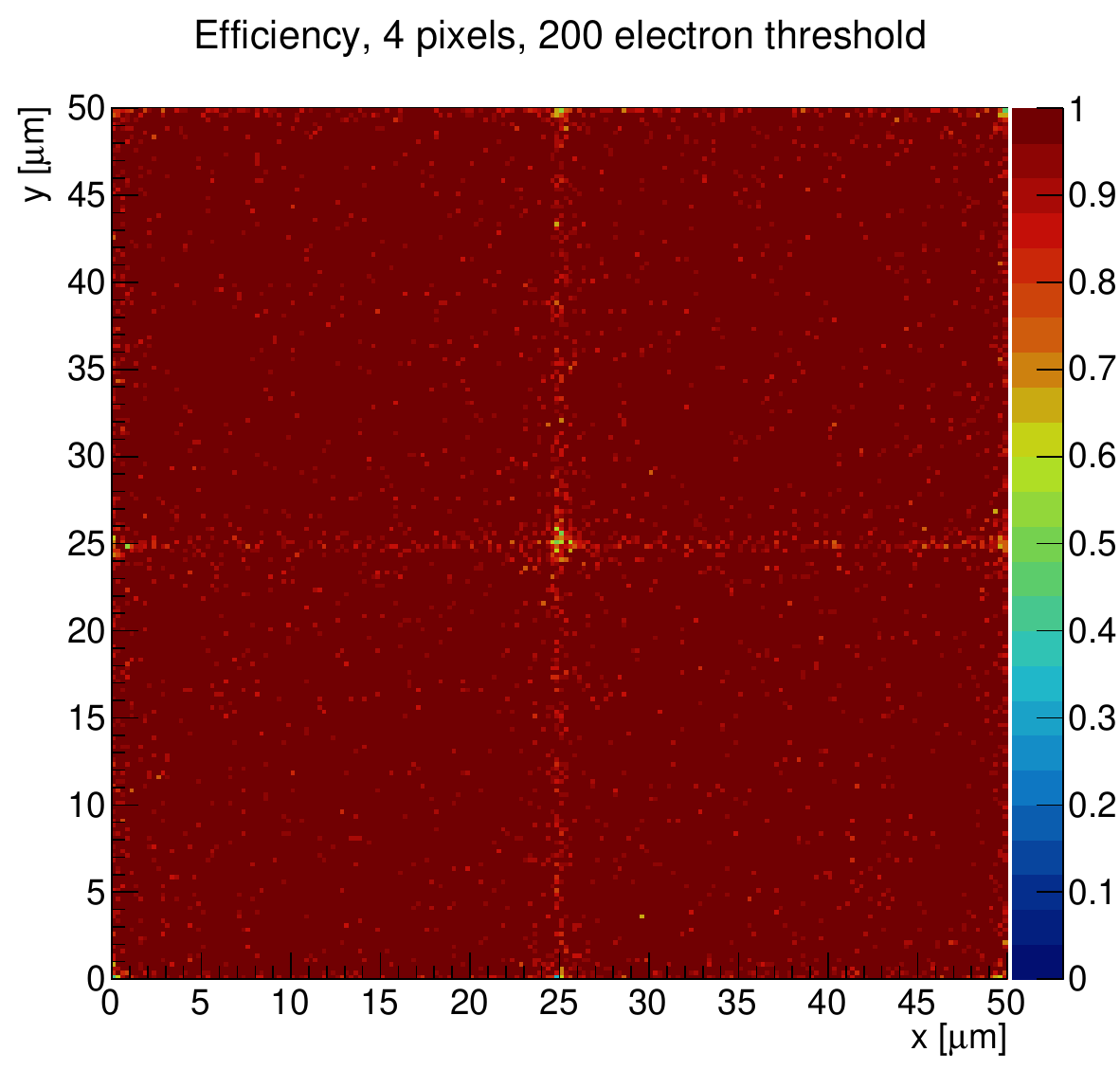}
    \caption{\emph{N-gap} layout}
    \label{fig::inPixelEfficiency_ngap_4p8V}
\end{subfigure}
\caption{In-pixel efficiency for four adjacent pixels with a pixel size of $25 \times 25$~\umsq{}, at a bias voltage of $-4.8$~V and a threshold of 200~electrons.}
\label{fig::inPixelEfficiency_4p8V}
\end{figure}

Figure~\ref{fig::inPixelEfficiency_4p8V} shows in-pixel maps of the mean efficiency, for four adjacent pixels in the \emph{standard} and \emph{n-gap} layouts at a threshold of 200~electrons.
The \emph{n-gap} layout shows a clear improvement of efficiency at pixel edges and corners, compared to the \emph{standard} layout.
The inefficient region in the \emph{standard} layout is expected, as the pixels in this layout are only depleted around the pixel centre. As charge moves predominantly by diffusion outside this region, the movement is slow and charge has a higher probability of being recombined before it reaches a collection electrode, or not being collected within the integration time window (25~ns in this case, corresponding to the LHC bunch crossing frequency). The increased charge sharing due to diffusion also leads to each pixel receiving less charge so that it is less likely that the total collected charge in a pixel will reach above the threshold value. In the \emph{n-gap} layout, the charge collection is more complete, and the efficiency is higher and more uniform. The charge sharing is also reduced, as the pixels are fully depleted and charge moves primarily by drift towards the collection electrodes.

\subsection{Impact of threshold value}
\label{sec::thrScan}

\begin{figure}[h]
    \centering
    \includegraphics[width=\columnwidth]{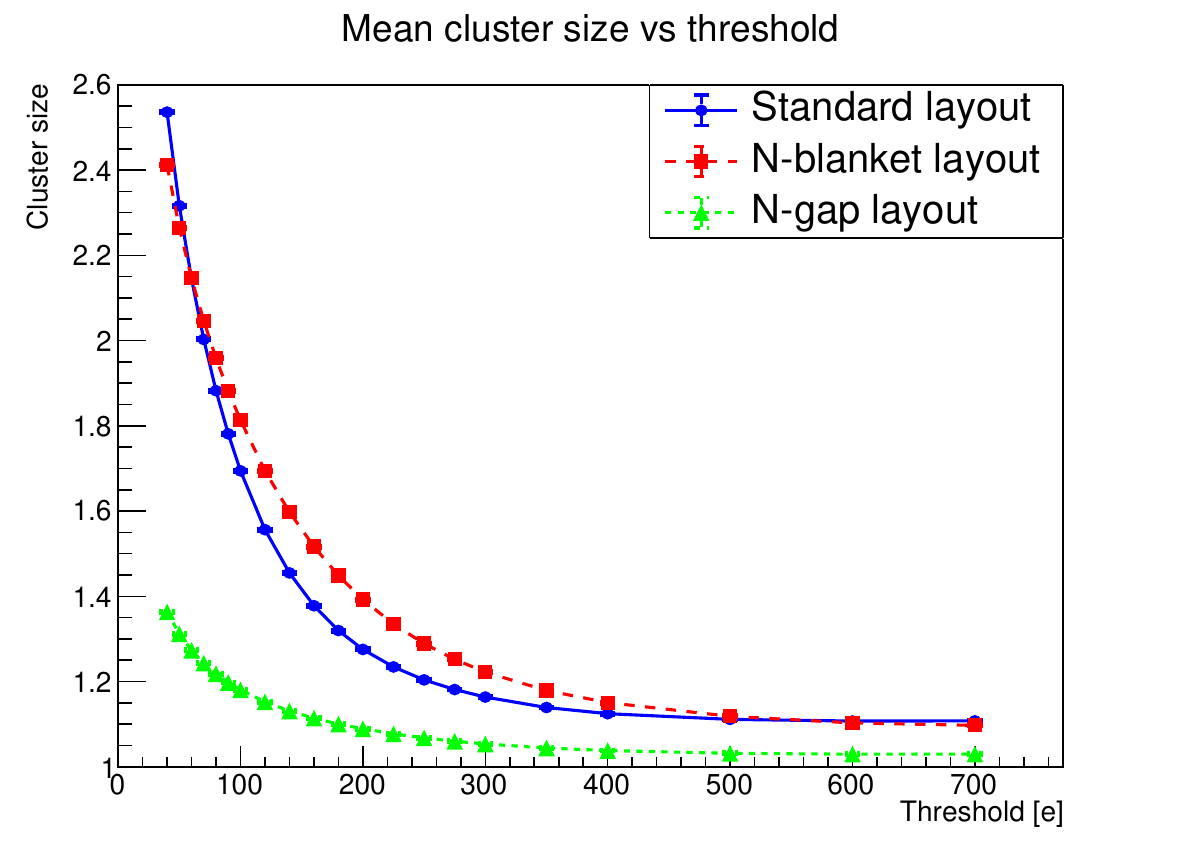}
    \caption{Threshold dependency of the mean cluster size, at a bias voltage of $-4.8$~V for a sensor with a $25 \times 25$~\umsq{} pixel size.}
    \label{fig::stdModNgap_25x25_4p8V_clusterSize_1}
\end{figure}

Studies of the dependence of observables on the detection threshold can be performed by running simulations with different detection threshold values in the digitisation stage.
Figure~\ref{fig::stdModNgap_25x25_4p8V_clusterSize_1} shows the mean cluster size of the sensor versus the threshold value, for the three investigated sensor layouts. The results are presented for a pixel size of $25 \times 25$~\umsq{} at a bias voltage of $-4.8$~V. Each data point in the graphs is the mean result of 500~000 single-electron events, so the statistical errors are small.

The mean cluster size becomes smaller as the threshold increases, as fewer pixels receive a signal above the threshold, especially when the charge is shared among multiple pixels. The cluster size of the \emph{n-gap} layout is significantly lower than for the other two layouts over all tested threshold values, as it has a smaller amount of charge sharing between pixels. The reduction in cluster size as the threshold increases for the \emph{standard} layout is affected by the loss of efficiency in this layout at higher thresholds. The efficiency reduction is largest at pixel corners and edges, which are the main particle incidence areas that lead to a higher cluster size, as can be seen in Figure~\ref{fig::inPixelClusterSize_4p8V}.

\begin{figure}[h]
    \centering
    \includegraphics[width=\columnwidth]{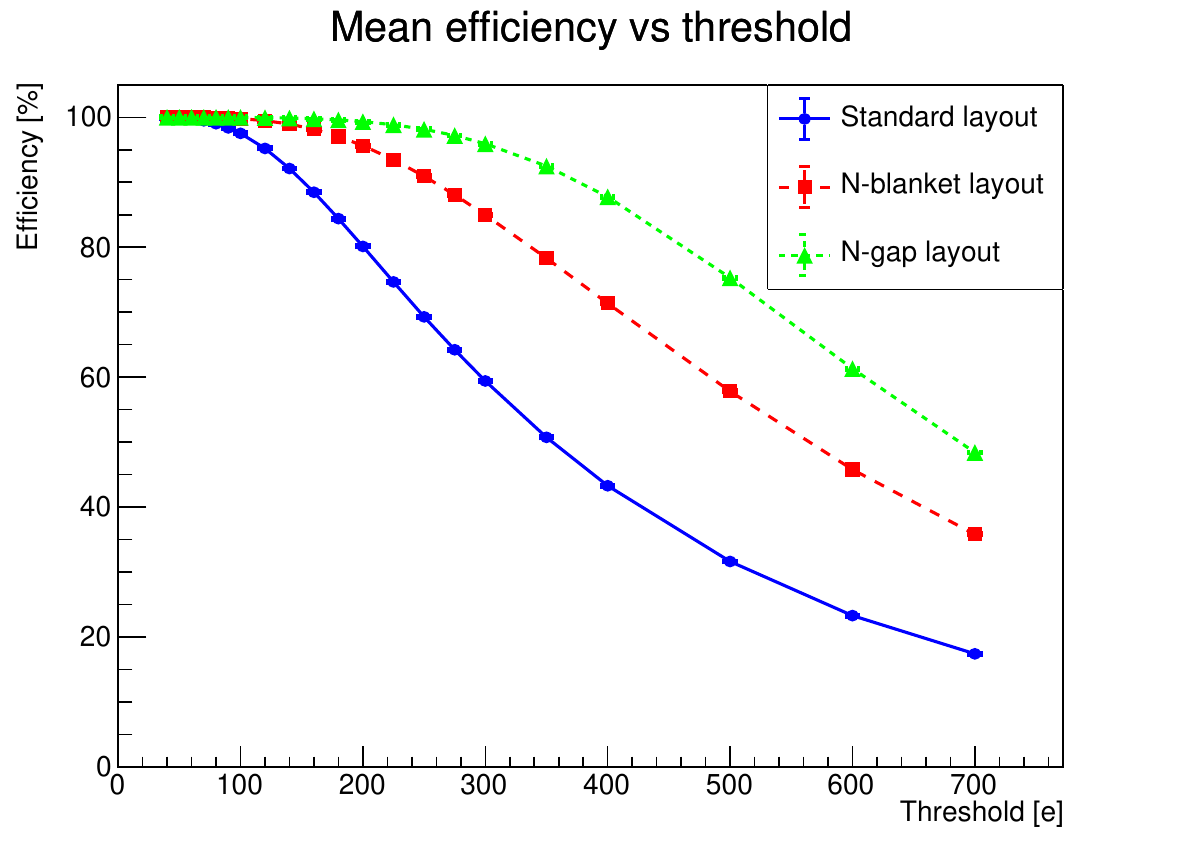}
    \caption{Threshold dependency of the detection efficiency, at a bias voltage of $-4.8$~V for a sensor with a $25 \times 25$~\umsq{} pixel size.}
    \label{fig::stdModNgap_25x25_4p8V_efficiency_1}
\end{figure}

\begin{figure}[h]
    \centering
    \includegraphics[width=\columnwidth]{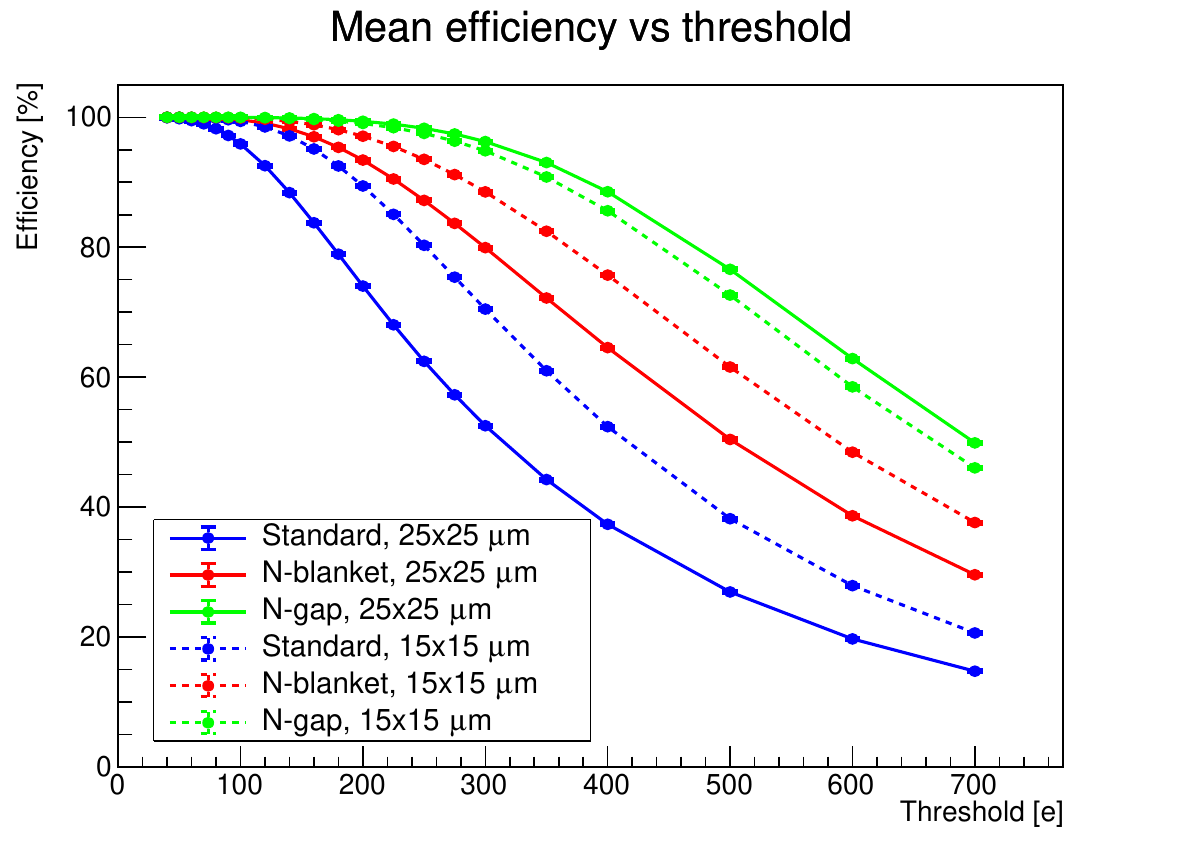}
    \caption{Threshold dependency of the detection efficiency, for sensors with two different pixel sizes at a bias voltage of $-1.2$~V.}
    \label{fig::pixelSizeComp_1p2V_meanEfficiency_1}
\end{figure}

The dependence of the detection efficiency on the threshold value is shown in Figure~\ref{fig::stdModNgap_25x25_4p8V_efficiency_1}, where the mean efficiency value of the sensor is plotted for the three layouts.
%
It can be observed that the \emph{n-blanket} and \emph{n-gap} layouts maintain efficiency over a larger threshold range than the \emph{standard} layout, which is consistent with what is shown in Figure~\ref{fig::inPixelEfficiency_4p8V}. The \emph{n-gap} layout thus enables the largest efficient operating margin. This trend is consistent for all tested bias voltages. For a pixel size of $15 \times 15$~\umsq{} the efficient threshold range is larger for the \emph{standard} and \emph{n-blanket} layouts, compared to the larger pixel size of $25 \times 25$~\umsq{}. This can be seen in Figure~\ref{fig::pixelSizeComp_1p2V_meanEfficiency_1}, where results are shown for a bias voltage of $-1.2$~V. For the \emph{n-gap} layout, the efficient threshold range is slightly smaller than for the larger pixel size at this bias voltage. This is due to a reduced efficiency in the gap in the n-blanket between pixels; in a smaller pixel this region takes up a larger fractional volume, as the gap size remains the same.
The trend between the different layouts remains the same, with the \emph{n-blanket} layout more efficient than the \emph{standard} layout, and the \emph{n-gap} layout more efficient than the \emph{n-blanket} layout.

\begin{figure}[h]
    \centering
    \includegraphics[width=\columnwidth]{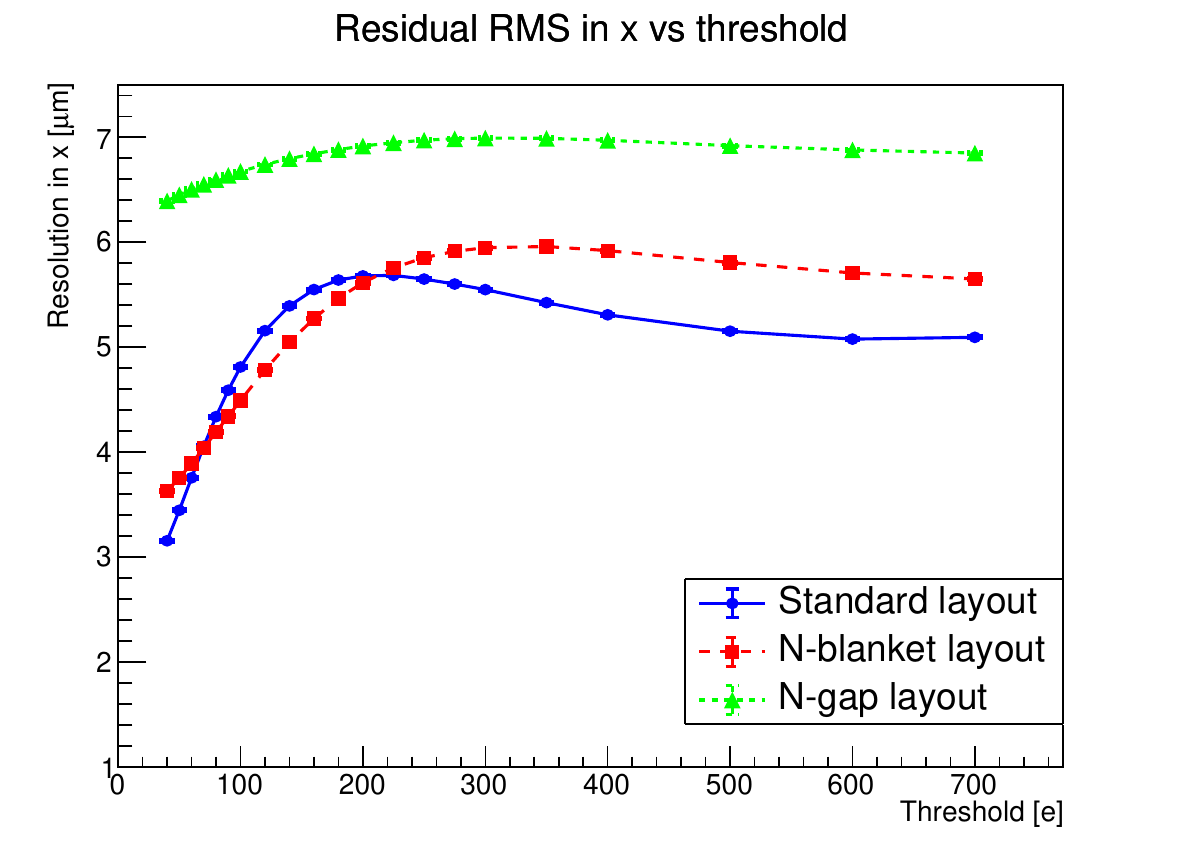}
    \caption{Threshold dependency of the spatial resolution, at a bias voltage of $-4.8$~V for a sensor with a $25 \times 25$~\umsq{} pixel size.}
    \label{fig::stdModNgap_25x25_4p8V_resolution_1}
\end{figure}

Figure~\ref{fig::stdModNgap_25x25_4p8V_resolution_1} shows the mean resolution of the sensor in the x-direction versus the threshold, for the same simulation setup as before. As the pixels are square and symmetric, the resolution is identical in the y-direction. The resolution is defined as the root mean square of the central $3 \sigma$ (99.73\%) of the residual distribution, i.e. the distribution of the difference of reconstructed particle position and Monte Carlo truth position for each event. The reconstructed position is taken as a charge-weighted mean position of all pixel hits in a cluster. In these simulations, the full charge information is used, rather than the value from a charge-to-digital converter with limited resolution.

The resolution deteriorates as the threshold increases, and for a large range of threshold values the \emph{standard} layout has the best (lowest) resolution. At low thresholds, this is due to the larger amount of charge sharing compared to the other two layouts. The reconstructed position is more accurate when the cluster size is larger, due to the charge-weighted position reconstruction occurring between more pixels. At high thresholds, the resolution for the \emph{standard} layout decreases as threshold increases. This is an effect of the reduction of the efficiency, as can be seen in Figures~\ref{fig::inPixelEfficiency_4p8V} and~\ref{fig::pixelSizeComp_1p2V_meanEfficiency_1}. As efficiency is reduced at pixel edges when the threshold increases, only particle hits close to the pixel centre can be reconstructed, and thus the effective pixel size is reduced.

The smaller cluster sizes of the \emph{n-blanket} and \emph{n-gap} layouts deteriorate their resolutions, but as can be seen in Figures~\ref{fig::stdModNgap_25x25_4p8V_efficiency_1} and \ref{fig::inPixelEfficiency_4p8V}, their efficiency is improved. The \emph{n-gap} layout has the highest efficiency, but the largest intrinsic resolution.




\subsection{Transient pulse studies}
\label{sec::apTransient}


Performing transient simulations as described in Section~\ref{sec::transientAP2} can considerably reduce simulation time in comparison to transient simulations using TCAD~\cite{transientMCstudies}. Using Monte Carlo simulations also allows inclusion of stochastic effects, such as Landau fluctuations and secondary particles. 
A validation between both approaches has been performed, utilising the same parameters and setup, to ensure that using \Ap{} mimics the results of TCAD transient simulations as described in Section~\ref{sec::tcadTransient}.
In these validation studies, only an epitaxial layer with a thickness of 10~\um{} was simulated. The mobility model parameter values in \Ap{} were changed to match the extended Canali mobility model used in TCAD~\cite{sdevice_tcad}. 

The simulated geometry consisted of a matrix of ${3 \times 3}$~pixels, with a pixel size of $20 \times 20$~\umsq{}. Charges were injected along a straight line in the corner between four pixels using the \texttt{[DepositionPointCharge]} module, with 63~electron-hole pairs deposited per~\um{}. Electric fields, doping concentrations, and weighting potentials from TCAD were imported into \Ap{} using their respective module readers.

 Using the \texttt{[TransientPropagation]} module as described in Section~\ref{sec::transientAP2}, an \texttt{integration\_time} of 40~ns was used for all simulations. A coarse value of the \texttt{timestep} parameter may lead to smaller pulses than expected, so a \texttt{timestep} of 15~ps was used in the presented results. As the charge is injected in the corner, pulses are expected to be induced in all four pixels sharing the corner. The total pulses were calculated as the average of the induced pulses in the four collection electrodes for each event.

\begin{figure}[hbt]
    \centering
    \begin{subfigure}{.5\textwidth}
        \includegraphics[width=1.0\linewidth]{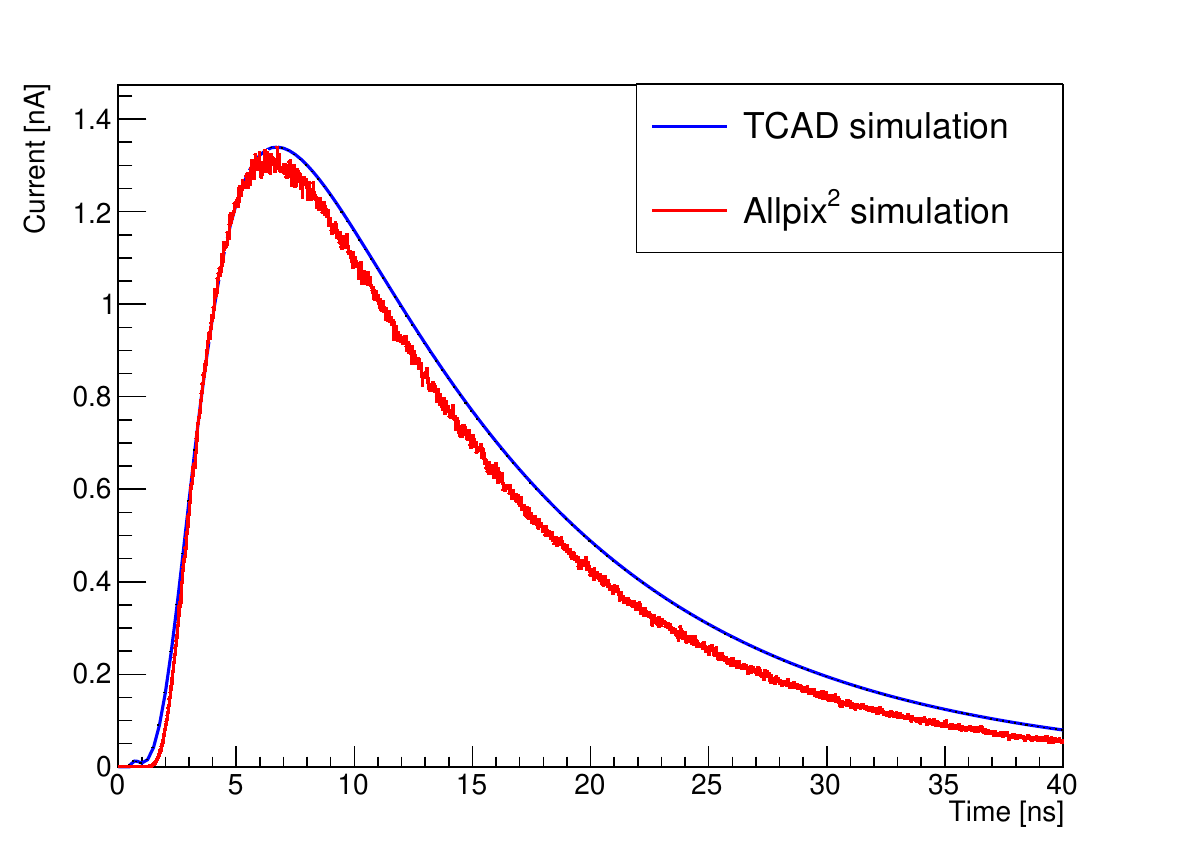}
        \caption{\emph{Standard} layout}
        \label{fig::Transient_Studies_Standard}
    \end{subfigure}
    \begin{subfigure}{.5\textwidth}
        \includegraphics[width=1.0\linewidth]{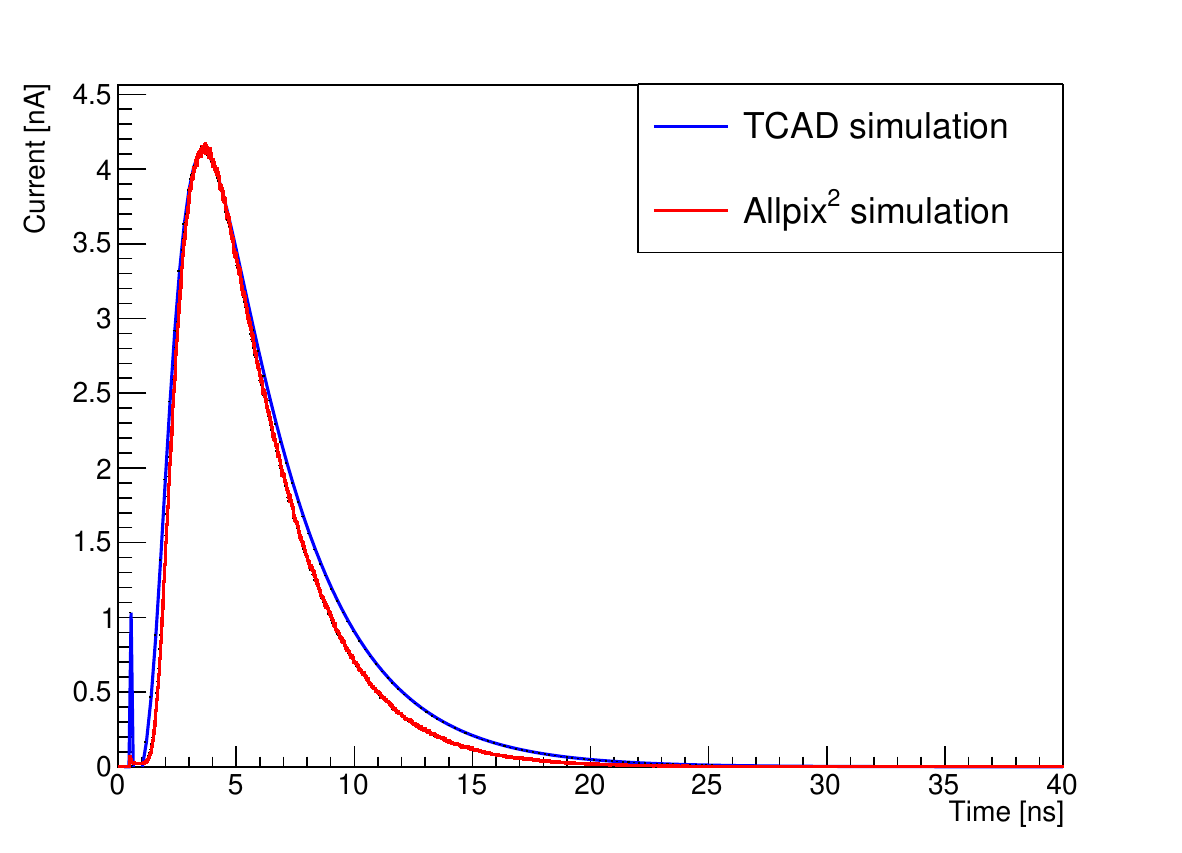}
        \caption{\emph{N-blanket} layout}
        \label{fig::Transient_Studies_NBlanket}
    \end{subfigure}
    \caption{Comparison between pulses obtained with TCAD (blue) and \Ap{} using TCAD fields (red). A TCAD pulse corresponds to a single event, while the \Ap{} pulse is the average of 10~000 events.}
	\label{fig::Transient_Studies}
\end{figure}

Figure~\ref{fig::Transient_Studies} shows the resulting pulses for both TCAD and \Ap{} simulations, for the \emph{standard} and \emph{n-blanket} layouts. The \Ap{} pulses are the average of 10~000~events, whereas the TCAD pulses come from single events.
The plots show that the two methods agree in terms of pulse height and peaking time, which indicates that the \Ap{} method largely yields compatible results with the TCAD method. At the falling edge of the pulses, there is a small difference between the approaches, however. The peak structure in the TCAD pulse for the \emph{n-blanket} layout between 0 and 1~ns is an artefact of the TCAD simulations from the initial charge deposition, and its integral is zero and does not affect the rest of the pulse.

A noticeable difference in the pulse rise time and duration is present between the shown \emph{standard} and \emph{n-blanket} layouts, with the \emph{n-blanket} pulse being faster, which is expected due to the larger depleted region and thus more charge collection by drift. This also increases the charge collection efficiency per pixel and results in a higher peak and higher integrated charge for the \emph{n-blanket} layout.

\subsection{Multi-sensor studies}

Several sensors can be simulated simultaneously in \Ap{}, in for example a beam telescope setup. By using the \texttt{[CorryvreckanWriter]} module, the results of the \Ap{} simulation can be exported in a format suitable for the Corryvreckan test beam reconstruction framework~\cite{corry_paper}. This framework can then be used to extract parameters such as telescope resolution at different positions for the setup. The simulation of multi-sensor setups enables studies of the tracking performance of different setups and sensor designs, and construction of a beam telescope represents a possible use case of the sensors described in this work.

\begin{figure}[h]
    \centering
    \begin{subfigure}{.5\textwidth}
        \includegraphics[width=1.0\linewidth]{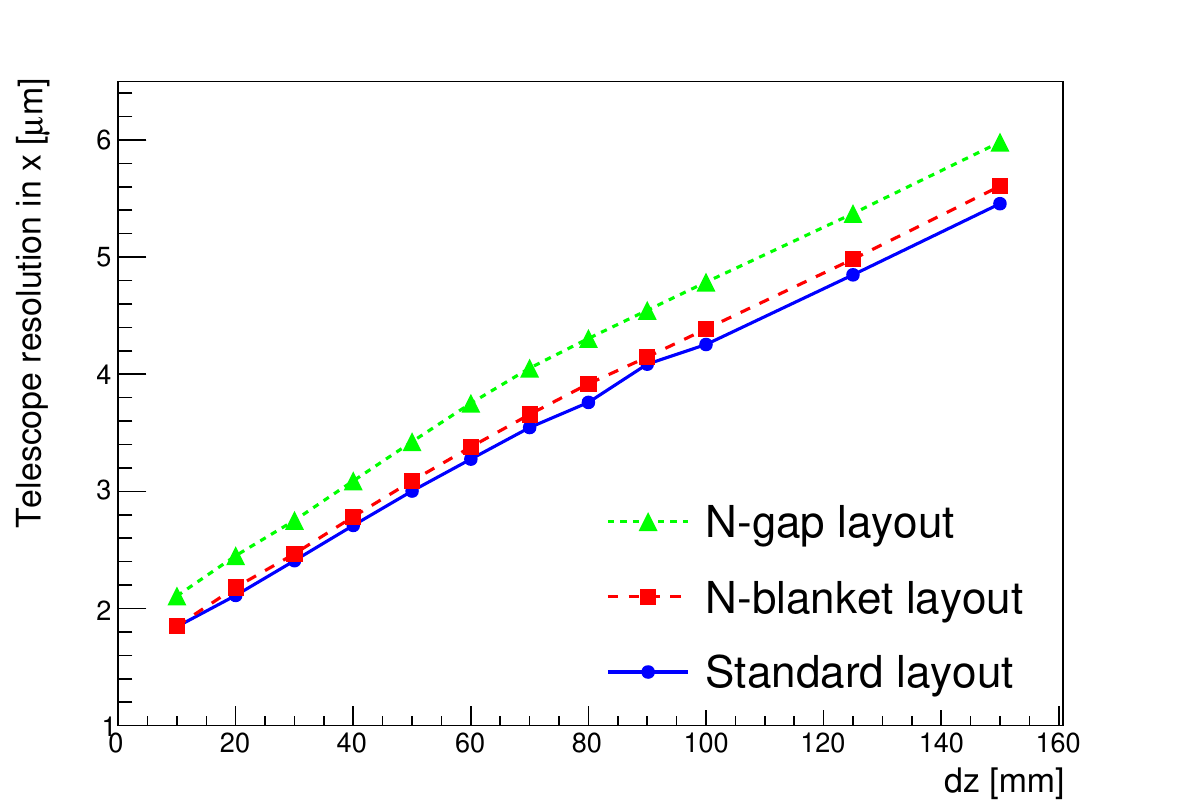}
        \caption{Resolution at the DUT position}
        \label{fig::telescope_layouts_comp_resolution}
    \end{subfigure}
    \begin{subfigure}{.5\textwidth}
        \includegraphics[width=1.0\linewidth]{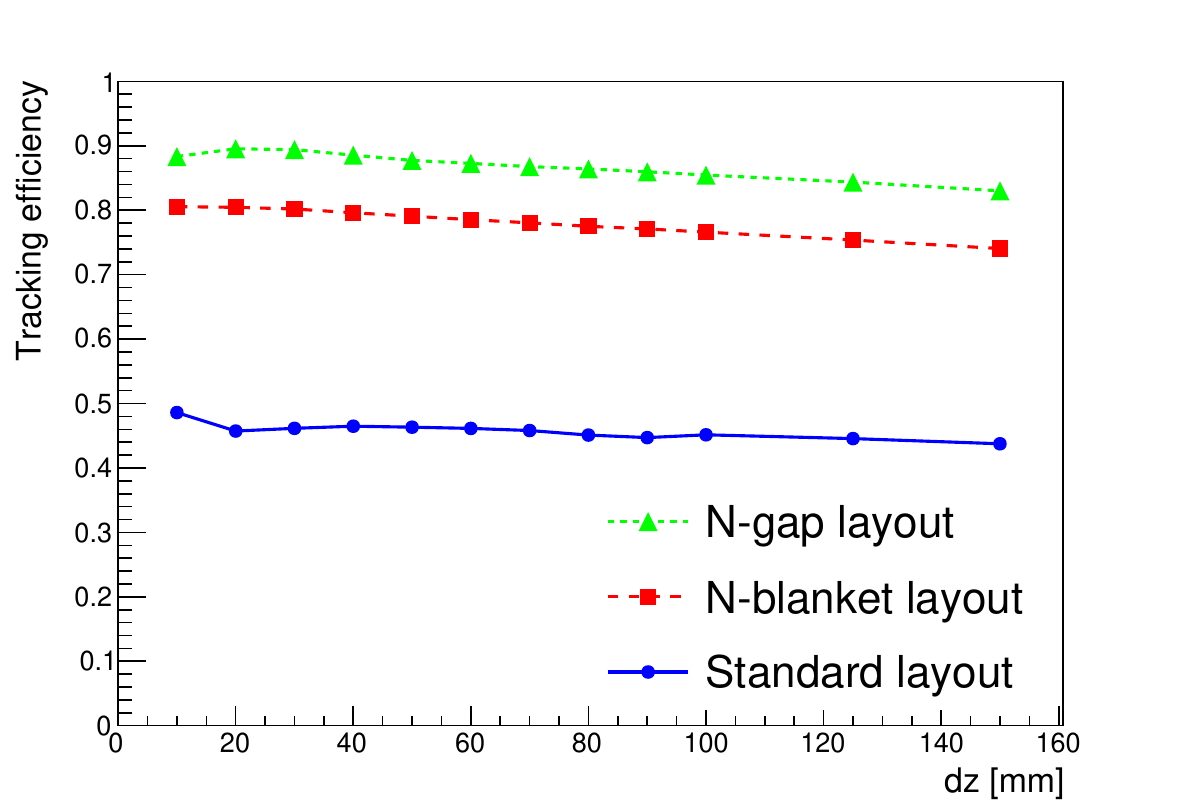}
        \caption{Tracking efficiency estimate}
        \label{fig::telescope_layouts_comp_efficiency}
    \end{subfigure}
    \caption{Telescope resolution at the DUT position, and tracking efficiency estimate, as a function of the distance between telescope planes for the three different layouts at a threshold of 200~electrons.}
	\label{fig::telescope_layouts_comp}
\end{figure}

Simulations were carried out with a six-plane beam telescope surrounded by air, using sensors in the three different layouts with a pixel size of $20 \times 20$~\umsq{}. A beam of electrons was fired at the setup, with a single electron per event. The distance between telescope planes was varied, and the spatial resolution at the device-under-test (DUT) position (in the middle of the setup) extracted. The resolution was determined from the distribution of the difference of track intercept locations and the true particle positions at the DUT, where the tracks were reconstructed using the telescope plane hits and the general broken lines method~\cite{gbl}.
Figure~\ref{fig::telescope_layouts_comp_resolution} shows the resulting telescope tracking resolution at the DUT position for different distances between telescope planes, for the three sensor layouts. The distance is the same between any two adjacent planes, and the presented results are at a threshold of 200~electrons for each of the six sensors.

An estimate of the telescope tracking efficiency is shown in Figure~\ref{fig::telescope_layouts_comp_efficiency}. The calculation is performed by dividing the number of reconstructed tracks by the total number of simulated events for each data point. At least five of the six telescope planes have to register hits for an event to be considered for track reconstruction, and not all such events will have a track successfully reconstructed due to scattering leading to hits outside of the spatial cut. A spatial cut equal to the pixel size (20~\um{}) is used in the final track reconstruction.

From the telescope tracking resolution, it can be seen that the \emph{standard} layout provides the smallest resolution, while the resolution for the \emph{n-blanket} and \emph{n-gap} layouts is slightly larger. This agrees qualitatively with the results shown for single-sensors in Figure~\ref{fig::stdModNgap_25x25_4p8V_resolution_1}, but the difference is smaller than indicated there. As the distance between telescope planes increases, the tracking resolution deteriorates. Both of these effects agree well with expectations; as multiple sensors are used, the resolution of the full system is better than that of an individual sensor, and when the distance between planes increases the scattering in air increases along with the uncertainty in deflection angle. The telescope tracking efficiency estimation qualitatively agrees with the single-sensor results shown in Figure~\ref{fig::stdModNgap_25x25_4p8V_efficiency_1}; at a threshold of 200~electrons, the tracking efficiency is low for the \emph{standard} layout due to the low efficiency of each individual sensor. The \emph{n-blanket} layout shows a significantly higher tracking efficiency, and the \emph{n-gap} layout is the most efficient. The tracking efficiency estimate has a weak dependence on the distance between telescope planes, with a decrease due to the increased scattering in air as the distance between planes increases. In conclusion, the spatial resolution of a beam telescope consisting of the investigated sensors is comparable to that of the EUDET-type beam telescopes~\cite{eudetTelescopePaper}. While the tracking efficiency is low for the \emph{standard} layout, it can be improved without significant loss of tracking resolution by utilising one of the other layouts.

\section{Comparisons to data and previous simulations}
\label{sec::dataComparisons}

Comparisons of the outlined simulation procedure to previously published data are performed, using results from a test beam carried out in the framework of the CLICdp Collaboration on the TowerJazz Investigator~1 sensor, with a pixel size of $28 \times 28$~\umsq{}~\cite{tcadMCcombination}. This sensor is designed in the \emph{standard} layout in a 180\,nm CMOS imaging process with an epitaxial layer thickness of approximately 25~\um{}, and the studies are made at a bias voltage of $-6$~V. The sensor investigated here is thus different from what was previously used as an example in developing the simulation procedure, demonstrating the versatility of the approach.

\begin{figure}[hbt!]
    \centering
    \begin{subfigure}{.45\textwidth}
        \includegraphics[width=1.0\linewidth]{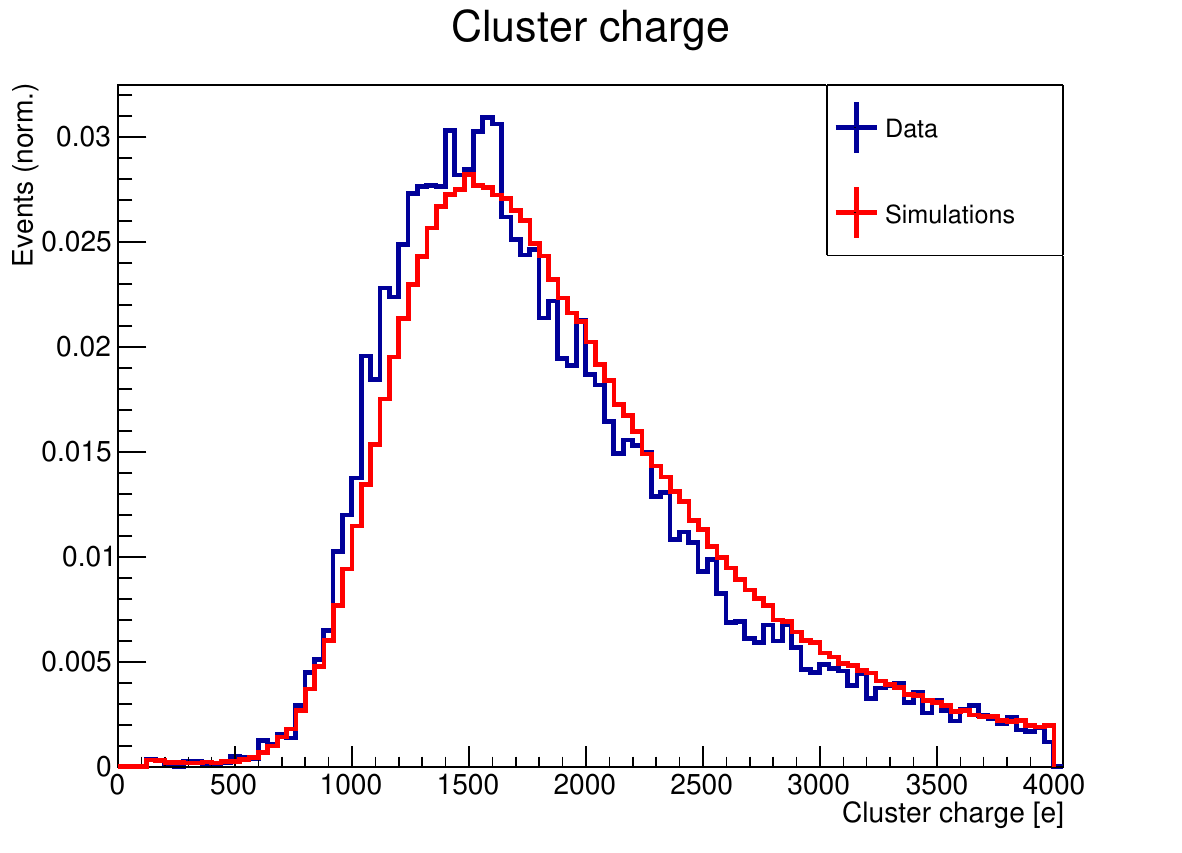}
        \caption{Cluster charge distribution at a threshold of 120~electrons}
        \label{fig::comparison_clustercharge}
    \end{subfigure}
    \begin{subfigure}{.45\textwidth}
        \includegraphics[width=1.0\linewidth]{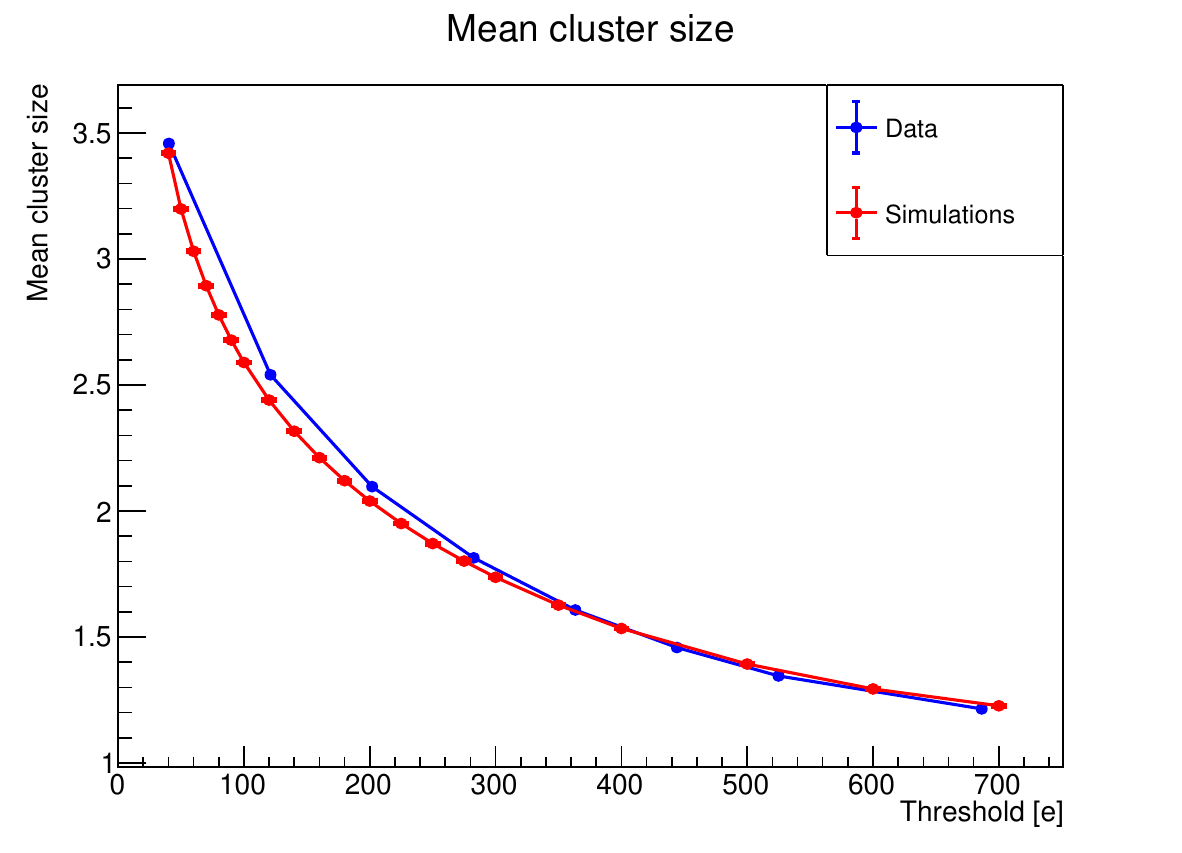}
        \caption{Threshold dependency of the cluster size}
        \label{fig::comparison_clustersize}
    \end{subfigure}
    \caption{Comparison between simulation results obtained using the method described in this paper and test beam data~\cite{tcadMCcombination}.}
	\label{fig::comparisons_TJInvestigator}
\end{figure}

Figure~\ref{fig::comparisons_TJInvestigator} shows comparisons between data taken with the sensor and results using the simulation procedure outlined in this paper. Figure~\ref{fig::comparison_clustercharge} shows the cluster charge at a threshold of 120~electrons, and Figure~\ref{fig::comparison_clustersize} shows the cluster size versus threshold.


In the figures, data are shown in blue and the results of simulations using the generalised procedure outlined in this paper are shown in red. Figure~\ref{fig::comparison_clustercharge} indicates that the simulation result cluster charge is shifted slightly higher compared to the data, while the rising and falling slopes of the distributions match in shape. A fit is performed using a convolution of a Gaussian and Landau function, which gives a most probable cluster charge value of 1.47~kiloelectrons for the simulation results. For the data, the value is 1.42~kiloelectrons~\cite{tcadMCcombination}. The width of the Gaussian part is 0.22~kiloelectrons in the simulation results, and 0.21~kiloelectrons in the data.

In Figure~\ref{fig::comparison_clustersize} the simulations and data match across the full investigated threshold range, with a slight deviation at thresholds smaller than 300~electrons. The errors shown are purely statistical for both data and simulations, and the maximum deviation between the data and simulations is 4\%, at a threshold of 120~electrons. 

Comparative studies have also been carried out in the frame of the Tangerine project, using test beam data for sensors in a 65\,nm CMOS imaging process~\cite{tangerine4}. These studies show an agreement between data and simulations within 1\% for the n-gap layout.

In conclusion, the simulations using the method presented in this paper match data well. There is a maximum deviation of approximately 4\% in both the charge distribution and the cluster size. The qualitative trends agree and to a level sufficient to draw conclusions concerning sensor performance and its origins without use of any proprietary information. The results are also compatible with simulations carried out at CERN using more realistic fields from TCAD, which have been compared to the same data~\cite{tcadMCcombination}.


\section{Summary and outlook}
\label{sec:summary}

In this paper, a simulation procedure for silicon sensors with complex non-uniform electric fields has been described, starting from first principles of a simple pn-junction, and going to high-statistics simulations of a multi-sensor beam telescope. 
Three-dimensional electrostatic TCAD simulations were produced, based on generic doping profiles and first principles of sensor operation, without knowledge of proprietary information. Studies of the impact of varying different sensor parameters have been carried out, observing their impact on the electric fields. Three different sensor layouts have been tested and compared, in several different pixel sizes for both rectangular and hexagonal pixel geometries. The geometries used only describe the large-feature geometry of the sensors, and do not attempt to mimic the intricacies of a CMOS imaging process, but they are sufficient for modelling a signal response describing observed sensor behaviour to an accuracy within a few percent for key observables.

By importing the TCAD fields and doping profiles into \Ap{}, fast and complete simulations of particle interactions and charge transport can be performed, taking stochastic fluctuations stemming from the underlying physics processes into account. Through this process, sensor performance observables such as efficiency, cluster size, and resolution can be extracted. Example results of such simulations have been presented, and agree well with expectations and studies of similar sensors.

Transient simulations have been carried out in both TCAD and \Ap{}, and the results match well. Using the induced charge given by transient simulations is more accurate than using the notion of ``collected charge'', and when charge pulses are available more sophisticated digitisation simulation can be performed to extract accurate values for time-of-arrival and time-over-threshold. This work is foreseen to continue in the near future, also including more accurate simulation of the sensor front-end response.

The described simulation procedure is applicable in multiple different cases, and constitutes a generic toolbox for performing similar studies without using proprietary information. These simulations are able to provide accurate predictions of sensor behaviour and trade-offs with different designs, and can thus be used to inform decisions taken for future sensor designs.





\section*{Acknowledgements}
\label{sec::acknowledgements}
The presented simulation studies have been performed in the frame of the Tangerine project, and in collaboration with the CERN EP R\&D programme.

\section*{CRediT authorship statement}
\label{sec::credit}

\noindent
\textbf{Dominik Dannheim:} Conceptualisation, Resources, Writing - Review \& Editing.
\textbf{Manuel Del Rio Viera:} Formal analysis, Investigation, Writing - Original Draft, Visualisation.
\textbf{Katharina Dort:} Methodology, Resources, Writing - Review \& Editing.
\textbf{Doris Eckstein:} Resources, Funding acquisition.
\textbf{Finn Feindt:} Writing - Review \& Editing.
\textbf{Ingrid-Maria Gregor:} Resources, Funding acquisition. 
\textbf{Lennart Huth:} Methodology.
\textbf{Stephan Lachnit:} Software.
\textbf{Larissa Mendes:} Formal analysis, Investigation, Writing - Original Draft, Visualisation.
\textbf{Daniil Rastorguev:} Writing - Review \& Editing.
\textbf{Sara Ruiz Daza:} Formal analysis, Investigation, Writing - Original Draft, Visualisation.
\textbf{Paul Schütze:} Methodology, Software.
\textbf{Adriana Simancas:} Formal analysis, Investigation, Writing - Original Draft, Writing - Review \& Editing, Visualisation.
\textbf{Walter Snoeys:} Conceptualisation, Writing - Review \& Editing.
\textbf{Simon Spannagel:} Conceptualisation, Methodology, Software, Writing - Review \& Editing, Project administration.
\textbf{Marcel Stanitzki:} Funding acquisition.
\textbf{Alessandra Tomal:} Supervision.
\textbf{Anastasiia Velyka:} Methodology, Software, Formal analysis, Investigation, Writing - Original Draft, Writing - Review \& Editing, Visualisation.
\textbf{Gianpiero Vignola:} Writing - Review \& Editing.
\textbf{Håkan Wennlöf:} Methodology, Software, Formal analysis, Investigation, Writing - Original Draft, Writing - Review \& Editing, Visualisation.

\section*{Declaration of competing interest}
\label{sec::competingInterest}
The authors declare that they have no known competing financial interests or personal relationships that could have appeared to influence the work reported in this paper.
\section*{Funding}
\label{sec::funding}

This work has been carried out within Tangerine, a Helmholtz Innovation Pool Project.

This project has received funding from the European Union’s Horizon 2020 Research and Innovation programme under Grant Agreement No 101004761.

Part of this work has been sponsored by the Wolfgang Gentner Programme of the German Federal Ministry of Education and Research (grant no. 13E18CHA).

\bibliography{bibliography}

\end{document}